 \documentclass[twocolumn,showpacs,preprintnumbers,amsmath,amssymb]{revtex4}
 \usepackage{graphicx}% Include figure files
 \usepackage{dcolumn}% Align table columns on decimal point
 \usepackage{bm}% bold math

 \usepackage{latexsym}
 \usepackage{epsfig}

\def\omunit{{km sec$^{-1}$ kpc$^{-1}$}} 
\def\omun{\mr{km~sec^{-1}~kpc^{-1}}} 
\def\cal{\mathcal} \def\sun{\odot}

 \def\figurewidth{4.5in}
\def\smallfigurewidth{3.5in}

\def\citealtm{\citealt}
\def\citealtm{\citepp}
\def\mr{\mathrm}
\def\citepp{\cite}

\begin{document}

%% \begin{frontmatter}%\doublespace

  \title{\sc The Spiral Structure of the Milky Way, \\ Cosmic Rays, and Ice
 Age Epochs on Earth}

 \author{Nir J. Shaviv}   

% \thanksref{email}
 %\thanks[email]{E-mail: shaviv@phys.huji.ac.il}

 \email{shaviv@phys.huji.ac.il}

% \affil
%\address{
\affiliation{
Racah Institute of Physics, Hebrew University, Jerusalem
 91904, Israel and \\ Canadian Institute for Theoretical Astrophysics,
 University of Toronto, \\ 60 St.~George Str., Toronto, ON M5S 3H8,
 Canada }

%%%\shorttitle{Ice Ages and the Milky Way}
%\shortauthor{N.~J.~Shaviv}

\begin{abstract}

The short term variability of the Galactic cosmic ray flux (CRF) reaching Earth
 has been previously associated with variations in the global low altitude cloud
 cover. This CRF variability arises from changes in the solar wind
 strength. However, cosmic ray variability also arises intrinsically from
 variable activity of and motion through the Milky Way. Thus, if indeed the CRF climate connection is real, the increased CRF witnessed while
 crossing the spiral arms could be responsible
 for a larger global cloud cover and a reduced temperature, thereby
 facilitating the occurrences of ice ages. This picture has been
 recently shown to be supported by various data \citepp{Shaviv2001}. In
 particular, the variable CRF recorded in Iron meteorites appears to vary
 synchronously with the appearance ice ages.

 Here we expand upon the original treatment with a more thorough
 analysis and more supporting evidence. In particular, we discuss the cosmic ray diffusion model which considers the motion of the Galactic spiral arms.  
 We also elaborate on the
 structure and dynamics of the Milky Way's spiral arms. In particular, we bring forth  new
 argumentation using HI observations which imply that the galactic 
spiral arm pattern speed appears to be that which 
 fits the glaciation period and the cosmic-ray flux record extracted
 from Iron meteorites.  In addition, we show that apparent peaks in the star
 formation rate history, as deduced by several authors, coincides with 
 particularly icy epochs, while the long period of 1 to 2 Gyr before 
 present, during which no
 glaciations are known to have occurred, coincides with a significant
 paucity in the past star formation rate. 
\end{abstract}

%% \begin{keyword}
%\keywords 
%keywords from
%http://www.journals.uchicago.edu//ApJ/keywords_text.html
%%\PACS 98.35.Hj, 98.70.Sa, 96.40.Kk, 92.70.Gt, 92.40.Cy
% \PACS 98.35.Hj  92.40.Cy  92.70.Gt  98.70.Sa
\pacs{98.35.Hj  92.40.Cy  92.70.Gt  98.70.Sa \hfill To appear in {\em New Astronomy}}

 %%\end{keyword}

%% \end{frontmatter}

\maketitle

\section{Introduction}
\label{sec:intro}

 It has long been known that solar variability is affecting climate on
 Earth.  The first indication for a solar--climate connection can be
 attributed to William \citet{Herschel}, who found that the price of
 grain in England inversely correlated with the sunspot number.  He
 later suggested that it was due to changes in the solar irradiance
 \citepp{Herschel2}. The irradiance variability is probably not large enough to explain the climatic variability observed by Herschel, nevertheless, synchronous temperature and solar variations do exist. For example, typical surface  temperatures during  northern summers were found to differ by  $0.5^{\circ}$K to $1.5^{\circ}$K  between  solar minima and solar maxima  \citepp{Labitzke1992}.   

 Over the past century, Earth has experienced a gradual, though
 non-monotonic warming.  It is generally believed to be a result of a
 greenhouse effect by anthropogenic fossil fuel emissions.  However, a
 much better fit is obtained if part of the warming is attributed to a
 process, or processes, correlated with the solar activity, thus
 explaining for example, the non-monotonic global temperature change
 \citepp{Christ1991,Soon1996,Beer2000}.  Moreover, the part of the climatic
 variability which is synchronized to the solar activity is larger than
 could be expected from just the 0.1\% typical change in the solar
 irradiance \citepp{Beer2000,Soon2000}.  Namely, the variability in the
 thermal flux itself appears to be insufficient to explain, for
 example, the global temperature variations observed.

  If one goes further back in time, then climatic variability on the
  time scale centuries, is too correlated with solar activity.  Cold
  episodes in Europe such as the Maunder, Sp\"orer and Wolf Minima
  clearly correlate with peaks in the $^{14}$C flux, while warm
  episodes, such as the ``medieval warm'' period during which Vikings
  ventured across the Atlantic, correlate with minima in the $^{14}$C
  flux (e.g., \citealtm{CloudExperiment}).  This flux itself is
  anti-correlated with the solar activity through the solar wind which
  more effectively reduces the Galactic cosmic ray flux that reaches
  Earth (and produces $^{14}$C) while the sun is more active.  On a
  somewhat longer time scale, it was even found that climatic changes
  in the Yucat\'an correlate with the solar activity
  \citepp{Hodell2001} (and possibly with the demise of the Maya
  civilization).  While on even longer time scales, it was shown that
  the monsoonal rainfall in Oman has an impressive correlation with
  the solar activity, as portrayed by the $^{14}$C production history
  \citepp{Neff2001}.

 Two possible path ways through which the solar activity could be
 amplified and affect the climate were suggested. First, solar
 variations in UV (and beyond) are non-thermal in origin and have a much
 larger relative variability than that of the total energy
 output. Thus, any effect in the atmosphere which is sensitive to
 those wavelengths, will be sensitive to the solar activity. UV is
 absorbed at the top part of the atmosphere (at typical altitudes of
 50 km), and is therefore responsible for the temperature inversion
 in the Stratosphere. Any change in the UV heating could have effects
 that propagate downward. In fact, there is evidence that it can be
 affecting global circulations and therefore also climate at lower
 altitudes. For example, it could be affecting the latitudinal extent
 of the Hadley circulation \citepp{Haigh1996}.

 Another suggested proxy for a solar-climate connection, is through
 the solar wind modulation of the galactic cosmic ray flux, as first
 suggested by \citet{Ney1959}.  Ney pointed out that cosmic-rays (CRs)
 are the primary source for ionization in the Troposphere, which in
 return could be affecting the climate.

 First evidence in support was introduced by \citet{Tinsley1991} in the
 form of a correlation between Forbush events and a reduction in the
 Vorticity Area Index in winter months.  Forbush events are marked
 with a sudden reduction in the CRF and a gradual increase over a
 typically 10 day period.  Similarly, \citet{Pudovkin} reported a cloud
 cover decrease (in latitudes of 60N-64N, where it was measured)
 synchronized with the Forbush decreases.  Later, an effect of the
 Forbush decreases on rainfall has also been claimed \citepp{Lebvedev}
 -- an average 30\% drop in rainfall in the initial day of a Forbush
 event (statistically significant to 3$\sigma$) was observed in 47
 Forbush events recorded during 36 years in 50 meteorological stations
 in Brazil.  While in Antarctica, \citet{Egorova} found that on the first day after a Forbush event, the temperature in Vostok station dramatically increased by an average of 10$^\circ$K, but there was no measurable signal in sync with solar proton events. On the longer time scale of the 11-year solar cycle,  an impressive
 correlation was found between the CRF reaching Earth and the
 average global low altitude cloud cover \citepp{Sven1997,Marsh2000}.

 Although the above results are empirical in nature, there are several
 reasons to believe why the cosmic-ray route could indeed be
 responsible for a connection between the solar variability and cloud
 cover.  First, CRs are modulated by the solar activity.  On average,
 the heliosphere filters out 90\% of the Galactic CRs (e.g.,
 \citealtm{Perko1987}).  At solar maximum, this efficiency increases as
 the solar wind is stronger.  Since it takes time for the structure of
 the heliosphere to propagate outward to the heliopause at 50 to 100
 AU and for the CRs to diffuse inward, the CR signal reaching Earth
 lags behind all the different indices that describe the solar
 activity (e.g., the sunspot number or the 10.7 cm microwave flux
 which is known to correlate with the EUV flux).  The cloud cover
 signal is found to lag as well behind the solar activity and it
 nicely follows the lagging CRF. Second, both a more detailed analysis
 \citepp{Marsh2000} and an independent study \citepp{SecondAnalysis}
 show that the correlation is only with the low altitude cloud cover
 (LACC).  Among the different possible causes which can mediate between
 the solar variability and climate on Earth, only Galactic CRs can
 affect directly the lower parts of the atmosphere.  It is the
 Troposphere where the high energy CRs and their showers are stopped,
 and are responsible for the ionization.  EUV variability will affect
 (and ionize) the atmosphere at higher altitudes ($\gtrsim$ 100 km). 
 As mentioned, thermal heating by Ozone absorption could possibly
 affect also low altitudes, however, the CRF-cloud cover connection is
 seen only in low altitude clouds.  Solar CRs (which are less
 energetic than Galactic CRs) are not only stopped at similarly high
 altitudes, the terrestrial magnetic field also funnels them towards
 the poles.  On the other hand, the LACC-CRF correlation is seen
 globally.

 Last, \citet{Tinsley1991} who first found a correlation between
 Forbush events and a reduction in the Vorticity Area Index in winter
 months, showed that these events correlate significantly better with
 the cosmic ray flux than with the UV variations (which generally
 start a week before the Forbush events). They also suggested that the
 UV cannot be responsible for this Tropospheric phenomenon since the
 time scale for the Stratosphere to affect the Troposphere is longer
 than the Forbush--VAI correlation time scale.

 Although the process of how CRs could affect the climate is not yet
 fully understood, it is very likely that the net ionization of the
 lower atmosphere (which is known to be governed by the CRF, e.g.,
 \citepp{Ney1959}) plays a major role, as the ionization of the
 aerosols could be required for the condensation of cloud droplets
 \citepp{Dickinson1975,Kirkby2000,Harrison2000}.  An experiment is
 currently being planned to study the possible cosmic-ray flux --
 cloud-cover connection. It could shed more light and perhaps solidify
 this connection \citepp{CloudExperiment}. Moreover, some physical understanding appears to be emerging \citepp{Yu}. Interestingly, the latter work may explain why the apparent effect is primarily on the lower troposphere and why the global warming of the past century has been more pronounced at the surface than at higher altitudes.

 Hence, the evidence shows it to be reasonable that solar activity
 modulates the cosmic ray flux and that this can subsequently affect
 the global cloud cover and with it the climate.  Assuming this
 connection to be true, {\em we should expect climatic effects also
 from intrinsic variations in the CRF reaching the solar system}.

 With the possible exception of extremely high energies, CRs are
 believed to originate from supernova (SN) remnants (e.g.,
 \citealtm{Longair1994}, \citealtm{Berez1990}). This is also supported
 with direct observational evidence \citepp{SNRs}. Furthermore, since the
 predominant types of supernovae in spiral galaxies like our own, are
 those which originate from the death of massive stars (namely, SNe of
 types other than Ia), they should predominantly reside in spiral
 arms, where most massive stars are born and shortly thereafter die
 \citepp{Drag1999}. In fact, high contrasts in the non-thermal radio
 emission are observed between the spiral arms and the disks of
 external spiral galaxies. Assuming equipartition between the CR
 energy density and the magnetic field, a CR energy density contrast
 can be inferred. In some cases, a lower limit of 5 can be placed for
 this ratio \citepp{SNRs}.

 Thus, when the sun passes through the Galactic spiral arms, an
 increased CRF is expected.  If the CRF-LACC connection is real, this
 will increase the average LACC and reduce the average global
 temperature.  The lower temperatures will then manifest themselves as
 episodes during which ice ages can occur.  Moreover, if the Milky Way
 as a whole is more active in forming stars, more massive stars will
 die and produce CRs.  We show in this work that both these effects
 appear to be supported by various data. 

 \citet{Shaviv2001} studied this conjecture and found evidence which
 supports it, thereby strengthening the possibility of a CRF--climate connection.  In this
 work, we elaborate the original treatment by performing a more
 thorough analysis.  We significantly extend the discussion on the
 dynamics of the spiral pattern of the Milky Way as it is important
 for determining the reoccurrences of ice age epochs, and introduce a
 new argument that helps determine the pattern speed.  We also
 introduce more evidence in the form of an apparent correlation between
 the recorded Milky Way activity (as described by the star formation
 rate, SFR) and the occurrence of ice age epochs.  In particular, it is
 shown that the lack of glaciation activity on Earth between 1 and 2
 Gyr BP (before present) appears to correlate with a dip in the Star
 formation rate in the same period (at least, as obtained by several but not all authors!). 
 On the more speculative side, since the SFR activity may
 correlate with the activity in the LMC and with its estimated
 passages through perigalacticon, ice-ages could be attributed, to some
 extent, to fly-by's of the LMC. 

 It is also interesting to note that other mechanisms have been
 previously proposed to link the Galactic environment with climate
 variability on Earth.  The first such mechanism was proposed long ago
 by \citet{Hoyle1939} who argued that an encounter of the Solar System
 with an interstellar cloud might trigger an ice age epoch by {\em
 increasing} the solar luminosity, which produces an over compensating
 increase in cloudiness.  The increased luminosity is a result of the
 accretion energy released.  However, it is currently believed that
 radiation driving has a positive feedback (e.g., \citealtm{Rind1993}),
 not a strong negative one.  Namely, an increase in the solar
 luminosity will result with an increase of the temperature, not a
 decrease.  Nonetheless, encounters with interstellar clouds could
 still have a temperature reducing effect if sufficient quantities of
 dust grains are injected to the upper atmosphere to partially shield
 the solar radiation \citepp{Yabushita1985}.  These events are more
 likely to occur during spiral arm crossing, since it is there where
 dense molecular clouds concentrate.  However, since they require high
 density clouds, it seems unlikely that they can explain several
 $10^7$ yr glaciation epochs each spiral crossing.

 A second mechanism has to do with the shrinking of the heliosphere. 
 \citet{Begelman1976} have shown that while crossing moderately dense
 ISM clouds with densities of $10^2$ to $10^3$ cm$^{-3}$, the bow
 shock of the heliosphere will be pushed further in than 1 AU. As a
 consequence, the slowing down effect that the heliosphere has on
 Galactic cosmic rays, will cease to work and the flux of Galactic low
 energy CRs will be significantly increased.  On the other hand, the
 charged particles comprising the solar wind will not reach Earth. 
 Either way, the flux of low energy charged particles reaching Earth
 could be significantly altered.  Although these particles are not
 known to have a climatic effect at the moment, such an effect cannot
 be ruled out.  Unlike the previous mechanisms, if this route can
 work, it may require significantly less dense ISM clouds which are
 more frequent.  However, it is still unclear whether a several $10^7$
 yr long glaciation event can be obtained via this route.

 A third mechanism operating mainly during spiral arm crossing
 is the perturbation of the Oort cloud and injection of comets into
 the inner solar system. \citet{Napier1979} and \citet{Alvarez1980}
 discussed the effects that grains injected into the atmosphere by
 cometary bombardment will have on the climate by blocking the solar
 radiation. \citet{Hoyle1978} proposed that a cometary disintegration
 in the vicinity of Earth's orbit would similarly inject grains into
 the atmosphere.

 One should note that these mechanisms all predict ice-age epochs in
 synchronization with the spiral arm crossing. This is counter to the
 model described here in which a phase lag exists.

 There were also proposals that related the Galactic year (i.e., the
 revolution period around the galaxy) to climate on Earth
 (\citealtm{SG1973},\citealtm{Williams1975}, \citealtm{Frakes1992} and references therein).  For example, \citet{Williams1975} suggested that IAEs on Earth are
 periodic, and that this rough $\sim 150$ Myr period is half the Galactic
 year.  Williams raised the possibility that this Galactic-climate
 connection could arise if the disk is tidally warped (e.g., by the LMC), but did not mention a specific mechanism that can translate the warp into a climatic effect. On the other hand, \citet{SG1973} suggested that climatic variability may arise if the solar orbit around the galaxy is eccentric and if, for some unknown physical reason, the solar luminosity is sensitive to the galactic gravitational pull.

Another interesting suggestion for an  extraterrestrial trigger for the ice-age epochs, has been made by \citet{Dilke1972}; (see also \cite{Dilke1974}), who showed that the solar core may be unstable to convective instability under the presence of chemical inhomogeneities induced by the nuclear burning. These authors have argued that both the time scales and luminosity variations involved could explain the occurrence of IAEs. 

In addition to the extraterrestrial factors, there are also terrestrial factors which are in fact most often claimed by the paleoclimatological community to affect climatic variability on geological time scales. These are the continental geography, sea level, atmospheric composition, and volcanic, tectonic and even biological activity. It is likely that, at least to some extent, many of the aforementioned terrestrial and extraterrestrial factors affect the global climate. Therefore, one of the main questions still open in paleoclimatology is the relevant importance of each climatic factor.

 We begin by reviewing the observations and measurements.  These
 include a summary of the glaciation epochs on Earth, the dynamics and
 star formation history of the Milky Way, and the CRF history as
 derived from Iron/Nickel meteorites.  Some of these results are described
 here for the first time.  We then proceed to describe the model which
 relates the Galactic environment to climate on Earth though the
 variability in the CRF, assuming CRs do affect the climate, and follow with the predictions of the model.  The model's backbone is the solution of the problem of CR diffusion while
 incorporating that the CR sources reside primarily in the spiral arms,
 and adding the climatic effect that the CRs may have.  Then, we continue with a
 comparison between the proposed theory and observations.  We show
 that an extensive set of tests employing currently available data
 points to the consistency of the theory.

\begin{table}
\caption{Acronyms}
{ 
\vskip 0.5cm \begin{center}
\begin{tabular}{c l}
\hline
Notation & Definition \\
\hline
BP & Before Present \\
CR & Cosmic Ray \\
cR & Co-Rotation  \\
CRF & Cosmic Ray Flux \\
HI, HII & Atomic, Ionized Hydrogen \\
IAE & Ice-Age Epoch \\
LACC & Low Altitude Cloud Cover \\
LMC & Large Magellanic Cloud \\
MW & Milky Way \\
SFR & Star Formation Rate \\
SN & Supernova \\
\hline
%$D$ & CR diffusion coefficient \\
%$l_{H}$ & Half width of CR diffusion halo \\
%$R$ & Galactocentric radius \\
%$R_{\odot}$ & Solar galactocentric radius \\
%$\kappa$ & Epicyclic angular velocity (as a function of $R$) \\
%$\Omega$ & Galactic angular velocity (as a function of $R$) \\
%$\Omega_\odot$ & Solar galactic angular velocity \\
%$\Omega_p$ & Spiral arm's angular velocity \\
%\hline
\end{tabular}
\end{center}
}
\end{table}

\section{The Observations and Measurements}
\label{sec:observations}

 We begin by reviewing several seemingly unrelated topics: The
 evidence for climatic variability on Earth on a time scale of
 $10^{7}-10^{9}$ yr, as portrayed by the occurrence of ice ages, the
 dynamics of the Milky Way with its spiral structure in particular, as well as 
 the data on CR exposure ages from Fe/Ni meteorites. Some
 of the observational conclusions are essentially quoted ``as is''
 while several results are obtained by analyzing previously published
 data.

\subsection{Earth's glaciation history}

 During the course of Earth's history, the climate has been variable 
 on all time scales ranging from years to eons. Since clear 
 geological signatures are left from periods when Earth was cold 
 enough to have extensive glaciations, studying the occurrence of 
 ice-ages is a good method to quantify long term climatic variability, 
 though it is not the only way (for example, we could have studied 
 the occurrence of ``evaporates'' left during warm periods). We 
 therefore choose to look at the occurrence of ice-ages. 
 
 Before we continue, we should point out that ice-ages on Earth 
 appear on two time scales. Over ${\cal O}(10^{4})$ yr, ice ages come 
 and go. However, epochs during which ice ages can appear or epochs 
 during which ice ages can be altogether absent, exist on time scales 
 of ${\cal O}(10^{7})$ yr. In the rest of our discussion, the term 
 ice-age epoch (IAE) will correspond to these long epochs. Today, we 
 are in the midst of a long IAE, though specifically in a 
 mid-glaciation period between ${\cal O}(10^{4})$ yr long ice-ages.

 Extensive summaries describing the IAEs experienced by Earth
 are found in \citet{Crowell1999} and in \citet{Frakes1992}.  These mostly rely on geological 
evidence of ice ages for the occurrence of glaciations, but not only.   
The nature
 of the glaciations in the Phanerozoic ($0 - 545$ Myr BP) are to a
 large extent well understood.  Partially it is because more recent
 data is more readily available and partially because dating layers
 with fossils is easier.  Moreover, the analysis of \citet{Veizer} who measured the tropical sea surface temperatures over the Phanerozoic serves as an independent analysis from those studying the occurrence of glaciations. As can be seen in figure \ref{fig:epochs} and table \ref{table:iceages}, the different analyses are quite consistent with each other.

The Neo-Proterozoic (1000 - 550 Myr BP) was
 probably an intrinsically cooler period in Earth's history, and
 glaciation was more abundant than in the Phanerozoic.  However, two
 epochs stand out as particularly more glaciated.  These are
 \cite{Crowell1999} the Marinoan and Varangian Glaciations (545 -
 585, 590 - 640 Myr BP) and the Sturtian Glaciations (700 - 750 Myr
 BP).  In the former, the extent of glaciations was particularly
 impressive, with evidence of low latitude sea level glaciations,
 which triggered ideas such as the `Snowball Earth'
 \citepp{Hoffman1995}.  A third, earlier epoch around ca.~900 Myrs BP
 is still very questionable, with some less firm indications pointing
 to it ($\sim 940$~Myr according to \citet{Williams1975}, and ca. 
 $900$ Myr according to others, \cite{Hambrey1985},
 \citealtm{Crowell1999}).  To be conservative, we will not take this epoch in our analysis
 (though it does nicely correlate with a spiral arm crossing if it existed).  Before 1000
 Myr BP, there are no indications for any glaciations, expect for
 periods around 2.2 - 2.4 Gyr BP and 2.9-3.0 Gyr BP. The lack of
 glaciations could be attributed to a changed solar orbit within the
 Galaxy.  However, since the probability for the solar system to
 abruptly change its Galactic orbit is very small, this change which
 occurred at 1 Gyr BP, is more likely to be attributed to intrinsic
 variations in the climate---for example, due to a slow reduction in
 greenhouse gases, or to variations in the MW's average SN rate.  

 The paleoclimatological data  of \citet{Crowell1999}, \citet{Frakes1992}
and  \citet{Veizer} is summarized in table \ref{table:iceages}, together with our adopted
 age for the mid point of the ice-age epochs and its error.  Panel C,D and E in fig.~\ref{fig:epochs} depict a graphical summary of the
 appearance of glaciations in the past 1 Gyr.  

As a big word of caution, one should note that the glaciation data does not come without its caveats. For example, unlike \citet{Frakes1992}, Crowell believes that the data is insufficient to claim periodicity in the occurrence of IAEs. See \S\ref{sec:caveats} and fig.~\ref{fig:epochs} for a detailed summary of the caveats.

\begin{table}
\caption{Ice Age Epochs from geological records (in Myr BP)}
\vskip 0.5cm \begin{center}
\begin{tabular}{c c c c c c }
\hline
\multicolumn{3}{c}{Midpoint of IAEs$^{1}$} & Adopted Age & Spiral Arm & Total \\
Crowell & Frakes    & Veizer         &   and Error$^{2}$ &      Error$^{3}$ & Error \\
\hline
% $\lesssim 15$  & $\lesssim 28$  &  $\sim 144$ & $20 \pm 10$  &   11       &  15   \\
% $\sim 155$     & $\sim 144$         &  $\sim 144$ & $150 \pm 10$  &   11       &  15   \\
% $\sim 325$     & $\sim 293$         &  $\sim 144$ & $309 \pm 20$  &   12       &  24   \\
% $\sim 450$     & $\sim 440$         &  $\sim 144$ & $445 \pm 10$  &   13       &  17   \\
$\lesssim 15$  & $\lesssim 28$  &  $\sim 30$ & $20 \pm 10$  &   11       &  15   \\
$\sim 155$     & $\sim 144$         &  $\sim 180$ & $160 \pm 10$  &   11       &  15   \\
$\sim 325$     & $\sim 293$         &  $\sim 310$ & $310 \pm 20$  &   12       &  24   \\
$\sim 450$     & $\sim 440$         &  $\sim 450 $ & $446 \pm 10$  &   13       &  17   \\
$\sim 595$     & $\sim 588$         & -- & $592 \pm 15$  &   14       &  21   \\
$\sim 730$     & $\sim 765$         & -- & $747 \pm 20$  &   15       &  25   \\
$\sim 900$     & $\sim 940$         & -- & $920 \pm 20$  &   16       &  26   \\
\hline
\end{tabular}
\end{center}
\vskip 0.5cm 
{\small

\noindent 
$^{1}$ \citet{Crowell1999} is the mid point of the epochs with glacial activity. 
\citet{Frakes1992} is the midpoint of the ``ice house" periods, while \citet{Veizer} is the time with the coldest tropical sea temperatures, for which evidence exists in the past 550 Myr.

$^{2}$ The adopted age is the average of \citet{Crowell1999}, \citet{Frakes1992} and \citet{Veizer},
except for the present IAE. Since it is ongoing, its midpoint is
likely to be more recent.  The adopted error is a rough estimate.  It
considers that the error in recent IAEs is smaller and that short
IAEs are much easier to pin point in time.

\noindent
$^{3}$ The ``Spiral Arm Error'' is the error arising from the
epicyclic motion of the solar system. That is, it arises from the non
circular motion that it can have around the galaxy. It effectively
introduces a `jitter' in the predicted location of the spiral arms 
(see appendix B).}
\label{table:iceages}
\end{table}

\subsection{Spiral Structure and Dynamics of the Milky Way}

 The exact pattern speed of the spiral arms of Milky Way and in fact
 the spiral structure itself, is still considered an
 open question.  This is primarily because of our internal vantage
 point inside the Milky Way.  Since these will soon be required, we review
 the current status and analyze the data available.

\subsubsection{Milky Way Spiral Structure}
A review of the different measurements for the spiral structure, and
in particular the number of arms is given by \citet{Vallee,Vallee2002} and by
\citet{Elmegreenbook}.  Vallee concluded that 4 arms are more favorable
than two. Elmgreen concluded that 4 arms appear to govern the
outer part of the Milky Way, while the inner part is much more
complicated.  The problem in the determination of the actual spiral
structure is that distances to objects are either not known accurately
enough or they do not trace the spiral arms unambiguously, if particular
objects are used (for example, HII regions, OB stars or Cepheids).  If
a smooth component is analyzed instead (such as the distribution of
molecular gas) then the distance, which is inferred from velocity
measurements and the Milky Way rotation curve, is a multi-valued
function of the gas velocity within the solar circle.  It is then hard
to disentangle the spiral structure from the observed $l-v$ (longitude
velocity) maps.  The main exception to the above is the HI (or
similar) measurements of gas {\em outside} the solar circle.  Since HI
traces the spiral arms nicely, and since outside the solar circle no
velocity-distance ambiguities exist, the spiral structure can be ``read
off'' the $l-v$ maps straight forwardly.  The result is a clear 4-arm
spiral structure \cite{NewMW,Dame2001}.  We therefore assume that at the solar
galactocentric radius and beyond, the spiral structure of the Milky
Way is that of 4 arms.  This does not imply that further inside the
Galaxy the same 4-arm structure exists.  In fact, we shall show that
there is a good reason for the two structures to be different (which
could explain why until now the picture was confusing).

\subsubsection{The Spiral Arm Pattern Speed -- Previous Results}
\label{sec:ptspeed}

Even less agreed upon and more confusing are the results for the
pattern speed $\Omega_p$ of the spiral arms in the Milky
Way\footnote{A common misconception is that the spirals are ``frozen''
in, such that material in the arms remains in them, and vice versa.  If
this would have been true, the spirals of galaxies should have been
much more tightly wound because of the differential rotation.}.  For
the sake of completeness, we first review previous determinations of
$\Omega_{p}$.  We will afterwards continue with a new analysis which
was previously applied to other galaxies but never to our own.

A survey of the literature reveals that quite a few different analyses
were performed to measure the Galactic spiral arm pattern speed.  Some
methods are local in the sense that they look at local age gradients
of young objects, such as OB stars or open clusters.  These are
presumably methods which rely on the least number of (Galactic)
assumptions.  For example, they should detect the correct pattern
speed irrespective of whether the spiral arm is a density wave or just
a star formation shock wave (without a density wave associated with
it), or irrespective of whether the MW has 2 or 4 spiral arms. 
Unfortunately, these methods tend to be inaccurate because of local
``dispersions'' and inaccuracies in age determinations.

A second type of methods looks at the birth place of objects not as 
young as before. These include, for example, the birth place of open 
clusters with a typical age of a few $10^{8}$ yr. This can in 
principle help place more accurate constraints on the pattern speed. 
However, unlike the previous methods, it requires a model for the 
spiral arms including their number,  their amplitude and pattern 
speed, all of which should be fitted for. In reality, one often 
assumes both that the number of arms and their amplitude are given 
within the context of the density wave theory. Then, several different 
guesses for $\Omega_{p}$ are guessed and the best fit is chosen.

A third type of methods relies on fitting the observed velocities of
stars to a spiral density wave, and in particular the non-circular
residue obtained after their circular component is removed.  The
advantage of this type of a measurement is that it does not rely on
age determinations at all, since it relies on the ``instantaneous''
configuration.  However, the residual kinematics are sensitive to the
rotation curve chosen as well as to the spiral wave parameters.

A fourth type of methods relies upon the identification of resonance 
features expected to arise from the spiral density wave theory. For 
example, \citet{Gordon1978} identifies the observed discontinuity in CO 
emission at about 4 kpc with the inner Lindblad resonance. 

 A summary of the various determinations of $\Omega_{p}$ in the literature is found in
 table~\ref{tab:pattern}.  The main result apparent from the table is
 that most of the values obtained for $\Omega_{p}$ cluster within two
 ranges ($\Omega_\odot -\Omega_p \sim 9 - 13.5$ \omunit and
 $\Omega_\odot-\Omega_p \sim 2.5-5$ \omunit), with a third range being
 either a tail for the second or a real ``cluster'' of results (for
 which $\Omega_\odot -\Omega_p \sim ($-$1) - ($-$4) $ \omunit). 
 Interestingly, the division between the clusters is not a function of
 the method used.  For example, \citet{Palous1977} have shown that two
 equally acceptable values are obtained from the same analysis. 
 Clearly, the results in the literature are still not converged, but
 possible values and unaccepted ranges can be inferred.

%-----------------------------

\subsubsection{Pattern Speed from HI Observations}
\label{sec:ptsHI}

 As previously mentioned, relying on particular objects to identify
 the spiral structure has the disadvantage of distance inaccuracies and
 that these objects do not always trace the spiral arms nicely enough. 
 One the other hand, mapping of various gas components has the
 disadvantage that the velocity-distance ambiguities can complicate
 the analysis significantly.  Therefore, analyzing gas at Galactic
 radii larger than the sun ($R\gtrsim R_{\odot}$) has a clear
 advantage, as it avoids the above ambiguities and uncertainties.

 \citet{NewMW} found that a four armed\footnote{Because of limited
 coverage in Galactic longitude, 3 spiral arms are seen.  2 {\em
 adjacent} arms end within the covered longitude, at the same inferred
 radius of $R_\mr{HI,out} \approx 2 R_{\odot}$.  The separation angles
 imply a $\sim 90^{\circ}$ separation.}  spiral structure in HI extends
 all the way to $R_{\rm HI,out}\approx 2 R_{\odot}$.  (Specifically, they
 found $R_{\rm HI,out}=20$ kpc when taking a rotation curve in which
 $R_\odot = 10$ kpc and $v_\odot = 200~{\rm km/s}$. For more up to date
 rotation curves, the value of $R_{\rm HI,out}$ is lower but still roughly
 twice our galactocentric radius).  Moreover, more upto date HI maps, as traced by CO reveal the  spiral arms outside the solar galactocentric radius even more clearly \cite{Dame2001}, thus reinforcing the results of \citet{NewMW}.
 This observation on the external radius of the galactic spiral arms can be proven useful to
 constrain the pattern speed, {\em if} this 4-arm structure is a spiral
 density wave. 

 According to spiral density wave theory, 4-armed spiral density waves
 can only exist within the inner and outer 4:1 Lindblad resonances 
 (e.g., \cite{Binney1988}). Otherwise, the waves become evanescent. The 
 inner and outer 4:1 Lindblad resonances, $R_{I4:1}$ and 
 $R_{O4:1}$, are defined through:
\begin{equation}
 \Omega_\odot(R_{I4:1}) - {\kappa(R_{I4:1}) \over 4} = \Omega_p =
 \Omega_\odot(R_{O4:1}) + {\kappa(R_{O4:1}) \over 4},
\end{equation}
 where $\Omega(R)$ and $\kappa(R)$ are respectively the rotational frequency and
 the epicyclic frequency at radius $R$.  Therefore, the constraint
 $R_{\rm HI,out} \leq R_{O4:1}$ that the arms should terminate before or at the
 outer Lindblad resonance can be rewritten as
\begin{equation}
    \label{eq:Constraint}
 \Omega_p \leq \Omega_\odot(R_{\rm HI,out}) + {\kappa(R_{\rm HI,out}) \over 4}.
\end{equation}

\begin{figure*}
\begin{center}
\epsfig{file=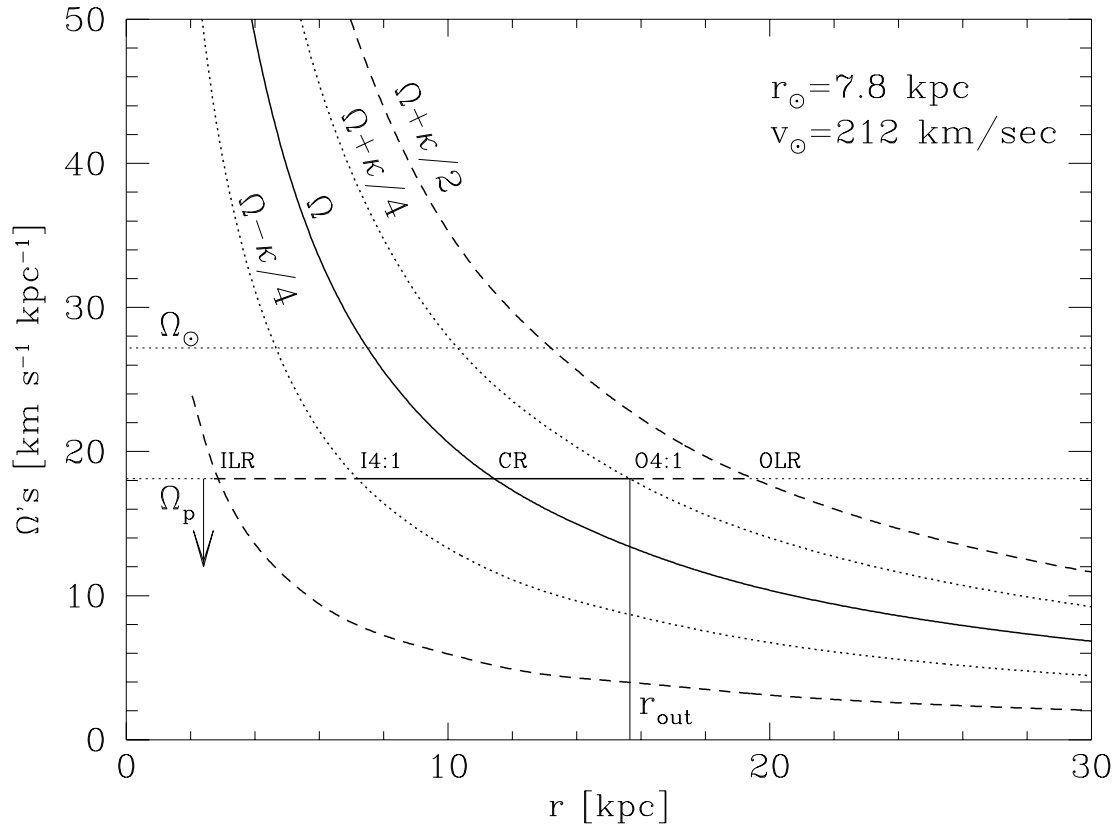,width=\figurewidth,angle=-90}
\end{center}
  \caption{
 %\small \sf 
  The dynamical frequencies of the Milky Way
  relevant to its spiral structure, plotted as a function of 
  galactocentric radius.  Using a
  typical rotation curve (taken from \cite{Olling1998}), one can
  calculate $\Omega$ and the epicyclic frequency $\kappa$. 
  Spiral density wave theory states that a four arm spiral cannot
  exist for $\Omega_p > \Omega+\kappa/4$.  Since a four arm spiral is
  observed in HI to extend out to $r_{\rm HI,out}\approx 2 R_{\sun}$, an
  upper limit on $\Omega_p$ can be placed.  This $\Omega_p$ implies
  that the inner edge of the arm structure should be located roughly
  at our galactocentric radius with only two arms possibly penetrating 
  further in (e.g. \cite{Amaral1997}), if an $m=2$ component exists with 
  the same $\Omega_p$.}
  \label{fig:rotationcurve}
\end{figure*}

 To obtain the numerical value of the r.h.s., we need to know the
 rotation curve of the Milky Way.  We use the summary given by
 \citet{Olling1998}, which includes the range of currently acceptable
 rotation curves for which $R_{\odot}$ ranges from 7.2 to 8.5 kpc. For
 each rotation curve, we recalculate the outer extent of the HI arms
 which the \citet{NewMW} result corresponds to.  We then calculate the
 location of the resonances and their constraint on the pattern speed
 (eq.~\ref{eq:Constraint}).  This is portrayed in figure
 \ref{fig:rotationcurve}. The results are given in table
 \ref{tab:HI}. They imply that:
\begin{eqnarray}
\label{eq:HIres}
    \Omega_{p} &\lesssim&16.9 \pm 2.5 ~\omun\\ \Omega_{\odot}
    - \Omega_{p} &\gtrsim& 9.1 \pm 2.4~\omun \nonumber.
\end{eqnarray}
 
 Again, it should be stressed that the results assume that the spiral
 arms are a density wave.  However, if they are, then the limit is
 robust since 4 spiral arms are clearly observed to extend to about twice
 the solar galactocentric radius.  Spiral density wave theory has thus
 far been the most successful theory to describe spiral features in
 external galaxies \cite{Binney1988}, therefore, it is only
 reasonable that it describes the spiral arms in our galaxy as
 well. Moreover, in alternative theories in which the spiral arms are
 for example shocks formed from stellar formation, one would not expect to see
 spiral arms beyond the stellar disk, which is ``truncated'' several
 kpc inwards from $R_{\rm HI,out}$ \cite{Robin1992,Ruphy1996}.
 
 This last point, combined with the fact that HI is seen beyond the
 outer extent of the arms (to $R \approx 3 R_{\odot}$) implies that
 their outer limit is probably the actual 4:1 resonance.  In other
 words, the relations given by eq.~\ref{eq:HIres} are probably not
 just limits but actual equalities.  The reason is that there is
 otherwise no other physical reason to explain why the arms abruptly
 end where they are observed to do so.  (They do not end because of
 lack of HI nor does the stellar population has anything to do
 with it).
 
 Further theoretical argumentation can strengthen the last point
 made.  The co-rotation (cR) point, has often been linked to spiral
 arm brightness changes in external galaxies.  The reason is that the
 spiral arm shocks are important at triggering star formation. 
 However, near co-rotation, the shocks are very weak.  This is often
 seen in external galaxies as an ``edge'' to the disk, external to
 which the surface brightness is much lower (e.g.,
 \cite{Elmegreenbook}).  If the 4-arm pattern has the aforementioned
 pattern speed, then the cR radius can be predicted.  This radius
 (relative to $R_\odot$) is given in the seventh column of table
 \ref{tab:HI}.  We find that $R_{cR} - R_\odot = 6.5 \pm 1.5$ kpc. 
 For comparison, observations of stellar distributions show that there
 is a sharp cutoff of stars at $R_{cR} - R_\odot = 5.5$ to $6$ kpc
 according to \citet{Robin1992} or, $R_{cR} - R_\odot = 6.5 \pm 2$ kpc
 according to \citet{Ruphy1996}.  These numbers are also in agreement
 with a sharp truncation of the CO mass surface density at $R_{cR} -
 R_\odot = 5 \pm 0.5$ kpc \cite{Heyer1998}.  Namely, observations are
 consistent with the cR radius predicted using the calculated pattern
 speed, provided that the sharp cut-off is related to the CR radius. 
 This is also in direct agreement of \citet{Ivanov1983} who finds
 a co-rotation radius at 11-14 kpc for $R_\odot=8$ kpc.

 The last argument for why the constraint given by eq.~\ref{eq:HIres}
 should be considered an equality and not a limit is the following. 
 If $\Omega_p$ is significantly lower than the limit given by
 eq.~\ref{eq:HIres}, then the inner extent of the 4-arms, as
 constrained by the inner 4:1 Lindblad resonance, should be outside
 the solar circle.  However, according to the \citet{NewMW} and \citet{Dame2001} data, the
 4-arms extend inward at least to the solar circle.  On the other
 hand, table \ref{tab:pattern} shows that the 4 arms cannot extend
 much further in.  Thus, not only should eq.~\ref{eq:HIres} be
 considered a rough equality, the spiral structure inside the solar
 circle has to have different kinematics than the structure of the
 external 4-arms.  This could explain the large confusion in the
 spiral structure and pattern speeds. This could also explain why several authors find that the solar system is located  near co-rotation. If we are near CR, then it would be the co-rotation radius of the {\em inner} spiral structure. Nevertheless, this point is still far from having a satisfactory explanation. See  \S\ref{sec:caveats} for more details.
 
 Considering now that the first range of results in table
 \ref{tab:pattern} appears to be consistent with the density wave
 theory and the observations of HI outside the solar circle, we
 average the results in this range to get a better estimate for the
 pattern speed.  We find $\Omega_{\odot} - \Omega_{p} = 11.1 \pm 1$
 \omunit.  This translates into a spiral crossing period of $134 \pm 25$
 Myr on average (taking into account the results of appendix B).

%\subsection{Spiral Arm dynamics of the Milky Way}

\begin{table*}
 \caption{The observational determinations of the Milky Way spiral arm pattern
 speed.} {  \vskip 0.5cm \small
\begin{center}
\begin{tabular}{l c c c l}
\hline \hline
\small First & $\Omega_\odot$ & $\Omega_p$ & $\Omega_\odot-\Omega_p$ & Method
/ Notes\\ Author & \hskip -5mm (km/s)/kpc & (km/s)/kpc & (km/s)/kpc & \\
\hline \hline 
Yuan (1969a)\nocite{Yuan1969a}    & 25  & $\sim13.5^{1}$    & $\sim11.5$   &  Arm dynamics fit to Lin \& Shu\\
Yuan (1969b)\nocite{Yuan1969b}    & 25  & $\sim13.5$    & $\sim11.5$   &  
Migration of young stars\\
Gordon (1978)\nocite{Gordon1978}  & 25  & $11.5\pm 1.5$ & $13.5\pm1.5$ &   CO discontinuity at 4 kpc is ILR\\
Palous (1977)\nocite{Palous1977}  & 25  & $\sim13.5$    & $\sim11.5$   & Cluster birth place \\ 
Grivnev (1983)\nocite{Grivnev1983} & 25  & 12-16         &  9-13        & Cepheid birth place            \\
Ivanov (1983)\nocite{Ivanov1983}   & 27.5 & 16-20        & 7.5-11.5     & Cluster age gradients     \\
Comer\'on (1991)\nocite{Comeron1991}& 25.9 & $16\pm 5$  & $10\pm5$     & Kinematics of young stars \\
\hline
% Marochnik (1972) see comment below
Creze (1973)\nocite{Creze1973}  &$\sim24.5$& $\sim 22$  & $ 2.5\pm1.5$ & Kinematics of young stars \\ 
Palous (1977)                     & 25  & $\sim 20$     &  $\sim 5$    & Cluster birth place \\
Nelson (1977)\nocite{Nelson1977}  & 25  & $\sim 20$     &  $\sim 5$    & 
Spiral Shocks profile \& 21cm line\\
Mishurov (1979)\nocite{Mishurov1979}  & 25  &$23.6\pm 3.6$  &  $1.4\pm 3.6$ & Kinematics of Giants \& Cepheids  \\
% Pavlovskaya (1980)  reference in http://link2.springer-ny.com/link/service/journals/00230/papers/7323003/2300775/biblist.htm#bibitem7
% unaccecible journal
Grivnev (1981)\nocite{Grivnev1981} & 25  &$21-23$        &   2-4        & Kinematics of HII regions \\
Efremov (1983)\nocite{Efremov1983} & 25  &$18-25$        &   0-7        & Cepheid age gradients \\
Amaral (1997)\nocite{Amaral1997}   & 23.3&$21\pm1$        & $2.3\pm 1$  & Cluster birth place \\
\hline
Avedisova (1989)\nocite{Avedisova1989} & 25.9  &$26.8\pm 2$  & $-1 \pm 
2$   & Age gradient in Sag-Car \\
Mishurov (1999)\nocite{Mishurov1999}   &  $26 \pm 2$ & $28.1 \pm 2$  & $-2 \pm 3$   &  Cepheids kinematics   \\
Fern\'andez (2001)\nocite{Fernandez2001} &25.9 &$30 \pm 7$     & $-4 \pm 7$   &  OB Kinematics \\ 
\hline \hline
Here (HI)              &\hskip -0.5 cm$26.5 \pm 1.2 $ &$16.9 \pm 2.5$ & $9.1 \pm 2.4$& $r_{out}\approx r_{O4:1}$ of 4 HI arms$^{2}$ \\
Here (CRF var.)               & --  & -- &  \hskip -10mm $10.5\pm0.8_{stat} \pm 1.5_{sys^{3}}$ & CRF variability in Fe meteorites \\
Here (ice-ages)                   & --  &    --         & \hskip -10mm
$10.4\pm0.35_{stat} \pm 1.5_{sys^{3}}$ & Fit to ice-age occurrence$^{4}$\\
\hline \hline
\end{tabular}
\end{center}
\vskip 0.5cm}
{\small
%\noindent
%$^{1}$ Note that some results assume on the galaxy has 2 arms. 
\noindent
$^{1}$ Some results have no quoted error, but they should typically be $\pm
1$ \omunit. For example, Palous et al (1977) checked specific pattern speeds:
11, 13.5, 15, 17.5, 20, 21.5 and found that only 13.5 and 20 agree
with cluster birth places.

\noindent
$^{2}$ In principle, the 4 HI arms can terminate at $r_\mr{HI,out} < r_\mr{O4:1}$ (in
which case $\Omega_p$ can be smaller and $\Omega_\odot - \Omega_p$
larger than the quoted numbers), however, as explained in the text, there is 
evidence pointing to $r_{\rm HI,out}$ actually being $r_{O4:1}$.

\noindent
$^{3}$ The origin of the systematic error is from possible diffusion
of the solar system (both radially and along its azimuthal trajectory), relative
to an unperturbed orbit. This is explained in appendix B. The values
include the expected correction ($0.54 \pm 1.5$ \omunit) due to the solar metallicity
anomaly.

\noindent 
$^{4}$ The agreement between the bottom three results form
 the basis for the spiral arms -- ice age epochs connection.

}
 \label{tab:pattern}
\end{table*}

\begin{table*}
\caption{Implications of the 4-arm HI$^{1}$ data using spiral density wave
 theory and a possible range for the Milky Way rotation curve.}
{ \vskip 0.5cm
\begin{center}
\begin{tabular}{ c c c  c c c c c c c c }
\hline
 $R_\odot$ & $v_\odot$ & $\Omega_\odot$ & $R_{\rm HI,out}$ & $R_\mr{in}$
 & $R_\mr{in}$ - $R_\odot$ & $R_\mr{cR}-R_\odot$& $\Omega_\mr{p,max}$ &
 $\Omega_\odot-\Omega_\mr{p,max}$ & $\Omega_\mr{b,min}^{(2)}$ \\ $\mr{kpc}$ & 
 $\mr{km\over s}$ & $\mr{km \over s~kpc}$ & $\mr{kpc}$& $\mr{kpc}$ & $ 
 \mr{kpc}$ & $\mr{kpc}$ & $\mr{km \over s~kpc}$ & $\mr{km \over s~kpc}$ & $\mr{km \over
 s~kpc}$ \\
\hline
7.1       &  184      & 25.92          & 16.2 & 8.3 &1.2& 5.1  &14.39            & 11.52    &  30.1                   \\
7.1       &  200      & 28.17          & 15.7 & 7.1 &0.0& 5.0  &17.64            & 10.59    &  29.5               \\
%.8       &  212      & 27.18          & 151 &      &   &     &18.65            & 11.52    &                   \\
8.5       &  220      & 25.88          & 18.7 & 6.4 &-2.1&7.7   &18.21            &  7.7    &  27.4             \\
8.5       &  240      & 28.24          & 18.3 & 7.4 &-1.1&7.9   &19.35            &  6.7   &  27.2                     \\
\hline
$7.8$   &$212$   &$26.5$   &  $17.2$  &$ 7.3$ &-0.5  & $6.5$  &    $16.9$&$    9.1$&     $28.6$\\
$\pm 0.7$&$\pm30$ &$\pm 1.2$&  $\pm1.5$&$\pm1$ &$\pm1.7$ & $\pm1.5$ & $\pm 2.5$&
$ \pm 2.4$& $\pm1.5$ \\
\hline
\end{tabular}
\end{center}
\vskip 0.5cm}
{\small
\noindent
$^{1}$ HI data is from \citet{NewMW}. Rotation curves are from \citet{Olling1998}.

\noindent
$^{2}$ The MW appears to have a bar which ends at 3 to 4.5 kpc. Since 
bars typically end between the I4:1 resonance and cR, a lower limit 
on the bar's pattern speed $\Omega_{b}$, can be placed (assuming the bar 
ends at 4.5 kpc and coinciding with I4:1 resonance). Since 
$\Omega_{\rm p,max} < \Omega_{\rm b,min}$, at least two different 
pattern speeds exist in the MW.
}
  \label{tab:HI}
\end{table*}

\subsection{Star Formation History of the Milky Way}
\label{sec:SFR}

 In general, the intrinsic flux of cosmic rays reaching the outskirts
 of the solar system (and which we will soon require) is proportional
 to the star formation rate (SFR) in the solar system's vicinity. 
 Although there is a lag of several million years between the birth
 and death of the massive stars which are ultimately responsible for
 cosmic ray acceleration, this lag is small when compared with the
 relevant time scales at question.  In the ``short term'', i.e., on
 time scales of $10^8$ yrs or less, this ``Lagrangian'' SFR should
 record passages in the Galactic spiral arms.  On longer time scales,
 of order $10^9$ yrs or longer, mixing is efficient enough to
 homogenize the azimuthal distribution in the Galaxy.  In other words, the SFR on
 long time scales, as recorded in nearby stars, should record long
 term changes in the Milky Way SFR activity.  This may arise for
 example, from a merger with a satellite or nearby passages of one.

 \citet{Scalo1987}, using the mass distribution of nearby stars, found
 SFR peaks at 0.3 Gyr and 2 Gyr before present (BP).  \citet{Barry1988}
 and a more elaborate and recent analysis by \citet{RochaPinto2000}
 (see also references therein), measured the SFR activity of the Milky Way
 using chromospheric ages of late type dwarfs.  They found a dip
 between 1 and 2 Gyrs and a maximum at 2-2.5 Gyrs BP. 
As a word of caution, there are a few authors who find a SFR which in contradiction to the above. More detail can be found in the caveats section \S\ref{sec:caveats}.
 
\begin{figure*}
\begin{center}
\epsfig{file=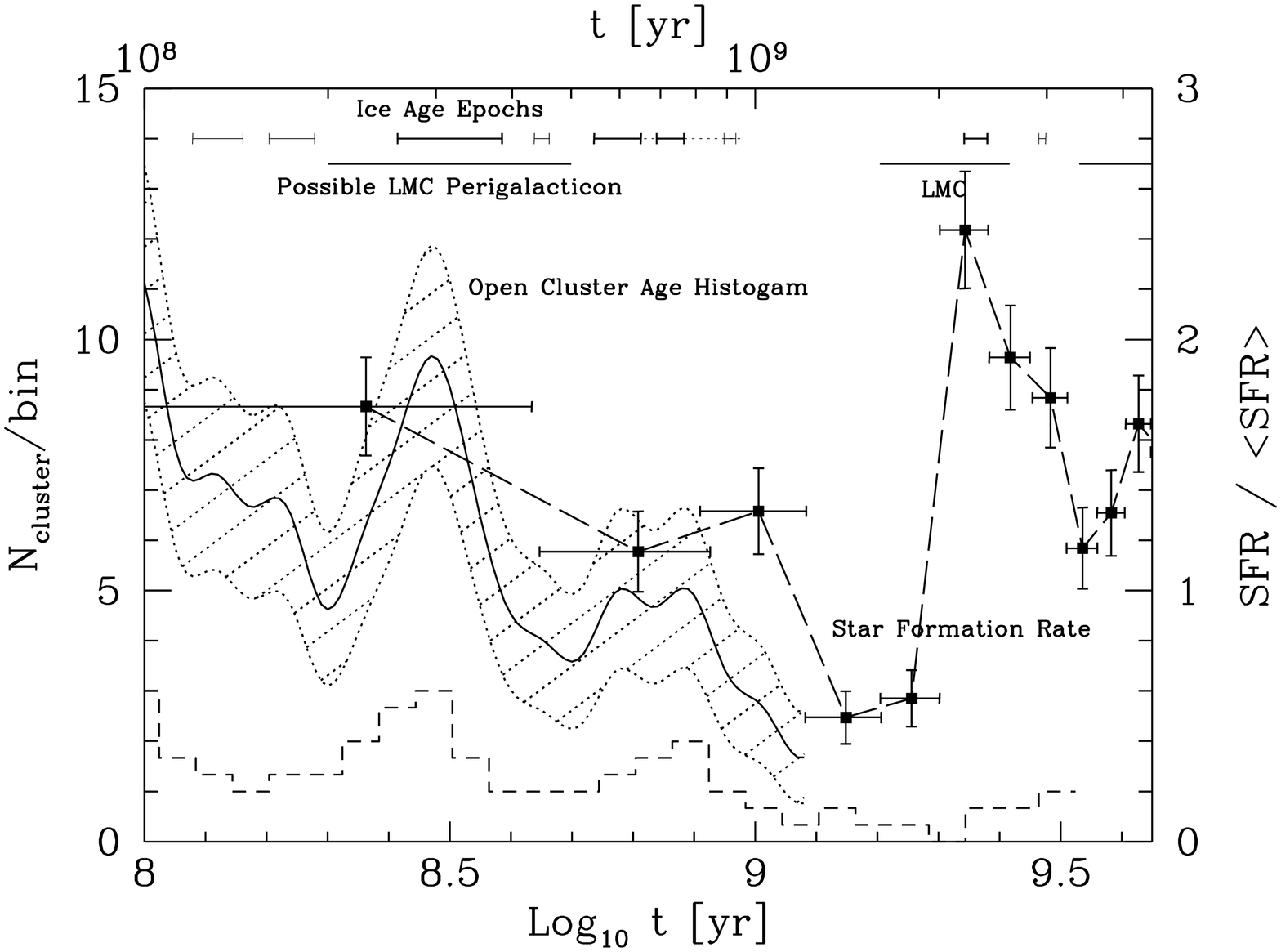,width=\figurewidth,angle=-90}
\end{center}
 \caption{
 %\small \sf 
 The history of the Star Formation Rate (SFR). 
 The squares with error bars are the SFR calculated using
 chromospheric ages of nearby stars \cite{RochaPinto2000}.  These
 data are corrected for different selection biases and are binned into
 0.4 Gyr bins.  The line and hatched region describe a 1-2-1 average
 of the histogram of the ages of nearby open clusters using the
 \citet{OpenClusters} catalog, and the expected 1-$\sigma$ error bars. 
 These data are not corrected for selection effects (namely, the
 upward trend with time is a selection effect, favorably selecting
 younger clusters).  Since the clusters in the catalog are spread to
 cover two nearby spiral arms, the signal arising from the passage of
 spiral arms is smeared, such that the graph depicts a more global SFR
 activity (i.e., in our Galactic `quadrant').  On longer time scales
 (1.5 Gyrs and more), the Galactic stirring is efficient enough for
 the data to reflect the SFR in the whole disk.  The dashed histogram
 underneath is the same as the histogram above it, though only with
 clusters having a better age determination ($w>1.0$, as defined in
 \citet{OpenClusters}).  There is a clear minimum in the SFR between
 1 and 2 Gyr BP, and there are two prominent peaks around 0.3 and 2.2
 Gyr BP. Interestingly, the LMC perigalacticon should have occurred
 sometime between 0.2 and 0.5 Gyr BP in the last passage, and between
 1.6 and 2.6 Gyr BP in the previous passage.  This might
 explain the peaks in activity seen.  This is corroborated with
 evidence of a very high SFR in the LMC about 2 Gyrs BP and a dip at
 0.7 - 2 Gyr BP. Also depicted are the periods during which
 glaciations were seen on Earth: The late Archean (3 Gyr) and
 mid-Proterozoic (2.2-2.4 Gyr BP) which correlate with the previous
 LMC perigalacticon passage \cite{Gardiner1994},
 \citealtm{Lin1995}, and the consequent SFR peak in the MW and LMC. The
 lack of glaciations in the interval 1 - 2 Gyr before present
 correlates with a clear minimum in activity in the MW (and LMC). 
 Also, the particularly long Carboniferous-Permian glaciation,
 correlates with with the SFR peak at 300 Myr BP and the last LMC
 perigalacticon.  The late Neo-Proterozoic ice ages correlate with a
 less clear SFR peak around 500-900 Myr BP. }
    \label{fig:SFR}
\end{figure*}

 These SFR peaks, if real, should also manifest themselves in 
 peaks in the cluster formation rate. To check this hypothesis, the 
 validity of which could strengthen the idea that the SFR was not 
 constant, we plot a histogram of the ages of nearby open clusters. 
 The data used is the catalog of \citet{OpenClusters}. From the 
 histogram apparent in figure~\ref{fig:SFR}, two peaks are evident. 
 One, which is statistically significant, coincides with the 0.3 Gyr 
 SFR event. The second peak at 0.6 Gyr, could be there, but it is not 
 statistically significant. Thus, we can confirm the 300 Myrs event. Note that this cluster histogram is not corrected for many systematic errors, such as the finite life time of the clusters or finite volume effects. As a result, the secular trend in it is more likely to be purely artificial. The same cannot be said about the non-monotonic behavior of the peaks.

 One source for a variable SFR in both the LMC and MW could be the
 gravitational tides exerted during LMC perigalactica.  According to
 the calculations of the perigalacticon passages, these should have
 occurred within the intervals: 0.2-0.5 Gyr BP, 1.6-2.6 Gyr BP and
 3.4-5.3 Gyr BP (with shorter passage intervals obtained by
 \citet{Gardiner1994}, and the longer ones by \citet{Lin1995}; the
 further back, the larger the discrepancy).  Interestingly, the 0.3
 Myr and 2 Gyr BP SFR events are clearly located in the middle of the
 possible LMC perigalacticon.

 Apparently, the MW activity is also correlated with SFR activity
 in the LMC. At 2 Gyr BP, it appears that there was a significant increase
 in the SFR in the LMC. Photometric studies of the HR diagram by
 \citet{Gallagher1996} show a prominent increase in the SFR somewhat
 more than 2 Gyrs before present.  An increase in the SFR 2-3 Gyrs
 before present was also found by \citet{Vallenari1996}.
 \citet{Dopita1997} analyzed planetary nebulae and found that the LMC
 metallicity increased by a factor of 2 about 2 Gyrs BP, and
 \citet{Westerlund1990} has shown that between 0.7 and 2 Gyrs BP, the
 LMC had a below average SFR.

\subsection{Cosmic Ray Flux history from Iron Meteorites}
\label{sec:IronCRF}

When meteorites break off from their parent bodies, their newly formed
surfaces are suddenly exposed to cosmic rays, which
interact with the meteorites through spallation.  The spallation
products can be stable nucleotides, which accumulate over time, or
they can be unstable.  In the latter case, their number increases but
eventually reaches saturation on a time scale similar to their half
life.  The ratio therefore between the stable and unstable nucleotides
can be used to calculate the integrated CRF that the meteorite was
exposed to from the time of break up to its burning in the atmosphere. 
Generally, it is assumed that the CRF is constant, in which case the
integrated flux correspond to a given age through a linear relation 
\cite{Singer1954,Lavielle1999}.

\begin{figure*}
\begin{center}
 \epsfig{file=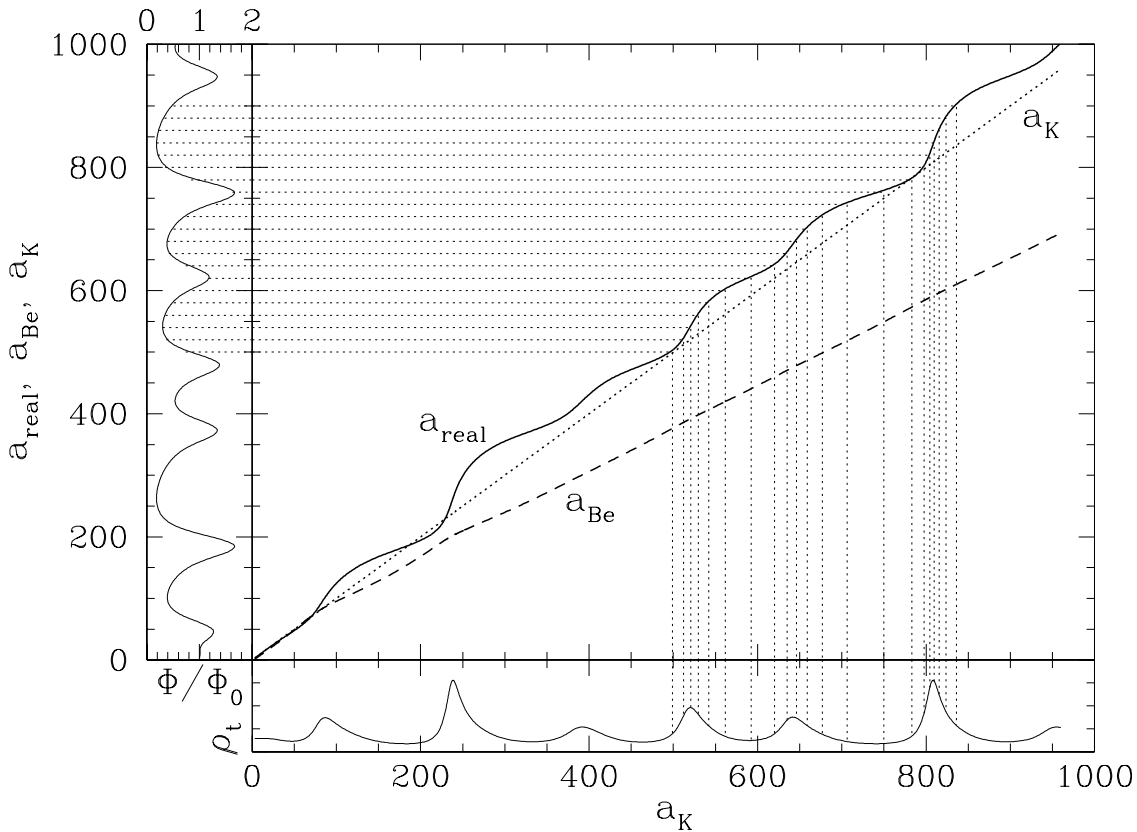,width=\figurewidth,angle=-90}\vskip -0.5cm
\end{center}
 \caption{ Theoretical comparison between different exposure ages of
 Iron meteorites and their real age.  Plotted as a function of the
 Potassium exposure age ($a_{\rm K}$) are the real age ($a_{\rm
 real}$, in solid line) and a non-Potassium exposure age ($a_{\rm
 Be}$, such as using $^{10}{\rm B}/^{21}{\rm Ne}$ dating, with a
 dashed line), and $a_{\rm K}$ (using a dotted line, with a unit
 slope).  Also plotted are the predicted CRF relative to the present
 flux ($\Phi/\Phi_{0}$) as a function of $a_{\rm real}$, and
 $\rho_{t}$--the (unnormalized) expected number of Potassium exposure
 ages per unit time, as a function of $a_{\mr K}$.  A histogram of
 $a_{\mr K}$ should be proportional to $\rho_{t}$.  The horizontal and
 vertical dotted lines describe how $\rho_{t}$ is related to the
 relation between $a_{\rm real}$ and $a_{\mr K}$---equally spaced
 intervals in real time are translated into variable intervals in
 $a_{K}$, thereby forming clusters or gaps in $a_{\mr K}$.  The graph
 of $a_{\rm Be}$ vs.~$a_{\rm K}$ demonstrates that comparing the
 different exposure ages is useful to extract recent flux changes
 (which determine the slope of the graph).  On the other hand, the
 graph of $\rho_{t}$ demonstrates that a histogram of $a_{K}$ is
 useful to extract the cyclic variations in the CRF, but not for
 secular or recent ones.  }
\label{fig:methodscompare}
\end{figure*}

A twist on the above, however, takes place in measurements employing 
the $^{41}{\rm K}/^{40}{\rm K}$ ratio.  Since the unstable isotope in the pair
has a half life slightly longer than 1 Gyr, it does not reach saturation.  By
comparing the age of meteorites using this method to methods which
employ unstable isotopes with a short half life (decay time $\lesssim$
a few Myr), it was found that the CRF in the past several Myr has been
higher by about 30\% than its average over 150 to 700 Myr BP
\cite{Lavielle1999}.  In principle, if the measurements were accurate
enough, the slight inconsistencies between the two types of methods,
as a function of time, could have been translated into a CRF history. 
However, a simulation shows that except for the flux variation over
the past several Myrs, this method becomes unfeasible.  This can be
seen in figure \ref{fig:methodscompare}.

To extract the CRF, another method should be used.  If we look at
figure \ref{fig:methodscompare}, we see that points on the graph which
are separated by equal {\em real-time} intervals, tend to bunch near the minima of the
CRF signal, when plotted as a function of potassium {\em exposure age}. 
This statistical clustering effect can be used to extract the
CRF, if a large sample of dated meteorites is used.

As previously mentioned, CR exposure dating assumes that the flux had
been constant in history.  However, if it is not, the assumed CR
exposure time will not progress linearly with the real time.  During
epochs in which the CRF is low, large intervals of time will elapse
with only a small increase in apparent CR exposure age.  As a result,
all the meteorites that broke off their parent body during this
interval, will cluster together.  On the other hand, epochs with a
higher CRF will have the opposite effect.  The CR exposure clock will
tick faster than the real clock such that meteorites that broke off
during this period will have a wide range of exposure ages.  Thus, the
number of meteorites per ``apparent'' unit time will be lower in this
case.

We use the data of \citet{Voshage1} and \citet{Voshage2}.  Together, we
have a sample of 80 meteorites which were $^{41}{\rm K}/^{40}{\rm K}$
dated.  In principle, we could also use meteorites which where exposure dated using unstable isotopes other than $^{40}{\rm K}$, however, this will introduce more complications. Because the other isotopes used for exposure dating have a relatively short life time, the relation between the real age and the exposure age is sensitive to the ratio between the recent CRF (over a few Myr) and the average over the past  1 Gyr. This is not the case with $^{40}{\rm K}$.  The second advantage of using the $^{40}{\rm K}$ data is that it encompasses more meteorites, which is important for the statistics. In the future, a more extensive and detailed analysis should clearly consider the additional data and the complications that it introduces.

Since we are assuming that meteoritic surfaces are formed
homogeneously in time, we should be careful not to be fooled from the
effects that real clustering can have.  To minimize this, we removed
all meteorites which have the same classification and are separated by
less than 100 Myr.  These are then replaced by their average. 

\begin{figure*}
\begin{center}
 \epsfig{file=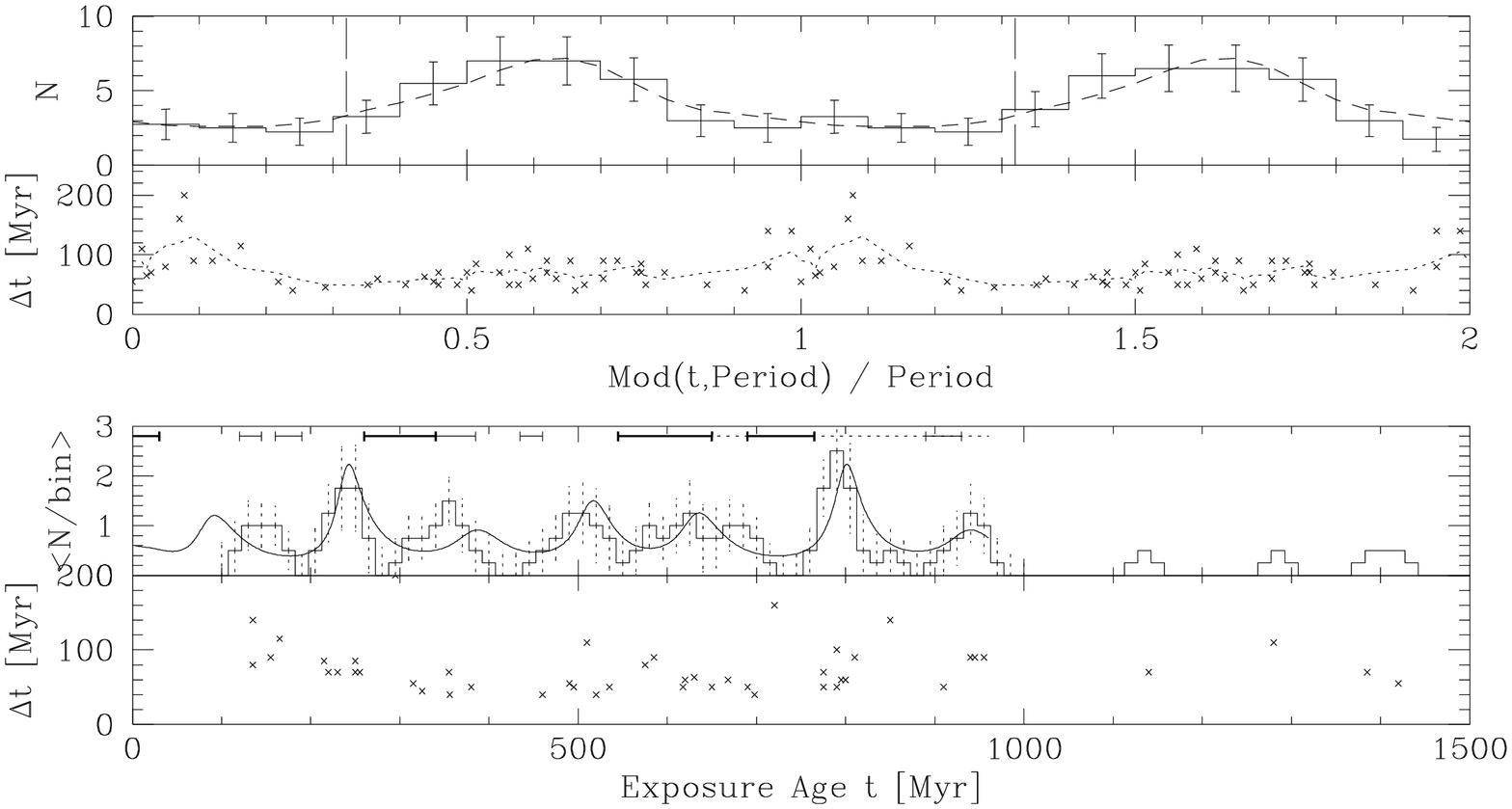,width=3in,angle=-90}
\end{center}
  \caption{Histogram of the Iron meteorites' exposure ages.  The
  lowest panel marks the $a_{\rm K}$ ages on the x-axis and the quoted
  age error on the y-axis.  Even by eye, the ages appear to cluster
  periodically.  The second panel is a 1:2:1 averaged histogram of
  meteorites with a quoted age determination error smaller than 100
  Myr, showing more clearly the clustering peaks.  Altogether, there
  are 6 peaks from 210 to 930 Myr BP. The period that best fits the
  data is $143 \pm 10$ Myr.  The third panel is similar to the first
  one, with the exception that the data is folded over the periodicity
  found.  It therefore emphasizes the periodicity.  A
  Kolmogorov-Smirnov test shows that a homogeneous distribution could
  generate such a non-homogeneous distribution in only 1.2\% of a
  sample of random realization.  Namely, the signal appears to be
  real.  This is further supported with the behavior of the exposure
  age errors, which supply an additional consistency check.  If the
  distribution is intrinsically inhomogeneous, the points that fill in
  the gaps should on average have a larger measurement error (as it is
  `easier' for these points to wonder into those gaps accidentally,
  thus forming a bias).  This effect is portrayed by the dotted line
  in the panel, which plots the average error as a function of
  phase---as expected, the points within the trough have a larger
  error on average.}
    \label{fig:Kages}
\end{figure*}

The result is a data set containing 50 meteorites.  We then plot a
histogram of the CR exposure age of the meteorites in figure
\ref{fig:Kages}.  Immediately apparent from the figure is a periodicity
of $143\pm10$ Myr.  Moreover, if we plot a scatter plot of the error
as a function of age, we see a tendency to have a higher error in
points falling in the gaps between the clusters.  This can be expected
if the clustering signal is real since points with a smaller error
will more easily avoid the gaps, thereby generating a bias.  This
consistency check helps assure we are looking at a real signal.  If the
{\em apparent} periodic signal would have been a random fluctuation, there would have
been no reason to have larger errors in the random gaps, except from
yet another unrelated random fluctuation.

Next, we fold that data over the apparent period to overcome the 
systematic error arising from the slowly changing selection effects.  
Once we fold the data over the period, we find an even clearer signal. 
We perform a Kolmogorov-Smirnov test on the folded data and find that
the probability for a smooth distribution to generate a signal as non
uniform as the one obtained, is only 1.2\% in a random set of
realizations. Therefore, it in unlikely that the meteoritic distribution was generated from a purely random process. Conversely, however, the K-S test can only say that the distribution is {\em consistent} with a periodic signal.

Furthermore, we can also predict the distribution using the diffusion
model described in \S\ref{sec:diffusionmodel}. 
 In the graphs we take the nominal case of $D=10^{28}~{\rm
cm}^{2}/{\rm sec}$ and $l_{H}=2$~kpc (as will be shown in 
\S\ref{sec:model}), and find that the distribution
obtained fits the predicted one.  In particular, the phase where the
obtained distribution peaks agrees with observations.  This implies
that the significance is of order 1:500 (instead of 1:80) to have a random signal accidentally produce the apparently periodic signal and {\em in addition} has the correct phase.

The agreement between the amplitude predicted and
observed is less significant.  The reason is that by changing the
diffusion model parameters, the contrast can be changed (as is
apparent in table \ref{table:results}).

Another point to consider is the error in the CR ages.  Although
errors were quoted with the exposure age data, their values are quite
ad hoc \cite{Voshage0}.  By comparing the Potassium age to ages
determined using other methods, it is evident that the quoted errors
over estimate the actual statistical error.  The {\em scatter} in the
difference between the Potassium age and the age in other methods is
typically 30 Myr.  This implies that the actual statistical error in
the Potassium age determinations is at most 30 Myrs, though it should
be smaller since some of the error should be attributable to the second age
determination.  Since the error will have the tendency to smear the
distribution, the obtained contrast between the minimum and maximum
flux is only a lower limit.

To summarize, the Iron/Nickel meteoritic exposure age distribution appears to have a periodicity of $143\pm 10$ Myr. The signal is not likely to arise from a random process, and it also has the correct phase to be explained by spiral arm crossings.  
The CRF contrast obtained (in the folded distribution) is $\left<
\max(\Phi) / \min(\Phi)\right> \gtrsim 3$. In addition it has been previously  
obtained that the flux today is higher by about 30\% than the average 
flux over the past billion years. Both these constraints should be 
satisfied by any model for the CR diffusion in the galaxy.

%--------------------------------

%-----------------------------------------------------------

\section{The Model}
\label{sec:model}

 We now proceed with a detailed description for how the dynamics of
 the Milky Way is expected to affect climate on Earth.  We begin with
 the large astronomical scale and work our way down the chain of
 physical links.  Namely, we begin with a model for the CR behavior in
 the Galaxy and its relation to the spiral arms in particular.  This
 will enable us to quantitatively predict the expected variability in
 the CRF. We then proceed to estimate the effects that this
 variability will have on cloud cover and the effects that the latter
 will have on the average global temperature.

\subsection{Dynamics of the Milky Way}

 The key ingredient which will be shown to be the ``driving'' force
 behind climatic variability on the long $\sim 10^8$ yr time scale, is
 our motion around the Galaxy. Variability arises because the Milky
 Way, like other spiral galaxies does not have cylindrical
 symmetry---it is broken with the presence of spiral arms.

 The basic parameters that determine this variability are the solar system's distance 
 from the center of the Galaxy $R_\odot$, its Galactic
 angular velocity $\Omega_\sun$, the angular pattern speed of the
 spiral arms $\Omega_p$, and their number $m$.
 Given these parameters, a spiral crossing is expected to occur 
 on average every time interval of
\begin{equation}
 \Delta t_{\rm  spiral} \approx {2 \pi \over m |\Omega_\odot - \Omega_p|}.
\end{equation}

 For typical values (as obtained in \S\ref{sec:observations}), a
 spiral crossing is expected to occur every ${\cal O}(10^8~{\rm
 yrs})$.  The next step, is to understand and predict the CRF
 variability that will arise from these spiral arm crossing events.

\subsection{A Diffusion Model for the long term Cosmic Ray Variability}
\label{sec:diffusionmodel}

 Qualitative theoretical and observational arguments were used to
 show that the CR density should be concentrated in the Galactic
 spiral arms.  The next step is to construct an actual model with
 which we can quantitatively estimate the variability expected as the 
 solar system orbits the Galaxy.

 The simplest picture for describing the CR content of the
 galaxy is the ``leaky box'', which assumes no spatial
 structure in the CR distribution ``inside'' the galaxy, and that the
 probability for a CR particle to remain inside the galaxy falls
 exponentially in time, with a time constant $\tau_{e,lb}$. Namely,
 the general equation describing the density of a specie $i$ is given
 by (e.g., \cite{Berez1990}):
\begin{eqnarray}
\label{eq:leakybox}
 {d N_i \over dt} &=& Q_i +\sum_{j>i}{P_{ij}\over \tau_j} N_j - N_i
 \left( {1\over \tau_{e,lb}} + {1\over \tau_{s}} + {1\over
 \tau_{r}}\right) \nonumber \\ &\equiv& {\tilde Q}_i - {N_i\over
 \tau_{i,\mathrm{eff}}}.
\end{eqnarray}
$\tau_\mathrm{i,eff}$ is the effective time scale for decay for specie $i$
which includes loss from the galaxy, spallation destruction and
radioactive decay.  ${\tilde Q}_i$ is the sum of all sources for
specie $i$ which include its actual formation $Q_i$ and the result of
spallation of more massive species $j$ (which has a spallation time
scale of $\tau_{j}$ and a branching ratio $P_{ij}$ to form specie $i$).

Eq.~\ref{eq:leakybox} can be generalized to the case when spatial and energy
homogeneity are not assumed.  One then obtains:
\begin{equation}
    \label{eq:diffusion}
  {\partial N_i \over \partial t} = D \nabla^2 N_i + {\partial \over
  \partial E} \left[ b(E) N_i \right] -{ N_i \over \tau_{i}} +
  {\tilde Q}_i.
\end{equation}
The $b(E)$ term corresponds to the slowing down of the CR rays. For the
energies at interest, we can safely neglect it as its time scale is
very long. Here, $\tau_i$ is the effective time scale for spallation
or radioactive decay. The leaky box $\tau_{e,lb}$ should not be
included explicitly in $\tau_i$, since proper losses from the galaxy are
implicitly included through losses from the boundary conditions.

The simplest of such models are 1D diffusion models which assume a
slab geometry for the galaxy.  More complicated models take more
careful consideration of the structure of the galaxy by including the
radial behavior, making them 2D in nature.  However, these models
do not consider that the sources reside primarily in the spiral arms
(cf.  \cite{Berez1990}).  We therefore construct a simple model in which
the spirals are taken into account.

\begin{figure*}[tbh]
\begin{center}
\epsfig{file=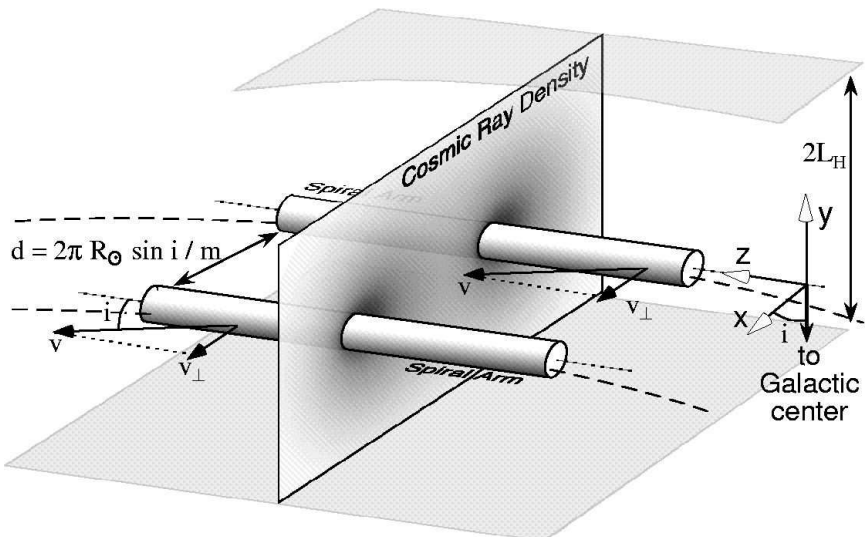,width=\figurewidth}
\end{center}
 \caption{The components of the diffusion model constructed to
 estimate the Cosmic Ray flux variation. We assume for simplicity that
 the CR sources reside in gaussian cross-sectioned spiral arms and that
 these are cylinders to first approximation. This is permissible since
 the pitch angle $i$ of the spirals is small. }
     \label{fig:model_cartoon}
\end{figure*}

A cartoon describing the geometry of the problem solved in given in
figure \ref{fig:model_cartoon}.  In particular, we assume the
following assumptions:
\begin{enumerate}
   
 \item The Galaxy is a slab of width $2 l_H$. Within it, a diffusion
 coefficient $D$ exists for the cosmic rays at the relevant energy
 $\sim 10~{\rm GeV}$. Out of this region, the diffusivity is much larger
 such that  CRs can effectively escape in a negligible time. This
 can be described with a boundary condition of the form $\Phi_{CR} =
 0$ at $z=\pm l_H$.

 \item The CR sources are located in cylinders with a Gaussian
 cross-section.  This deserves some explanation.  First, in the
 \citet{Taylor1993} model for the free electron distribution (which we
 use), the best fit to the free electron density is obtained with
 $\sigma_x = h_z= 300~\mr{pc}$ where $\sigma_x$ is the width of the
 horizontal Gaussian distribution $\propto \exp (-x^2/2 \sigma_x^2)$
 and $h_z$ is the width in the function ${\rm sech}^2 (z/h_z)$ which
 was used for the fit.  Namely, we simplify the problem by assuming
 that the vertical distribution is Gaussian as well, but with the same
 length scale as in \citet{Taylor1993}.  Second, the assumption that the
 spiral arms are straight cylinders is permissible, since the spirals
 are tightly wound: $i\approx 12^\circ$ with $i$ being their pitch
 angle).  That is, on the typical distance separating the arms, the
 radius of curvature of the arms is much larger.  The actual distance
 taken between them is:
\begin{equation}
  d_{ij} = \Delta \phi_{ij} R_\odot \sin i ,
\end{equation}
 where $\Delta\phi_{ij}$ is the angular separation between arms $i$
 and $j$ according to the Taylor and Cordes model.  The distribution
 of the free electrons and of the cosmic ray sources should be similar
 since the massive stars that undergo supernovae and produce CRs are
 also the stars that ionize the interstellar medium. More careful
 consideration later on will actually show that there is a small but
 noticeable temporal lag between the two.

 \item The arms are moving at a speed $v_\perp =
 (\Omega_p-\Omega_\odot) R_\odot \sin i$ in a direction perpendicular
 to their axes. This is permissible since under the above geometry,
 there is symmetry along the arms' axes, implying that their motion
 along their axis has no effect on the CR distribution.
\end{enumerate}

 We should consider that we currently reside in the ``Orion Arm'',
 which is only a spur or armlet \cite{Georgelin1976}.  Although it is not
 part of the global structure of our galaxy, we are required to take
 it into account in the calculation of the recent CRF variation.  As a
 consequence, we should expect because of it to witness a cosmic ray
 flux that is higher than predicted in the minimal four arm model. 
 Since the density of HII regions in this spur is roughly half of the
 density in the real nearby arms \cite{Georgelin1976}, we assume it to
 have half the typical CR sources as the main arms.  
 
 Since the lifetime of these ``spurs'' is expected to be of order the 
 spiral arm crossing period (e.g., \cite{Feitz82}), they are not 
 expected to repeat themselves after a whole revolution, nor can we 
 predict other possible ``transient'' spiral armlets that we might have 
 crossed in the past.

 The free parameters therefore left in the model are the diffusion
 coefficient $D$, the halo half width $l_H$, and the pattern speed
 of the spiral arms $\Omega_p$. Typical values obtained in diffusion
 models for the CRs yield $D \sim 10^{28} {\rm ~cm^2/s}$ and $R_h \sim
 2 \mr{~kpc}$ \cite{Berez1990,Webber1998,Lisenfeld1996}.

 We assume that the heliospheric structure does not vary over 
 intervals of
 a few 100 Myr.  Even if it does, since the cosmic ray energies in
 question are large ($\gtrsim 10$ GeV/nucleon), the heliosphere will
 not affect the CRF by more than $\sim 10\%$ \cite{Perko1987}.  Thus,
 the change in the flux reaching the solar system is also the change
 in the flux reaching the Earth. 

 We first solve for the CR flux itself and later solve explicitly for
 the Be nuclei which are used to date the ``age'' of the CRs.  For the
 CR flux, we can neglect the effect of spallation.  We also neglect
 the ionization energy losses.  We therefore solve a simplified
 version of eq.~\ref{eq:diffusion}:
\begin{equation}
    \label{eq:simpdiff}
 {\partial \Phi \over \partial t} = D \nabla^2 \Phi  + {S}.
\end{equation}
$S$ is the source distribution given by the moving cylindrical arms:
\begin{equation}
  S %= \sum_i S_i({\bf x},t)
    = \sum_i A_{i} \exp \left\{- {
  (x-x_{i,0}-v_{\perp}t)^{2} \over 2 \sigma_x^2 } - {z^{2}\over 2
  \sigma_z^2 } \right\},
\end{equation}
 where $A_{i}$ and $\sigma_{x,z}$ are the amplitude of arm $i$ (with a
 global normalization that is yet to be determined) and the horizontal
 and vertical width of the arms. $x_{i,0} = \phi_{i,0} R_\odot \sin i$
 is the current day ``location'' of these arms in the model geometry
 in which the spirals are straightened. $\phi_{i,0}$ is the
 galactocentric angle of the arms at the solar galactocentric radius.
 These numbers, together with $\Delta \phi_{i}$ are taken from \citet{Taylor1993}.

To solve the diffusion problem, we take full advantage of the fact
that eq.~\ref{eq:simpdiff} is linear. The problem is time dependent. 
Hence, we begin
by writing the solution in the form:
\begin{equation}
 \Phi({\bf x},t) = \int_{-\infty}^{t} dt' \varphi({\bf x}-(x_{i,0}+v_\perp
 t){\hat {\bf x}},t;t'),
\label{eq:5}
\end{equation}
 where $\varphi({\bf x},t;t')$ is the solution (at ${\bf x}, t$) for
 the diffusion for the cosmic rays that where emitted at time
 $t'$. This function can be written as:
\begin{equation}
 \varphi ({\bf x},t;t')  = \sum_i \varphi_i({\bf x}-(x_{i,0}+v_\perp
 t){\hat {\bf x}},t;t',i)
\label{eq:6}
\end{equation}
namely, as a contribution from each separate arm. We can then further
use the method of mirror images which ensures that on the boundaries
at $\pm l_H$ the CR flux will vanish. We do so by writing:
\begin{equation}
 \varphi_i ({\bf x},t;t',i) = \sum_{m=-\infty,\infty}
 \varphi_{i,0} ({\bf x}-2 ml_h {\bf \hat z},t;t',i) (-1)^m
\label{eq:mirror}
\end{equation}
where $\phi_{i,0}$ is the flux at ${\bf x}$ and $t$ of the CRs emitted
at $t'$ by arm $i$ assuming it is centered at $x'=0$ at $t'$, when no
boundaries are present. This has the form:
\begin{eqnarray}
 \varphi_{i,0} ({\bf x}-2 ml_h {\bf \hat z},t;t',i) = &
~~~~~~~~~~~~~~~~~~~~~~~~~~~~~~~~~~~~~
\nonumber
\end{eqnarray}
\vskip -0.5 cm
\begin{eqnarray}
\nonumber ~&=& A_i
 \int_{-\infty}^{\infty} \hskip -4pt dx' dz' {\exp\left( - {(x-x')^2 + (z-z')^2
 \over 4 D \Delta t} \right) \over 4 \pi D \Delta t} {\exp\left(-{{x'}^2
 \over 2 \sigma_x^2} - {{z'}^2 \over 2 \sigma_z^2}\right) \over 2 \pi
 \sigma_x \sigma_z} \\ & = & A_i{\exp \left( - {x^2 \over 2 (2D \Delta t +
 \sigma_x^2)} - {z^2 \over 2 (2D \Delta t + \sigma_z^2) } \right) \over 2 \pi
 \sqrt{2 D \Delta t + \sigma_x^2} \sqrt{2 D \Delta t + \sigma_z^2} }.
\end{eqnarray}
with $\Delta t \equiv t-t'$. 
From equation \ref{eq:mirror}, we have that on the $z=0$ plane (i.e.,
for $|z| \ll l_H \sim {\cal O}(1~kpc)$)
\begin{eqnarray}
 \varphi_{i} (x,z=0,t;t',i)
 = &
~~~~~~~~~~~~~~~~~~~~~~~~~~~~~~~~~~~
\end{eqnarray}
\vskip -0.5 cm
\begin{eqnarray}
\nonumber
\nonumber ~&=& A_i {\exp \left( - {x^2 \over 2 (2D\Delta t +
 \sigma_x^2)} \right) \vartheta_4 \left(0,\exp \left[-2 l_H^2 / (2 D
 \Delta t + \sigma_z^2)\right]\right) \over 2 \pi \sqrt{2 D \Delta t +
 \sigma_x^2} \sqrt{2 D \Delta t + \sigma_z^2} } \\ &\approx & {\exp
 \left( - {x^2 \over 2 (2D\Delta t + \sigma_x^2)} \right) \over
 \sqrt{2 \pi (2 D \Delta t + \sigma_x^2)}}{1\over l_H} \left[ \left( l_H^2
 \over 2 \pi (2 D \Delta t + \sigma_z^2) \right)^{1.8} + 1 \right]
 \nonumber \\ & & \times \exp\left[ - {2.4674\over l_H^2} \left( D
 \Delta t + {\sigma_z^2\over 2} \right) \right] \nonumber
\end{eqnarray}
where $\vartheta_4$ is the Elliptic Theta function of the fourth kind
and the approximation is accurate to $\pm 3\%$. We find that the
effect of the boundary condition is to introduce an exponential
decay. This shows that the diffusion model with boundaries is very
similar to the leaky box model, with a decay time constant of $\tau_e
= l_H^2 / (2.47 D)$. In fact, for $t-t'+\sigma_z^2/(2 D) \gtrsim
\tau_e / 3$ the error between a simple exponential fit and the Elliptic
theta function is less than 10\%.

If we go back steps \ref{eq:5} and \ref{eq:6}, we finally obtain:
\begin{eqnarray}
\label{eq:difres}
\hskip -9pt
 \Phi(x,z=0,t) &=& \sum_i A_i \int_{-\infty}^t dt' {\exp \left( -
 {\left(x-(x_{i,0}+v_\perp
 t)\right)^2 \over 2 (2D(t-t') + \sigma_x^2)} \right) \over 2 \pi \sqrt{2 D
 (t-t') + \sigma_x^2} } \nonumber \\ &  \times & { \vartheta_4
 \left(0,\exp \left[-2 l_H^2 / (2 D (t-t') + \sigma_z^2)\right]\right)
 \over \sqrt{2 D (t-t') + \sigma_z^2} }. \nonumber \\&&
\end{eqnarray}
 The global normalization of the amplitudes $A_i$ is determined by
 calculating the time average $\left< F_{CR} (x,y=0,t) \right>$ which
 from record in Iron Meteorites should be 28\% less than todays flux 
 \cite{Lavielle1999}.

\subsubsection{CR lag after spiral arm crossing}

An interesting point worth particular note is that the CR distribution 
is expected to lag behind the spiral arm passages. Two separate 
physical mechanisms are responsible for such a lag. 

\noindent {\bf CR distribution skewness}: Because the CR distribution
is skewed towards later times, the CRF is higher at a given time after
the spiral arm crossing than before the crossing.  This is because the
CRs diffuse in the interstellar medium, but the spiral arms which are
their source are moving, thus leaving behind them a wake of slowly
diffusing CRs.  Before an arm reaches the region of a given star, the
CR density is low since no CRs were recently injected into that region
and the sole flux is of CRs that manage to diffuse to the region from
large distances.  After the spiral arm crosses the region, the CR
density is larger since locally there was a recent injection of new
CRs which only slowly disperse. 

 This lag is intrinsic to the the diffusion model and therefore need
 not be considered separately.  It can be seen in
 fig.~\ref{fig:CRhistory}, in the skewed distributions seen around
 each spiral arm crossing.  Although the peak flux is lagged by only a
 small amount due to this effect, the skewness implies that the
 mid-point of epochs that are defined by the CRF being larger than a
 threshold value, will be lagged.  Table \ref{table:results} shows
 that this lag can range from 6 to 19 Myrs, depending on the model
 (while considering only those that fit the observed $^{10}\mr{Be/Be}$ 
 and CRF variations) and assuming a threshold that will soon be
 shown to correspond to ice-age epochs.

\noindent {\bf SN-HII lag}: A second delay between the CRF
distribution and the midpoint of the spiral arm passage arises because
the definition of the spiral arm used does not have to coincide with
the actual CR source distribution, which are SN remnants.  In our
case, the spiral arms location is defined through the free electron
distribution as fitted for by \citet{Taylor1993} using the pulsar
dispersion data.  The free electron density is primarily affected by
luminous OB stars that are blue enough to ionize the interstellar medium. 
Therefore, the midpoint of the free electron distribution will occur
around half the life span of the stars which dominate the ionization,
namely, a few Myr since it is dominated by the most massive O stars
that form.  SNe on the other hand, occur at the end of the life of
stars which can have lower masses and therefore have longer life spans than
the stars that dominate the ionization.  Since the least massive stars
that undergo SN are about 8$M_{\sun}$ and have a life span of
typically 35 Myr, the average SN occurs at typically half this
life span (as a result of the contribution of more massive stars).

\begin{figure*}[tbh]
\begin{center}
\epsfig{file=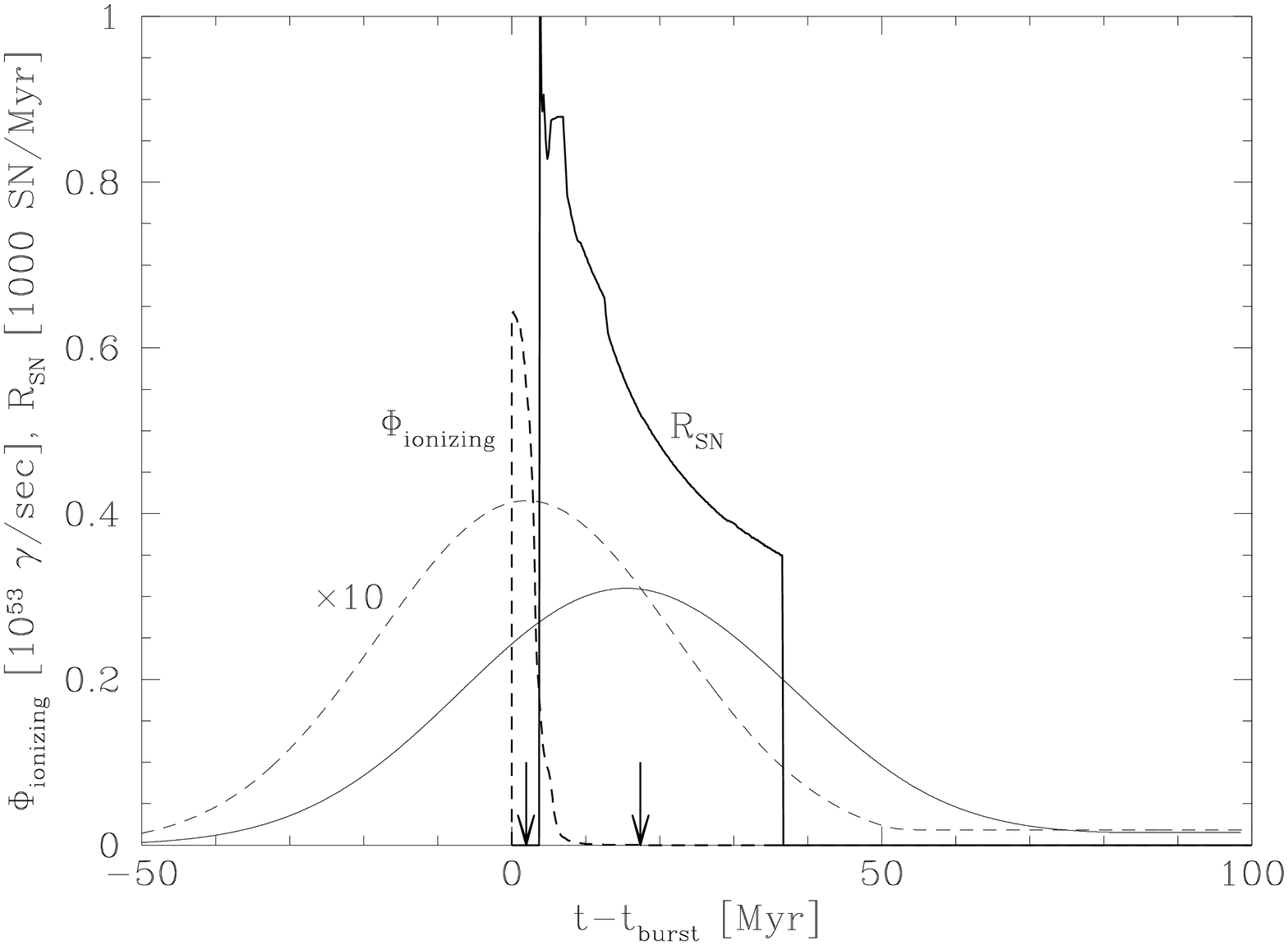,width=\figurewidth,angle=-90}
\end{center}
 \caption{ The ionizing radiation flux and SN rate as a function of
 time for a starburst of $10^6 M_{\sun}$ at time $t=0$, assuming
 standard parameters \cite{starburst}.  The heavy weight lines
 describe the total H ionizing radiation flux (dashed) and the SN rate
 (solid).  Since SNe are dominated by less massive stars than those
 responsible for the ionization, and since SNe occur only at the end of
 the stellar life, the SNe distribution is lagging behind the
 ionization distribution by 15.4 Myr (the average ionizing
 radiation is emitted at 2.0 Myr after the starburst, while the
 average SN explodes 17.4 Myr after it).  The light weight lines are
 the same as before only the instantaneous starburst is replaced with
 a Gaussian distribution with $\sigma = 16.5$~Myr, which reproduces the
 HII arms seen by Taylor and Cordes.  The lag remains the same, but
 the distribution for the SNR widens to 20 Myr.  }
    \label{fig:SNlag}
\end{figure*}

To better quantify this lag, we use the program of \citet{starburst}
which calculates various properties of a starburst population.  In
particular, it can calculate the flux of ionizing photons and the SN
rate, both of which we require.  We use standard parameters for a
starburst \cite{starburst}.  Namely, $10^{6}M_{\sun}$ are formed at
$t=0$ with a Salpeter IMF (i.e., a slope of -2.35), with a lower
cutoff of 1 $M_{\sun}$, an upper cutoff of 100 $M_{\sun}$ and a minimum
mass of 8 $M_{\odot}$ needed to trigger a SN explosion.  It also
calculates the wind loss using a theoretical estimate and using the
Kurucz-Schmutz spectra.  The results are depicted in figure
\ref{fig:SNlag}.  It shows that the average ionizing radiation is
emitted at 2.0 Myr after the starburst, while an average SN takes
place at 17.4 Myr after the starburst, giving an average delay of 15.4
Myr.  The single most important parameter to which this lag is
sensitive to is the lower cutoff mass needed to form a SN. On the
other extreme, the global normalization does not affect the lag at
all, implying that the total mass of stars formed and the lower cutoff
for star formation are both unimportant.  The figure also shows a more
realistic distribution obtained not for an instantaneous starburst,
but one that has a Gaussian distribution with $\sigma_{t}=16.5$ Myr. 
This gives a Gaussian distribution for the ionizing radiation expected
for an arm of width $\sigma_{arm} \approx 300$ kpc (for the pattern
speed that will later be shown to agree with the various data).  This
width is the value obtained by the \citet{Taylor1993} model.  The
result for the SN rate in this case, is similar to a Gaussian
distribution with a new variance of $\sigma_{t}\approx 20$ Myr.

If we wish to incorporate these two results to the CR diffusion, we 
need to correct the following:
\begin{enumerate}

 \item The spiral arm locations, as defined by the CR sources, are lagged 
 by 15.4 Myr after the spiral arms as defined through HII.
\item The width of the CR distribution is spatially (or temporally) 
larger by a factor of 19.5/16.4 = 1.20 than the one obtained from the 
\citet{Taylor1993} model.    

\end{enumerate}

\subsubsection{Effects of inter-arm SNe}

We have assumed thus far that SNe occur only within the spiral
arms. There are however two SN sources outside the spiral arms. These
are from infrequent `field' OB stars and from SNe of type Ia, of which
the progenitors are not massive stars.

\noindent
{\bf ``Field'' SNe:} Most star formation in spiral galaxies occurs
inside the spiral arms, as a result of gas being shock excited by
spiral arm passages. Nevertheless, some star formation occurs outside the
arms. To estimate the fraction of the latter, we look at two types of
observations -- the distribution of giant HII regions and of O
stars. HII regions are formed from the ionizing radiation of the
massive stars that later explode as SNe. According to
\citet{Evans1991}, about 15\% of these regions reside in the
inter-spiral regions. This can give an upper estimate for the inter-arm
SNe rate (of types others than Ia), since 15\% is the number fraction
of regions---it is not weighed by the actual number of OB stars in
each complex (which is expected to be higher in the arms, where HII
clouds are larger on average). A lower limit for the SNe rate in the
inter-arm regions would be given by the fraction of O stars that reside
in the inter-arm regions. O stars are more massive of the stars that
undergo SNe explosions; they are therefore expected to be more
concentrated in the spiral arms. According to the data in
\citet{Lynds1980}, about 20 of the 400 or so nearby O stars reside
outside the nearby spiral arms, or 5\%. A reasonable estimate for the
field SNe rate would therefore be 10\% of the spiral arm rate. It is
interesting to note that the discrepancy between SNe rate estimates
from historical records and from observations in external galaxies has
been attributed to SN being distributed mainly in the spirals of the galaxy
\cite{Drag1999}.

\noindent
 {\bf SNe Type Ia:} SNe Type Ia are believed not to originate from the
 core collapse of a single massive star. As a result, they can be
 found everywhere in the disk, not limited to within the spiral
 arms. For the same reason, they are also found in Elliptical
 galaxies. Their fractional rate in Spiral galaxies, depends on the
 spiral type and to a larger extent, the analysis used to derive the
 numbers. Typical numbers obtained are 23\% for Sab and Sb's, and 9\%
 for Sbc-Sd \cite{Tammann1994}, or, 8\%-23\% for Sab and Sb's, and
 10\%-19\% for Sbc-Sd \cite{vandBergh1994}. i.e., A typical number
 expected for the Milky way is 15\%.

 Both the above inter-arm SN sources add up to typically 25\% of the
 rate in arms. (And it is unlikely to be more than 35\% or less than
 15\% of the rate in the arms). Does this ratio correspond to an
 additional constant in the CR source that is 25\% of the CRs generated in 
 the arms? It would if CR acceleration efficiency was the same
 in spiral arms and outside of them. However, theoretical
 argumentation points to an efficiency proportional to the ambient
 density $n$ \cite{Bell1978}. There are also consistent
 observations to corroborate this. Radio emission from SNRs in M33, which is
 presumably from CRs synchrotron emission, is seen only in SNRs
 located in apparently denser environments \cite{SNRs}. This would
 imply that the inter-arm SNe will accelerate less efficiently the CRs,
 by the typical density ratio between arm and inter-arm regions.

 Thus, the average inter-arm SN to arm SN CR generation ratio, is
 expected to be:
\def\armdisk{\cal S}
\begin{equation}
 \armdisk \equiv {\left<\Phi_{\rm inter-arm}\right> \over
\left<\Phi_{\rm arm}\right>}
\approx {\left< \rho_{\rm inter-arm}\right> \over \left< \rho_{\rm
arm}\right>} {\left(f_\mr{Ia} + f_\mr{II,Ib}f_\mr{field~OBs} \right) \over
f_\mr{II,Ib}(1- f_\mr{field~OBs)} },
\end{equation}
 where $f_\mr{Ia}$ is the fraction of SN Ia's, $f_\mr{II,Ib}=1-f_\mr{Ia}$ is
 the fraction of SNe of other type, and $f_\mr{field~OBs}$ is the
 fraction of massive stars that form outside of the Galactic spiral arms. 
 According to \citet{Lo1987}, the typical arm/inter-arm density
 contrast in $H_2$, is at least 3 but can be as high as 15.  While it
 is typically 6 in $H\alpha$ \cite{Lees1990}.  Thus, for the nominal
 values for the different fractions and a density contrast of 6, we
 obtain that $\armdisk \sim 5\%$.  For the extreme values for the
 different fractions and density contrast, it can range from 1\% to
 15\%.  This is the ``background'' constant CRF that one has to add to
 the spiral arm source of CRs. That is, when calculating the flux, we have to add a constant to it:
\begin{equation}
 \Phi_\mr{total} = \Phi_\mr{arm} + \armdisk  
 \left< \Phi_\mr{arm} \right>.
\end{equation}

\subsubsection{Results for the CR diffusion}

\begin{figure*}[tbh]
\begin{center}
\epsfig{file=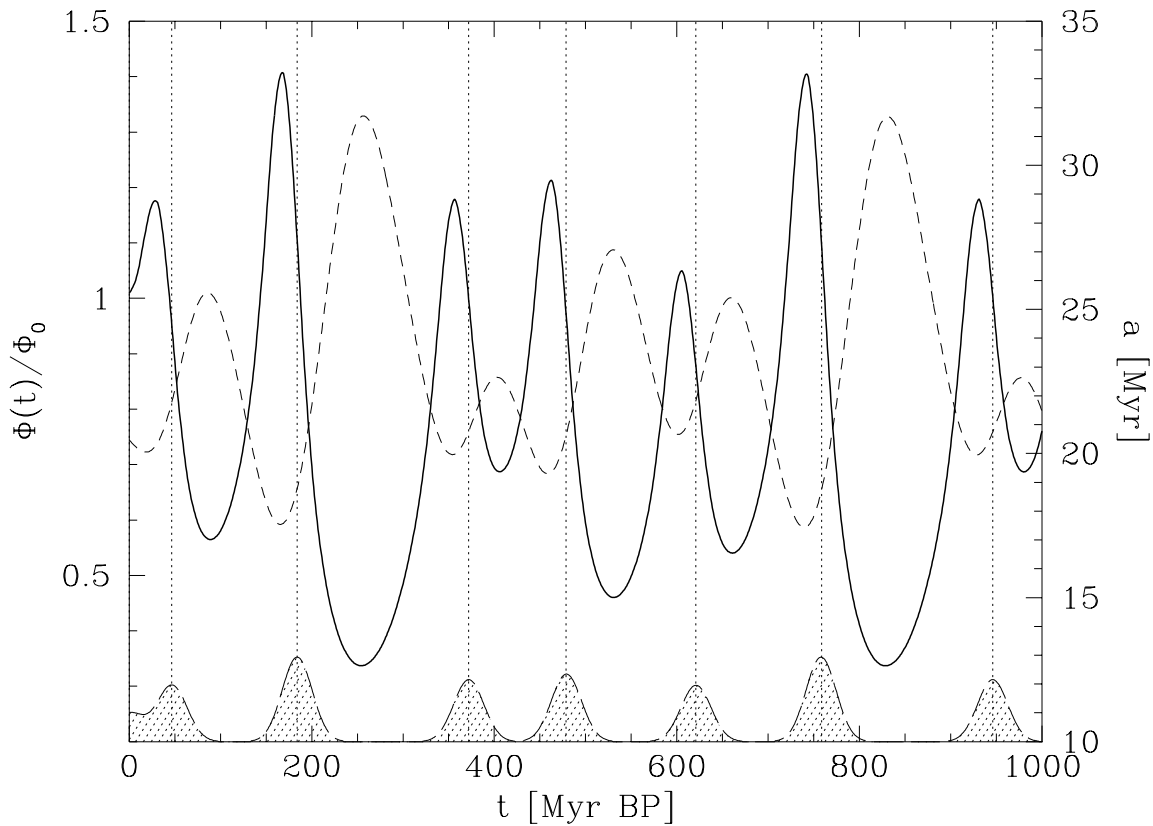,width=\figurewidth,angle=-90}
\end{center}
 \caption{The cosmic-ray flux variability and age as a function of
 time for $D=10^{28}{\rm ~cm^2/s}$ and $l_H=2$ kpc. The solid line is
 the cosmic-ray flux, the dashed line is the age of the cosmic rays as
 measured using the Be isotope ratio. The shaded regions at the bottom
 depict the location, relative amplitude (i.e., it is not normalized)
 and width of the spiral arms as defined through the free electron
 density in the Taylor and Cordes (1993) model. The peaks in the flux
 are lagging behind the spiral arm crosses due to the SN-HII
 lag. Moreover, the flux distribution is skewed towards later times. }
 \label{fig:CRhistory}
\end{figure*}

 After incorporating the effect of the inter-arm SNe and the lag
 generated by the SN delay, results for the CRF can be obtained.
 Given $D$, $v$ ($\leftrightarrow \Omega_p$) and $l_H$, we can
 calculate the variability in the CR flux. We can also calculate the
 ``age'' of the CRs as will be measured by the $^{10}{\rm Be}/^{9}{\rm
 Be}$ ratio. This is described in Appendix A.
 Sample variations of the CRF and the Be age as a function of time are
 given in figure \ref{fig:CRhistory}.

\begin{table*}
\caption{Results for the spiral arm CR diffusion model$^{1}$}
{\tiny 
\vskip 0.5cm 
\begin{center}
\begin{tabular}{c c c c c c c c c c c c}
\hline
$D_{28}$ & $l_H$ & $\tau_e$ &  $\Phi_\mr{min}/$ & $\Phi_\mr{max}/$ & 
$\Phi_\mr{max}/$ & $\left<\Phi\right>/$ &
 Mid Point & $a_0$ & $(a)_\mr{min}$  &  $(a)_\mr{max}$ &$\Delta T$ \\
         & kpc   & Myr      &  $\Phi_0$   & $\Phi_0$                    & 
         $(\Phi)_\mr{min}$  & $\Phi_0$             &
 Lag [Myr] & [Myr] & [Myr]     & [Myr] & [$^{\circ}$K] \\
\hline
0.32 & 0.50 & 10.54 & 0.05 &2.33 & 44.66 &0.65 & 5.32 &6.54 & 4.75 &16.89 & 32.40 \\
0.10 & 0.50 & 33.74 & 0.07 &1.96 & 27.92 &0.62 & 12.06 &24.98 & 13.59 &75.74 & 26.95 \\
0.03 & 0.50 & 105.43 & 0.13 &1.68 & 12.93 &0.66 & 17.19 &40.39 & 17.61 &114.69 & 22.13 \\
0.01 & 0.50 & 337.38 & 0.17 &1.60 & 9.18 &0.71 & 18.17 &41.45 & 14.73 &128.06 & 20.29\\ 
1.00 & 1.00 & 13.49 & 0.22 &1.76 & 7.87 &0.73 & 6.64 &9.14 & 7.19 &15.02 & 21.75 \\
0.32 & 1.00 & 42.17 & 0.23 &1.58 & 6.97 &0.70 & 8.62 &30.90 & 22.23 &58.88 & 19.25 \\
0.10 & 1.00 & 134.95 & 0.22 &1.43 & 6.62 &0.69 & 17.37 &49.52 & 29.77 &101.00 & 17.21 \\
0.03 & 1.00 & 421.73 & 0.21 &1.48 & 6.99 &0.72 & 18.09 &50.10 & 22.91 &121.00 & 17.98\\ 
3.20 & 2.00 & 16.87 & 0.56 &1.27 & 2.26 &0.82 & 5.95 &12.08 & 11.59 &15.20 & 10.02 \\
1.00 & 2.00 & 53.98 & 0.54 &1.23 & 2.28 &0.80 & 9.20 &38.12 & 33.80 &50.68 & 9.74 \\
0.32 & 2.00 & 168.69 & 0.43 &1.22 & 2.86 &0.76 & 10.71 &56.75 & 45.18 &83.92 & 11.22 \\
0.10 & 2.00 & 539.81 & 0.29 &1.26 & 4.36 &0.73 & 18.80 &59.73 & 37.74 &107.39 & 13.79 \\ 
10.00 & 4.00 & 21.59 & 0.90 &1.16 & 1.29 &0.97 & 3.70 &15.76 & 15.76 &17.95 & 3.66 \\
3.20 & 4.00 & 67.48 & 0.84 &1.12 & 1.32 &0.92 & 5.54 &42.52 & 42.45 &48.92 & 3.85 \\
1.00 & 4.00 & 215.92 & 0.69 &1.08 & 1.56 &0.84 & 9.28 &60.86 & 57.54 &72.78 & 5.46 \\
0.32 & 4.00 & 674.76 & 0.49 &1.13 & 2.30 &0.78 & 11.79 &65.62 & 54.16 &90.75 & 9.03 \\
\hline
\end{tabular}
\end{center} 
\vskip 0.5cm 
}
Note: $^{{1}}$ The results summarized in this table are a subset of the all the 
models calculated (and appearing in 
fig.~\ref{fig:CRFmodelconstraints}).
  \label{table:results}
\end{table*}

\begin{figure*}
\begin{center}
\epsfig{file=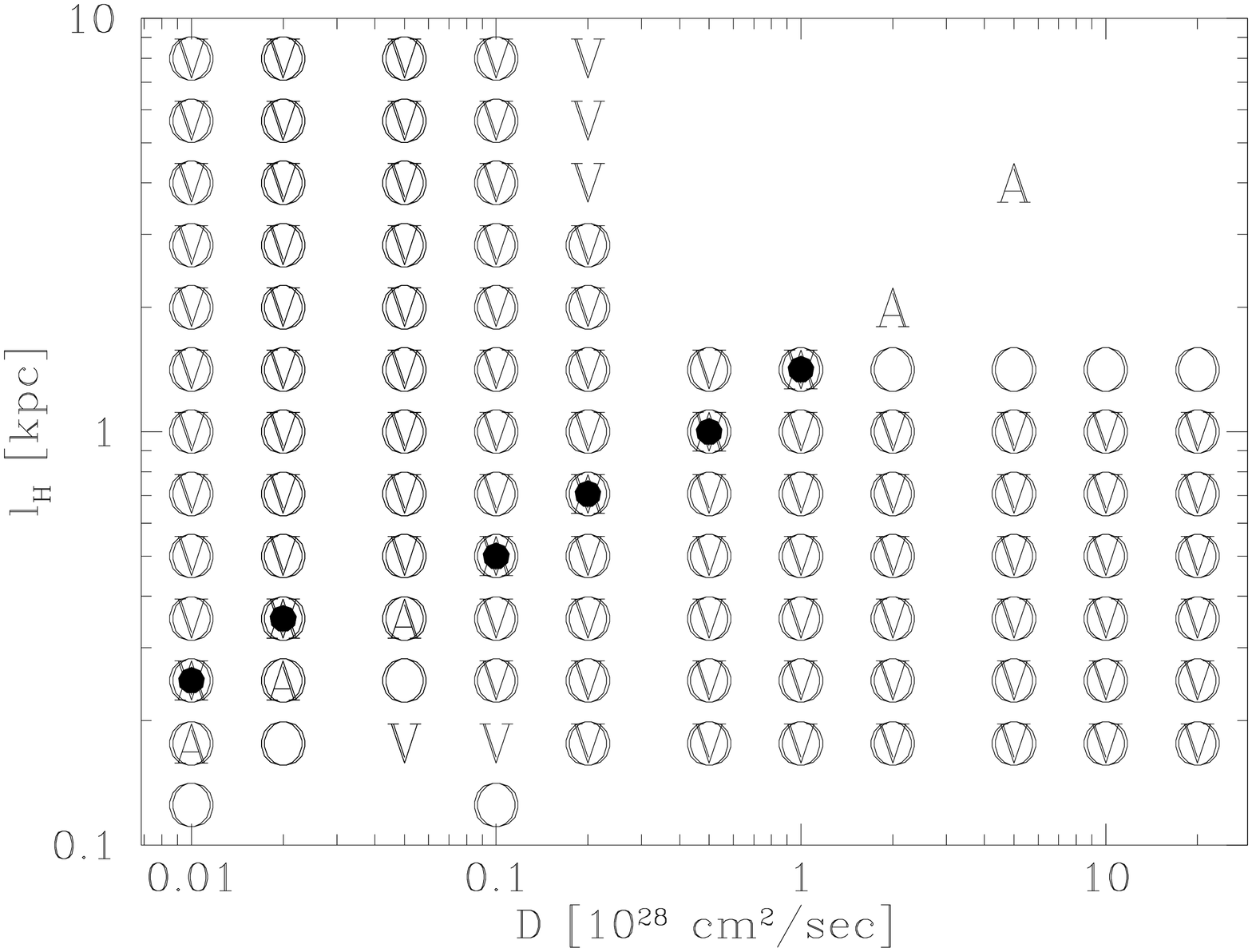,width=\figurewidth,angle=-90}
\end{center}
 \caption{
%  %\small \sf 
 Models which fit the observational constraints.  The free parameters
 in the diffusion model are the Diffusion coefficient $D$ and the half
 width of the Galactic halo (relevant for the diffusion).  ``O''
 denote models which fulfill the constraint on the ratio between the
 maximum flux and the minimum flux (once folded over the 143 Myr
 period).  ``V'' denotes models which satisfy the constraint on the
 ratio between today's flux and the average over the past billion
 years.  ``A'' denote models which satisfy the ``age'' constraint of
 the CRs, as measured by the $^{10}{\rm B}/{\rm B}$ ratio.  }
 \label{fig:CRFmodelconstraints}
\end{figure*}

\subsubsection{Constraints on the diffusion model}

The primary constraint used to place limits on CR diffusion models is
the $^{10}\mr{Be}/\mr{Be}$ ratio observed.  This ratio depends on the
typical duration the CRs spend between their formation and their
detection in the solar system, i.e., on the CR's typical age.  This is
because Be isotopes are basically formed as spallation products. 
$^{9}$Be is stable and therefore accumulates with time. On the other 
hand, $^{10}$Be has a decay time scale of several Myr. It therefore saturates.

In appendix A, the survival fraction of $^{10}\mr{Be}/\mr{Be}$ is
calculated using the spiral arm diffusion model.  This ratio can then
be translated to an effective confinement age that a ``leaky box''
model is required to have for it to yield the same Be ratio.  The results are
summarized in table \ref{table:results}.  This results should be
compared with the measured $^{10}\mr{Be}/\mr{Be}$ ratio.  The various
observed values for this number once translated to an effective CR age in
the leaky box model yield $24^{+12}_{-6}$~Myr (see appendix
A).  Namely, values ranging from 18 to 36 Myr are reasonable, and 
should be recovered by any consistent diffusion model, including the 
present one.

Unlike other diffusion models, the spiral arm diffusion model can have 
more constraints placed on it using the observed CRF variability. 
Since previous models are all in steady state, they cannot under any 
condition, satisfy these constraints.

Constraints on the CRF variability can be placed using CR exposure
ages of Iron meteorites.  Comparison between ages measured using
$^{41}\mr{K}/^{40}\mr{K}$ dating in which the unstable isotope
$^{41}\mr{K}$ has a very long half life (1.27 Gyr) and other
measurements employing a ``short lived'' isotope (decay time $\lesssim$ few~Myr),
have been consistently inconsistent.  The only viable explanation
being that the CRF in the past several Myrs has been roughly 30\%
higher than the average over the past billion years
\cite{Hampel1979,Sch1981,Lavielle1999}. We take 
$\Phi_{0}/\left<\Phi\right> = 0.7 \pm 0.1$. An even more interesting
constraint can be placed using the actual CRF variation, as extracted
from the distribution of CR exposure ages (see \S\ref{sec:IronCRF}).  This
gives $\max(\Phi)/\min(\Phi) \gtrsim 3$ .  For various reasons given in
\S\ref{sec:IronCRF}, an upper limit cannot be placed.

Table \ref{table:results} and fig.~\ref{fig:CRFmodelconstraints}
summarize the results.  Values consistent with the different 
constraints on the effective leaky box age, with the ratio between the
flux today and the average flux, and with the minimum to maximum flux contrast
are emphasized in the figure.  Since models are required to satisfy all 
three observed constraints, we
can place limits on the diffusion coefficient $D$ and the halo half
width $l_{H}$.  These are $10^{26}\mr{~cm}^{2}\mr{/s}
\lesssim D \lesssim 10^{28} \mr{~cm}^{2}\mr{/s}$ and $0.5 \mr{~kpc}
\lesssim l_{H} \lesssim 2 \mr{~kpc}$ (with the values being
correlated).

If we return back to table \ref{table:results}, we see that the models satisfying the 
constraints have a maximum variation of the CRF of $\delta \equiv (\max 
\Phi-\min \Phi)/\Phi_\mr{today} \sim 1.0$ to $1.9$ and a lag $a_{0}$ of 
6 to 19 Myr. These numbers will be needed later on when estimating 
the global temperature variations and the lag between spiral 
crossings and the occurrence of IAEs.

\subsection{The Cosmic-Ray Flux Temperature relation}

We proceed now to calculate the last link in the model, which is the
relation between the CRF variations and temperature changes on Earth. 
We first calculate the effect directly and assume it takes place
through LACC variations, as was empirically found by \citet{Marsh2000}. 
We then continue with an independent calculation of this relation
using global warming.  This indirect method does not assume that the
CRF-climate connection is through LACC variations, and it therefore
serves as a consistency check.

\subsubsection{$\Delta T / \Delta(CRF)$ through cloud cover variations}

 We first calculate $\Delta T/ \Delta(CRF)$ by breaking the relation
 into two parts, into $\Delta(LACC)/ \Delta(CRF)$ and $\Delta
 T/\Delta(LACC)$.  

 {\bf The $\Delta(LACC)/\Delta(CRF)$ relation}: Since the CRF-LACC
 effect is on low altitudes clouds ($<3.2$~km, \cite{Marsh2000}),
 the effect should arise from relatively high energy CRs ($\gtrsim 10$
 GeV/nucleon) which can reach equatorial latitudes, in agreement with
 observations showing a better CRF-LACC correlation near the equator
 \cite{Marsh2000}.  Thus, when estimating the forcing that the CRs
 have on cloud cover, the relevant flux is that of CRs that can reach
 a low magnetic latitude observing station and that has a high energy
 cut-off.  The flux measured in the University of Chicago Neutron
 Monitor Stations in Haleakala, Hawaii and Huancayo, Peru is probably
 a fair measurement of the flux affecting the LACC. Both stations are
 at an altitude of about 3~km and relatively close to the magnetic
 equator (rigidity cutoff of 12.9 GeV).

 The relative change in the CRF for the period 1982-1987 at Haleakala
 and Huancayo is about 7\% \cite{Bazil2000}, while the relative change
 in the LACC is about 6\% \cite{Marsh2000}. Namely, to {\em first
 approximation}, there is roughly a linear relation between the
 relevant CRF (i.e., the equatorial or low altitude CRF) and the cloud
 cover. Based on these observations, we now assume for simplicity that
 a linear relation exists between the CRF and the absolute LACC (with
 a current global average of 28\%) throughout a range of $\sim 0 -
 60\%$ in the LACC.

 This assumption of linearity is at this point unavoidable since at present, we don't have any theory to explain the physical process involved nor do we have any lab measurements to quantify its nonlinear regime. Thus, when a temperature estimate is obtained, we should keep in mind that this assumption of linearity was assumed, and that in principle, we can expect any  correction factor of order unity. Moreover, at this point, no one can promise us that the cosmic ray effect on climate does not saturate at small corrections, which would 	quench any interesting "galactic" effect.

{\bf The $\Delta T / \Delta(LACC)$ relation}:
 A change of $1\%$ in the global LACC corresponds to a net effective
 reduction of the solar radiative flux of $\Delta F_\odot\sim - 0.6 $
 W/m$^2$ \cite{Marsh2000}.  Thus, a change of $\pm 28\%$ in the LACC
 corresponds to roughly a $\mp 17$ W/m$^2$ change in the radiation
 flux.  Without any feedback (such as cloud formation), an effective
 globally averaged change of $1$ W/m$^2$ can be calculated to
 correspond to a global temperature change of $\Delta T =0.30^\circ$K.
 This rate of change $\lambda_{0}=\Delta T/\Delta F_\odot$ assumes
 that an average surface area receives 70\% (because of the 30\%
 albedo) of a quarter (because Earth is a sphere) of the solar
 constant of $F_\odot=1367$ W/m$^2$.

 More detailed models give rates $\lambda$ which range from $\lambda =
 0.3$ to $1.1^\circ$K/(W~m$^{-2}$) \cite{IPCC1995}, with typical
 values being more like $0.7-1$ \cite{Rind1993}. Namely, there are
 positive feedbacks that increase the zeroth order relation by a
 factor of $\lambda/\lambda_0 = 1$ to $3.7$. Therefore, an assumed
 maximum change of $\Delta F=\mp 17$~W/m$^2$ will correspond to a
 change of $\Delta T=\mp 5^\circ$K for $\lambda/\lambda_0 = 1$, or to
 $\Delta T=\mp 18^\circ$K for $\lambda/\lambda_0 = 3.7$. We shall
 take a nominal value of $\lambda = 0.85^\circ$K/(W~m$^{-2}$).  This
 value corresponds to a decrease of 0.14$^\circ$K for an increase of
 1\% in the CR flux.

 Note that the value of $\lambda$ is a function of the time scale at
 question.  On shorter time scales, the large heat capacity of the
 oceans acts as a moderator and a lower value of $\lambda$ is
 expected.

\subsubsection{$\Delta T/\Delta(CRF)$ measured using global warming}

  A second method for estimating the climatic driving force of CRF
  variations, is to estimate directly the relation between CRF changes
  and measurements of the global temperature change.  When we do so,
  we should bear in mind that greenhouse gases too contribute to the
  global temperature variation.  To decouple the two, we look at a
  period in which the global temperature decreased.  Since such a
  decrease cannot be explained by greenhouse gases, we can be safe
  that we are not measuring their contribution.

 During the period from the early 50's until the early 70's, the solar
 activity (averaged over the solar cycle) declined from a maximum to a
 minimum.  According to the CRF-cloud picture, this resulted with a
 weakening of the solar wind which increased the Galactic CRs reaching
 Earth, increased the cloud cover and reduced the average global
 temperature.  And indeed, during this period, the average land and
 marine temperature in the Northern hemisphere has dropped by about
 0.15$^\circ$K. A more detailed analyses \cite{Soon1996,Beer2000}
 which decomposes temperature trends into solar effects, anthropogenic
 and residuals shows that the component attributable to solar variability
 is actually larger -- a reduction of about 0.2$^\circ$K, where an
 increase of about 0.05$^\circ$K is a result of human activity. 

 To relate to Cosmic Ray flux variations, we use the ion-chamber data
 and the neutron monitor data in Climax quoted in \citet{Sven1998}. 
 Together with the ratio of the fluxes in Climax and in
 Huancayo/Haleakala \cite{Bazil2000}, one can obtain that the
 0.2$^\circ$K drop correlated with an increase of about 1.5\% in the CR
 flux at Huancayo/Haleakala that presumably is responsible for the
 cloud cover effect.  Namely, a 1\% increase in the CR flux is
 responsible for a 0.13$^\circ$K drop in global temperature.  This
 result is surprisingly close to the estimate (of -0.14$^\circ$K per
 1\% change in the CRF) that we obtained using the nominal value of
 the cloud-climate forcing.  This is probably a coincidence since realistically, we should expect an uncertainty of a factor of 2 or so. 
 Moreover, the long term effect of CRs on the temperature is probably
 somewhat higher since the data used includes marine air temperature. 
 Since the oceans have a high heat capacity (and therefore moderate
 the temperature), the long term temperature effects of CRs is
 expected to be higher.  That is to say, -0.13$^\circ$K per 1\% change
 in CRF is probably a lower estimate.
 
%-----------------------------------------------------------

\section{Theoretical Predictions vs.~Observations}

 We now proceed with the comparison between the model predictions as 
 derived in \S\ref{sec:model}, and the observational results 
 summarized (and derived) in \S\ref{sec:observations}.

\subsection{Spiral Structure vs.~Climate on Earth}

 For the nominal values chosen in our diffusion model, the expected
 CRF varies from about 25\% of the current day CRF to about 135\%. 
 This should correspond to an average absolute LACC of 7\% to 38\%
 respectively.  For the nominal value of $\lambda / \lambda_{0} =
 2.8$, it should correspond to a temperature change of $+10^{\circ}$K
 to $-5^{\circ}$K, relative to today's temperature.  This range is
 definitely enough to significantly help or hinder Earth from entering
 an ice-age.  (The typical difference between today, an interglacial
 period and the recent ice ages is of order $5^{\circ}$K).  The fact
 that we are currently in a glaciation epoch and that the observed CRF
 is higher than the long term average, over which Earth was mostly out
 of glaciation epochs, is consistent with this picture.

\begin{figure*}
\begin{center}
 \epsfig{file=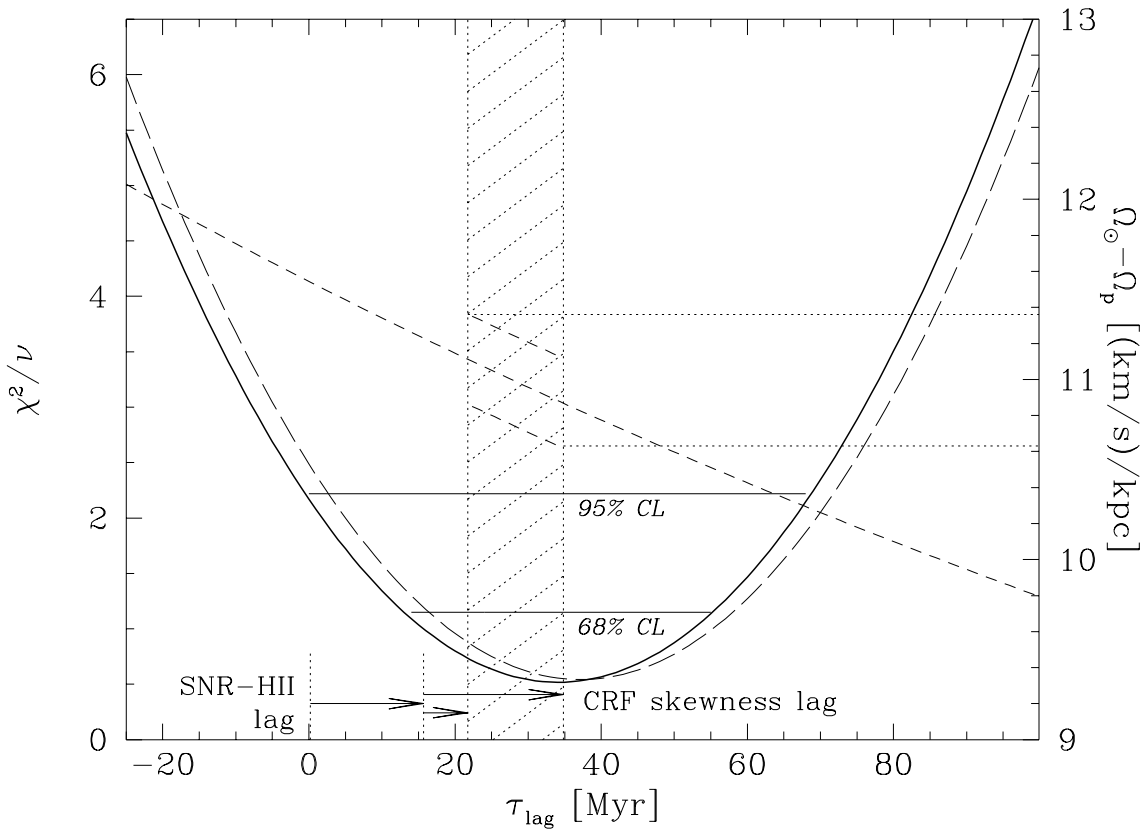,width=\figurewidth,angle=-90}
\end{center}
 \caption{
 %\small \sf 
 $\chi^{2}$ fitting of the glaciation epochs to Galactic spiral arm
 crossings as a function of the assumed lag between the HII arms and
 the middle of the glaciation epoch (solid line).  The hatched region
 describes expected lag from theory (including both SNR-HII lag and
 CRF skewness).  The short dashed line is $\Omega_\odot-\Omega_p$ (and its
 error, when in the shaded region) which best fits the glaciations as a
 function of the lag.  The long dashed line is a repeat for
 $\chi^{2}/\nu$ when assuming a symmetric arm location}
 \label{fig:Lag}
\end{figure*}

\begin{figure*}
    \vskip 0.5cm
\begin{center}
\epsfig{file=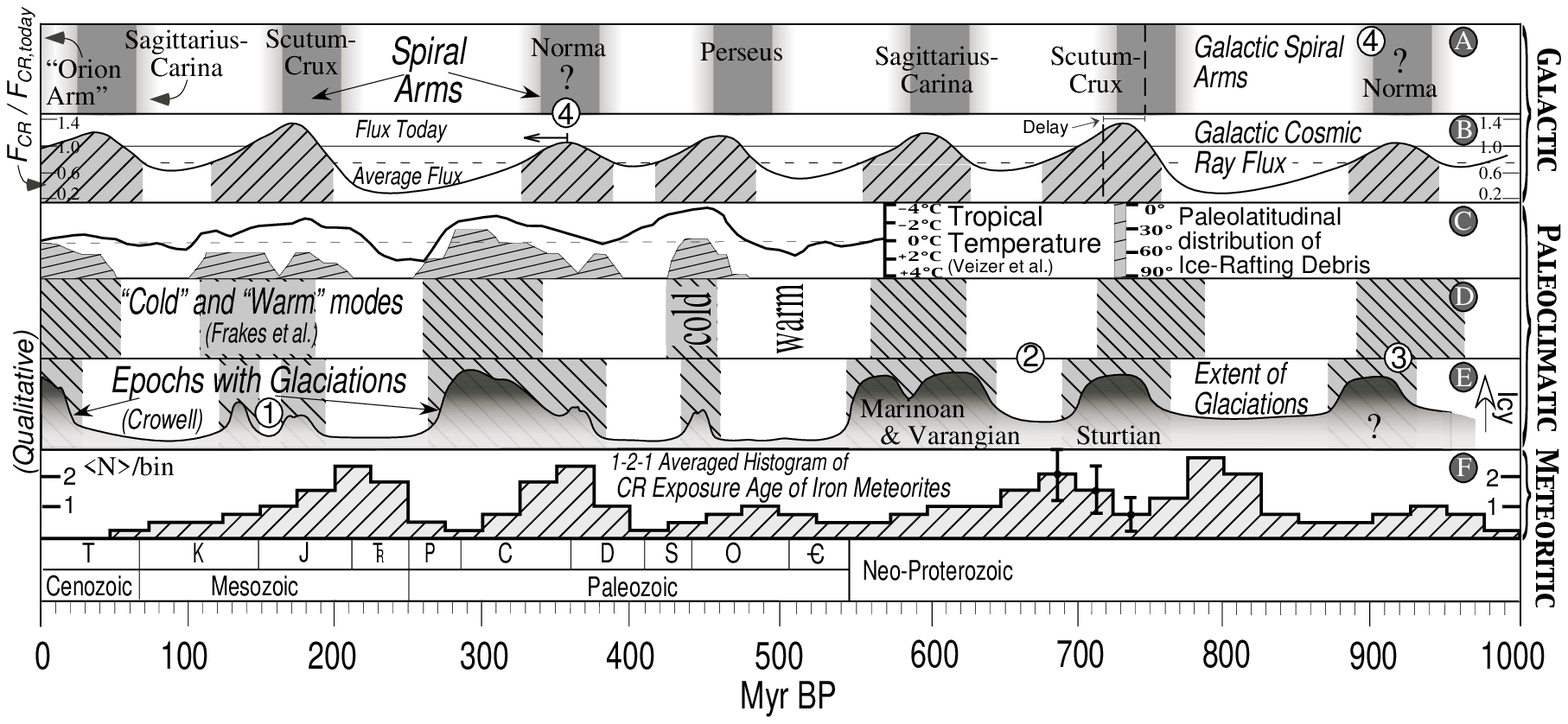,width=6in}
\end{center}
 \caption{
 %\small \sf 
 Earth's recent history.  The top panel (A) describes past crossings of
 the Galactic spiral arms assuming a relative pattern speed of
 $\Omega_{p}-\Omega_{\odot}=-11.0$ \omunit (which best fits the IAEs). 
 Note that the Norma arm's location is actually a logarithmic spiral
 extrapolation from its observations at somewhat smaller Galactic
 radii \cite{Leitch1998,Taylor1993}.  The second panel (B) describes the Galactic
 CRF reaching the solar system using the CR diffusion model, in units
 of the current day CRF. An important feature is that the flux
 distribution around each spiral arm is lagging behind spiral arm
 crossings.  This can be seen with the hatched regions in the second
 panel, which qualitatively show when IAEs are predicted to occur if
 the critical CRF needed to trigger them is the average CRF. Two
 dashed lines mark the middle of the spiral crossing and to the expected
 mid-glaciation point.  Panels (C) (D) and (E) describe the
 paleoclimatological record of the past Eon. The solid line in panel (C) depicts the tropical sea surface temperatures relative to today, as inferred from calcite and aragonite shells in the past $\sim 550$ Myr \cite{Veizer}. The filled areas describe the paleolatitudinal distribution of ice rafted debris \cite{Veizer}.   
Panel (D) and (E) qualitatively describes the
 epochs during which Earth experienced ice-ages, the top part as described by \citet{Frakes1992}, while the bottom one by \citet{Crowell1999}. The Phanerozoic part of
 panel E is directly taken from \citet{Crowell1999}.  The
 Proterozoic part is its extension using the data in Crowell.  It has an
 elevated base level (since glaciations were more common) and it
 describes the main epochs of extended glaciations. 
By fine-tuning the
 actual pattern speed of the arms (relative to our motion) to best fit
 the IAEs, a compelling correlation arises between the two.  The
 correlation does not have to be absolute since additional factors may
 affect the climate (e.g., continental structure, atmospheric
 composition, etc.).  Panel (F) is a 1-2-1 smoothed histogram
 of the exposure ages of Fe/Ni meteors.  The meteor exposure ages are predicted to cluster
 around epochs with a lower CRF flux.  }
 \label{fig:epochs}
\end{figure*}

 Comparison between the cosmic ray flux and the occurrence of
 glaciations shows a compelling correlation for the pattern speed
 chosen (fig.~\ref{fig:epochs}).  To quantify this correlation, we
 perform a $\chi^2$ analysis.  For a given pattern speed of the arm,
 we predict the mid-point of the spiral arm crossings using the data
 of \citet{Taylor1993}, as described by \citet{Leitch1998}.  The
 mid-point of the glaciation epochs are then predicted to lag by 15.4
 Myr (from the SN-HII lag) + 6-19 Myr (from CRF skewness) $\approx
 21-35$ Myr after the mid-point of the spiral crossing. 
 Fig.~\ref{fig:Lag} describes the value of $\Omega_{\odot}-\Omega_{p}$
 which minimizes the $\chi^{2}$ fit between the predicted and measured
 mid-point of the IAEs, and the minimum value of $\chi^{2}/\nu$
 obtained (with $\nu=6$ is the number of degrees of freedom minus the
 number of free parameters).  This is done as a function of the
 assumed lag between the mid-points of the spiral crossing and the
 predicted IAE. We find that the minimum $\chi^{2}/\nu$ of 0.5 is
 obtained for $\tau_\mr{lag} = 33$~Myr.  This is consistent with the
 predicted lag.  On the other hand, the graph shows that the no lag
 case is rejected at the $2\sigma$ level (85\%).  This implies that
 IAEs appear to be lagged after spiral crossings as predicted.  On the
 other hand, the graph shows that placing the arms symmetrically (with
 a $90^{\circ}$ separation instead of the values obtained by
 \citet{Taylor1993}) does not improve or degraded the fit.  It only
 shifts the preferred lag to slightly larger values.  Hence, this
 cannot be used to learn about the possible spiral arm asymmetries
 
 To check the consistency, a second analysis was performed to find the
 probability that a random distribution of glaciation epochs could
 generate a $\chi^2$ result which is as low as previously obtained. 
 To do so, IAEs where randomly chosen.  To mimic the
 effect that nearby glaciations might appear as one epoch, we bunch
 together glaciations that are separated by less than 60 Myrs (which
 is roughly the smallest separation between observed glaciations
 epochs).  The fraction of random configurations that surpass the
 $\chi^2$ obtained for the best fit found before is of order 0.1\%. 
 (If glaciations are not bunched, the fraction is about 100 times
 smaller, while it is about 5 times larger if the criterion for
 bunching is a separation of 100 Myrs or less).  

%Spur intermittent effect

%\subsubsection{Pattern Speed from Cluster age distribution}
%
%\begin{figure*}
%\epsfig{file=theta_dist.eps,width=6.in,angle=-90} 
% \caption{The histogram of cluster ``spiral angle'' $\theta_{sp}$ for
% the clusters in the 300 Myr BP peak.  (i.e., with $8.4 <log_{10} t <
% 8.7$), for three different spiral pitch angles. The wider the spread
% is, the faster should the pattern speed be in order not to spread the
% peak in the SFR history.}
%\end{figure*}

\subsection{Comparison with CRF record in Fe/Ni meteorites}

 A smoking gun, which would unequivocally demonstrate that the
 spiral-arm $\rightarrow$ cosmic-ray $\rightarrow$ climate connection is
 real, would be a ``historic'' record which would reveal that the CRF
 was indeed variable.  Such an independent record could on one hand
 show that the CRF varied as expected from the spiral arm CR diffusion
 model, while on the other hand, that it varied synchronously with the
 appearance of ice-age epochs on Earth.  In \S\ref{sec:IronCRF}, a new
 method was developed with which we could extract the actual CRF
 history.
 
  It was found that the CRF record shows a clear variability signal 
  with a periodicity of $143 \pm 10$ Myr. This variability is 
  consistent both in periodicity and phase, with the predictions of 
  the CR diffusion model which incorporates the spiral arms, of which 
  the expected periodicity is $134 \pm 25$ Myr. This is clearly 
  demonstrated in fig.~\ref{fig:Kages}.
  
  Since the periodicity in the CRF is more accurately known than the 
  spiral arm pattern speed (because of both statistical and 
  systematic errors), it is a much stronger tool to use when comparing 
  the astronomical data with the geological one. Indeed, the 
  periodicity of the IAEs as can be obtained from table 
  \ref{table:iceages}, is found to be $145.5 \pm 7$ Myr, which nicely 
  agrees with the aforementioned CRF period of $143 \pm 10$ Myr. This 
  agreement in phase can also be seen in fig.~\ref{fig:Kages} where 
  the clusters of exposure ages tend to be anti-correlated with the 
  occurrence of IAEs. 
  
  The last result is of course trivial once we consider that the CRF
  as recorded in Fe/Ni meteorites was already shown to agree with the
  predicted CRF using the galactic model and that the latter was shown
  to agree both in phase and period with the occurrence of IAEs.
  
  The implications are not only statistical, but also qualitative,
  since the above agreements provide a ``missing link'' which
  explicitly point to the cosmic rays as being the culprit in climatic
  variability.  Namely, it looks as if we found the smoking gun we sought
  after.
 
\subsection{The past Star Formation rate history and ice-epochs}

A summary of the past star formation rate history was given in 
\S\ref{sec:SFR}. It is summarized again in figure \ref{fig:SFR},
together with the epochs during which Earth witnessed ice-ages.

The main correlation apparent from the graph is the paucity in the MW's
star formation rate between 1 and 2 Gyr BP (with the cautionary note of \S\ref{sec:caveats}),
 which coincides with a
long interval on Earth during which there were no apparent glaciations
whatsoever.  On the other hand, during the last eon, and the one
between 3 and 2 Gyrs BP, Earth did have glaciations and the SFR was
higher.  Although it is speculative, this variability in SFR appears
to correlate with perigalacticon passages of the LMC, and in fact, 
also with the SFR in the LMC.

For the SFR on shorter time scales, we have to rely on the cluster 
data (see fig.~\ref{fig:SFR}), or the mass distribution of nearby stars 
\cite{Scalo1987}, which have  a `finer' resolution than the SFR
calculated using chromospheric ages of nearby stars. These show that at 0.3
Gyr BP, there was a higher than average SFR. A second peak 
appearing in the cluster data, could have
occurred at 600 Myrs BP (though the data is not statistically
significant).  Interestingly enough, the IAE 300 Myrs and 600-700 Myr
BP appear to be correlated with more extensive glaciations: The recent 
peak correlates with the Carbonaceous-Permian IAE, while the peak at 600-700
Myrs coincides with the late Neo-Proterozoic IAE, both of which were 
relatively extensive. 

Although it is hard to quantify the statistical significance of the 
SFR-climate correlation, the fact that the results are consistent with that expected 
from the theory is reassuring (though see the possible caveats in \S\ref{sec:caveats}).
Namely, we expect and indeed find the 
relation: Higher star formation rate $\rightarrow$ more CR 
generation $\rightarrow$ colder climate.

\section{Discussion}

\begin{figure*}
\begin{center}
\epsfig{file=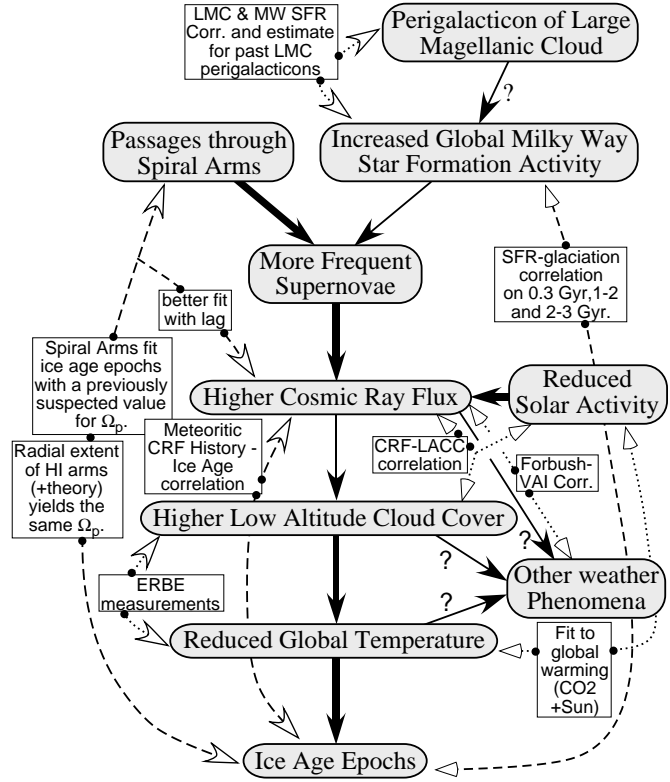,width=\smallfigurewidth}
\end{center}
\caption{The model for how the Galaxy affects the ice ages.  Solid
arrows reflect physical cause and effect.  Bold lines are hard to
dispute relations while thin lines are reasonable relations based on
observational correlations.  Arrows with question mark are unknown
though possible relations.  Dotted and dashed arrows reflect
observational correlations which relate two (or more) physical
ingredients.  Dotted lines are previously found correlations while
dashed lined are correlations obtained and described here.}
 \label{fig:model}
\end{figure*}

The basic evidence presented supports a picture in which climate on
Earth is affected by our changing location in the Milky Way, by way of
a variable cosmic-ray flux.  The various physical links and the
observational evidence which support these relations are charted in
fig.~\ref{fig:model}.  One should note the redundancy in the 
key results which support this conclusion. These are the temporal
correlations between the expected spiral arm crossing, the variable
CRF observed in Iron meteorites, and the appearance of ice-ages on
Earth.

For nominal values taken, we find an effect of +5 to -10$^{\circ}$K
relative to today, however, there are some large uncertainties.  A
factor of order unity arises from the uncertainty in the cloud --
temperature relation.  Another factor of order unity arises from the
uncertainty in the relation between the Cosmic Ray flux itself and the
cloud cover.  The normalization of the latter can be achieved using
the solar modulation with the complication however that it modulates CR fluxes of different
energies differently.  Nevertheless, it is clear that the low altitude 
nature of the CRF-climate effect,
requires the relevant energies to be high ($\gtrsim 10$~GeV). Thus, a 
normalization factor of 2 higher or
lower between the CRF and cloud cover connection is conceivable.  A consistency
check  which verifies that the net CR $\rightarrow$ LACC 
$\rightarrow$ temperature relations that we obtained are reasonable, 
is using global warming in the past century to directly calculate the 
CR $\rightarrow$ temperature relation. 

The last source for an inaccuracy in the CR $\rightarrow$ temperature
relation arises because we extrapolate a several percent effect to
several tens of percent.  Without knowing the physical process behind 
the effects of the CRs on the atmosphere.
We cannot know, for example, whether the effect does not saturate at
small relative changes.

Although the actual links presented here relate the CRF to the
changing global temperature, there is good reason to believe that the
link itself is through a variable cloud cover.  First, such a link was
already empirically observed in the form of a direct correlation
between the low altitude cloud cover and the CRF. Moreover,
the observed correlation yields a radiation driving that is more than
sufficient to explain the occurrence of ice-ages, but also the changes
in the past century which are attributable to solar activity.  In other
words, other pathways relating the CRF with temperature are not
required at this point, but clearly, we cannot rule them out.

If the link is not through the variable low altitude cloud cover (or
if an additional link exists), then lower energy CRs could be
affecting the climate as well. In this case, the effects that a dense ISM
cloud has on the heliosphere \cite{Begelman1976} could clearly be
affecting the climate as well since these events will drastically
change the low energy charge particle flux reaching Earth. In such a
case, the ice-age epochs could be dotted with relatively short events
during which the climate changes dramatically for the typical durations it takes to cross such ISM
clouds (either heating or
cooling, depending on whether the reduced flux of solar CRs or the 
increased galactic CRs are more important, and on the actual climatic 
mechanism). 

For example, if relatively low molecular cloud densitites of 100 cm$^3$ can significanty alter the heliospheric structure and with it the low energy cosmic ray flux reaching Earth, and if the latter do affect climate, then an effect could  actually take place for $10^7$ years, which is the typical duration it would take the solar system to cross a 100 pc cloud. In any case, 
These events are more likely to be synchronized with the spiral arm crossing itself and not lagged behind it.

One should also note that passing through young stellar
nurseries could also locally increase the intrinsic CRF reaching the solar
system. Through the low altitude cloud cover connection, this would
cause dramatic short term  reduction in the temperature. Namely, the
long ice-age epochs could be dotted with short events of even colder
climate.

The CR diffusion model developed here, has shown that the spiral 
structure should be incorporated in CR diffusion models since 
qualitative and quantitative corrections are otherwise missed. (The model 
described here is nowhere as complex as some CR models, so neither it 
nor previous models should be the final word).  Moreover, even the 
spiral structure itself deserves a more detailed description. For 
example, the spiral arms were assumed to be straight ``rods''. A 
much bigger effect could however arise from the approximation that the 
diffusion coefficient is the same everywhere. However it is expected 
that the diffusion coefficient will be lower in the denser spiral arms. 
This could have an effect on the entrapment of the CR in the spirals 
for a longer time, thereby increasing the CR density contrast.

To the extent that spiral density wave theory applies to the Milky Way's spiral arms,  the spiral structure of the outer parts of our galaxy ($R\gtrsim
R_{\odot}$) is apparently pinned down. Therefore, it should serve as an outer
boundary when Milky Way models are constructed. We found that the structure
between us and the center will however be different in nature. The reason is that the
outer 4-arm spiral cannot extend much further inside the solar
galactocentric orbit, though spiral arms are seen further inside the 
disk. This implies that models previously used to
describe the galaxy with only one pattern speed are not sufficient. It could explain why previous results for the number were in contradiction.

Another interesting observation  is that geology is now is a unique
position to serve as a tool in the study of galactic dynamics.  For
example, it now appears that the life time of spiral arms  is at least $\sim 900$
Myr or 1.5 rotations. This is expected whether the spiral arms
are a `transient' phenomenon or a more permanent one (an issue which
is still not settled). Should however the same periodicity be found in
the glaciations 2-3 Gyr BP, then a very interesting constraint could be
placed on the spiral dynamics. 

Moreover, using temporal variability (either in the CRF or the
glaciations), we have managed to measure the spiral pattern speed more
accurately than using pure `astronomical' methods. The accuracy with
these temporal methods is in fact large enough that one needs to
consider that the corrections from our diffusion in the galaxy results
in an offsetted pattern speed, since this correction is of order the error
bars themselves or even larger. This is the first time a temporal 
record (instead of a ``snapshot'') is used to study galactic dynamics.

%---------

 It is also worth noting that the effects of terrestrial magnetic
 reversals on climate are probably not dramatic as could be naively expected, even
 though the total CRF is significantly increased in such events.  The
 main reason is that the LACC-CRF connection is observed in low
 altitudes and low latitude clouds. This necessarily implies that
 high energy CRs are those mostly responsible for the effect.  These
 CRs reach the equator irrespective of the strength of Earth's
 magnetic field.  For example, if we look at the original results of \citet{Compton33}, we find that the change in atmospheric ionization at sea level varies by only $\sim 14\%$ between the magnetic equator and poles. Therefore, even if the Earth's magnetic field were to completely vanish, our estimates for the temperature change would correspond to at most a 1 to 2$^\circ$K reduction in temperature.  This is corroborated by the Laschamp event (36-41.5
 kyr BP) during which the terrestrial field was significantly reduced
 \cite{Guyodo1996}.  During the event, an increased $^{10}$Be and
 $^{36}$Cl flux was recorded in Greenland ice cores, but no statistically significant climatic effect was recorded in $^{18}$O or CH$_{4}$ \cite{Wagner1999}. This is expected considering that the temperature variability recorded in the ice cores, either before or after the event, is much larger than 1 to 2$^\circ$K.

% Although it is not clear to what extent glaciations played a role in
% mass extinctions, some mass extinctions could be due to a collision
% with a comet, as is suspected to be the case with the K/T extinction
% \cite{Zhao1989}\cite{Bergh1994}.  These events should also be clustered
% around spiral arm crossing as it is then when the Oort cloud is most
% likely to be perturbed. We find the best fit for the relative
% rotation speed of the arms is consistent with the major extinctions
% \cite{Raup1982} being related to spiral arm crossing, except for the
% Permian/Triassic.  Thus, if the latter is indeed related to a bolide
% collision \cite{Becker2001}, it would more likely in this picture be
% an asteroid than a comet.

\subsection{Caveats}
\label{sec:caveats}

The picture presented here includes many elements which add up to what appears to be a coherent picture. Unfortunately, most if not all observational data used here, whether astronomical, meteoritic, or paleoclimatological does not come without a grain of salt. The purpose of this section is to discuss and summarize the limitations of the data and possible caveats. Hopefully, by illuminating where the problems may rest, future research will try to address these points with the goal of resolving them.

{\bf Paleoclimatological Data}: By far, the least conclusive data used is the past evidence for the occurrence of glacial epochs. In particular, it is debated whether ice-age epochs in the past billion years have been periodic or not. Some claim that a periodicity exists \cite{Williams1975,Frakes1992}, while others claim that insufficient data exists to claim periodicity \cite{Crowell1999}. Besides the timing of the glaciations (which becomes poorer as we go back in time), there is also the problem that ice-age epoch severity clearly varies from one IAE to the next.
For example, the mid-Mesozoic ice-age epoch ($\sim 150-170$ Myr BP) was clearly very 'mild' compared to other IAEs. On the other hand, from 800 Myr BP until well after 600 Myr BP, it seems that Earth continuously had some sort of glaciations present, with colder episodes around 600 and 750 Myrs BP. In fact, around 750 Myrs BP, there is evidence for glaciations at sea level and equatorial latitudes. That is, glaciations more extensive than anything seen by Earth at any period \cite{Hoffman1995}. In addition to the problem of a varying severity or extensiveness, there is also the question of length. For example, the  Carboniferous-Permian glaciations were 3 or 4 times the duration of the short Silurian-Devonian glaciations that preceded.   These problems are also noted in figure \ref{fig:epochs}. Note however, that even though the above problems exist, the {\em timing} of the IAEs of Frakes and of Crowell are generally consistent with each other.

 Another point which one might note at first glance of fig.~\ref{fig:epochs} is
 that although the general agreement is good, there are some
 inconsistencies between spiral crossings and actual occurrence of
 IAEs.  As previously explained, this should not be a concern because
 of several reasons.  First, the actual spiral crossing signal is
 expected to have a `jitter' because of the epicyclic motion of the
 solar system.  Second, since the Milky Way is not a `grand design'
 spiral galaxy, the spiral arms are not the perfect geometrical
 objects they are assumed to be.  In particular, `gaps' in which the
 spiral arms are weaker or arm `spurs' which fill the inter-spiral-arm
 with young CR producing stars are not uncommon.  The lifetime of the
 spurs, however, is less than a galactic orbit. Therefore, they are not
 expected to repeat themselves after 4 spiral crossings.  Third,
 geological factors are also expected to be important in climate change. 
 For example, the changing continental structure might free the poles
 of land masses, and could even allow temperate ocean currents to
 reach the poles.  This can hinder Earth from triggering the formation
 of ice sheets.

 The above factors appear to be particularly important in the last
 IAE. Although the main spiral arm crossing took place about 50 Myr
 BP, the last IAE started only `recently', about 30 Myrs BP. The
 reason we are now in an IAE is easy to explain since we are in the
 Orion armlet.  This leaves the question of why didn't the IAE
 begin as the solar system entered the Sagittarius-Carina arm?  This could be a result
 of factors intrinsic to Earth.  The main factors often cited as the
 reason that the latter part of the Cretaceous
 was one of the warmest periods in Earth's history  (e.g., \cite{Huber1998}), are that the land
 mass did not occupy the poles and that the global sea level was
 highest than ever (thus moderating the temperatures).  Nevertheless,
 if careful attention is given to the temperature variations over the
 Cretaceous and Cenozoic (e.g., \cite{Huber1998},
 \cite{Zachos2001}) then it is apparent that the temperature
 variations were more complicated than a simple monotonic cooling.  At
 90 Myr BP, the average global temperature peaked.  But in the 10 Myr
 preceding the K/T transition (at 66 Myr BP), the temperature dropped
 by more than 5$^{\circ}$K. The temperature rose again by several
 degrees, and peaked at around 52 Myr BP, after which it declined
 (though non-monotonically) until present, when extensive glaciations
 are common.  It is tempting to attribute the first minimum to the
 crossing of the Sagittarius-Carina arm. Namely, Earth did experience
 a reduced global temperature however, because of intrinsic effects
 (and perhaps because the arm at the point of crossing was not
 prominent), the temperature did not fall enough to trigger
 glaciations. After the Sag.-Car. arm was crossed, the temperature 
 rose again. However, the Orion arm and the changing global 
 conditions caused a large cooling afterwards. 

Although the above explanation is far for satisfactory, it shows that the model developed in the paper is incomplete. In particular, it is probably the case that ``standard"  paleoclimatic factors intrinsic to Earth are as important at determining the global climate as the external effect of the cosmic rays.

{\bf Astronomical Data}: We have used in the analysis astronomical data describing the spiral structure, spiral pattern speed and star formation rate. Unfortunately, there is still no consensus on any one of them.

Although the analysis presented here, and in particular the meteoritic exposure age data, seems to be consistent with the various spiral arm indicators located outside the solar galactocentric radius, if we look inwards, the structure and dynamics appear significantly less clear. In particular, there are convincing arguments which indicate that the solar galactic orbit is close to co-rotation (e.g., \cite{Mishurov1999}). If true, it would imply a spiral arm pattern speed similar to $\Omega_\odot$, instead of a much larger $\Omega_\odot-\Omega_p$, as found here. Obviously there is a contradiction that should be resolved. One possibility raised in \S2, is that the Milky Way has more than one set of spiral arms with different pattern speeds. This is required if the outer parts have a low pattern speed, and it would explain the large confusion that exists in the field. Clearly, this issue is still not resolved and deserves more attention than it has received in recent years.

Interestingly, having two sets of spiral arms also has the potential of explaining the `staggered'
severity and length seen in the ice-age epochs. Namely, it could be that we are also affected by the spiral structure internal to our galactocentric radius, and that its period relative to us is about 300 Myr. Nevertheless, this is quite speculative, and it is not unlikely that this staggered behavior arises from terrestrial factors. For example, atmospheric CO$_2$ levels around the Carboniferous-Permian period were several times smaller than the other periods in global history with the exception of recent geological times.

 Another caveat has to do the location of the Norma arm at our
 galactic radius.  This location was obtained through a logarithmic
 extrapolation from where it is observed at smaller galactic radii
 (\`a la \citet{Leitch1998,Taylor1993}).  However, as discussed in \S\ref{sec:ptsHI}, there
 are good reasons to believe that the structure within the solar circle
 is different from the one at our radius and outwards.  This would imply
 that the Norma arm seen at smaller radii is not necessarily part of
 the same structure that should be localized between the Scutum-Crux
 and the Perseus external arms, on the other side of the galaxy. 
 If so, the preferred crossing of this arm
 would be later by $\sim 40$ Myr, as its preferred location would be symmetrically in between the adjacent arms.  If we look at
 fig.~\ref{fig:epochs}, we see the this will in fact increase the
 agreement between the predicted location of the spiral crossings
 around 300 and 900 Myr BP and both the time of occurrence of the
 relevant IAEs and the observed CRF variability. It would not explain though the extent the the Carboniferous-Permian IAE.

A third astronomical caveat has to do with the star formation rate. The SFR with which the glacial activity on Earth was compared with was that of \citet{RochaPinto2000a}, which is also consistent with various other authors. However, like most results quoted here, there are a few contradictions published in the literature, which, if verified will pose a problem for the model presented here. In particular, \citet{HippSFR} have used the {\sc Hipparcos} astrometric data combined with detailed  theoretical isochrones to obtain the SFR of nearby stars (which due to diffusion corresponds to the azimuthally averaged SFR on periods over a billion years). The gross features they obtained are inconsistent with other analyses (for example, with \cite{RochaPinto2000a}). In particular, the SFR they obtained decreased from 2 Gyr BP until today with only relatively small `wiggles'. Moreover, they did not find a dip between 1 and 2 Gyr BP.
Thus, an independent study of the {\sc Hipparcos} data should be pursued in order to resolve this discrepancy. 

{\bf Meteoritic Data}: 

Regarding the variable CRF from exposure age data, one should note the
following.  Just the data showing the clustering of meteorites by {\em
itself} can be interpreted in several ways.  The clustering can either
be a manifestation of a variable CRF, or, it can be a genuine
clustering of exposure ages.

Since the stable isotopes used in the currently employed
methods for exposure dating Fe/Ni meteorites (e.g., $^{36}$Ar,
$^{21}$Ne, $^{41}$K), are all formed primarily from Galactic CRs
\cite{Michel1995}, the observed data could not be a result of a
variable solar CRF. Thus, within the interpretation of a variable CRF,
the observed variability can arise from either an intrinsic
variability in the Galactic CRF (as assumed in this work) or a
variability induced by a variable heliosphere.  In the second
possibility, one can accredit the exposure data to a variable
solar wind (with the observed periodicity of $\sim 150$~Myr) and
attribute the observed correlated climatic variability to the solar
variations.  Although possible, this option which has far reaching
implications, seems less reasonable because of the correlation of the
CRF and climatic data with the observed galactic pattern speed with
the same period. Namely, it would be strangely coincidental if the 
sun were to have a variability that is synchronized with the spiral 
arm crossings.

The third possibility for interpreting the exposure age data is that
the CRF was constant and that the clustering in the exposure ages is
real.  Namely, episodes in the past in which asteroids or comets were
more likely to break apart existed periodically.  This is not an
unreasonable hypothesis.  For example, one could claim that the Oort
cloud is perturbed more often during spiral arm crossing, injecting
more comets to the inner solar system, thus breaking apart more
asteroids. Moreover the more numerous debris could reasonably have
climatic effects as well (see \S\ref{sec:intro}).  Within the context
of this interpretation, clusters of meteoritic exposure ages, as well
as the ice age epochs, should coincide with the spiral arm crossing
with no phase lag. The data however shows that the clusters are not in 
phase with either the spiral crossings or the IAEs, nor are the IAEs
in phase with the spiral arm crossings. 

Thus, the most consistent picture arises if we interpret the exposure
age clusters as a result of an intrinsically variable Galactic CRF.
One way to prove this, is to look at age distributions of Iron
meteorites which were CR exposure dated using a method in which the
stable isotope that accumulates is sensitive primarily to the solar
cosmic-ray flux.  The reason is that if the clustering is real (and
not due to a variable Galactic CRF) then the {\em same} clustering
will appear in solar sensitive dating methods.  This would also be
useful to rule out the possibility that the apparent variations are
not Galactic in origin, but are induced by solar variability instead. 
If this unlikely case is true (which would imply that IAEs are linked
with solar variability, since the CRF-climate correlation exists 
irrespectively), then the opposite clustering will be present in solar
sensitive data.  With the most reasonable interpretation, though, no
clustering is predicted in the solar sensitive data, once it could be
obtained.

\section{Summary and Conclusions}

\newcounter{myenum}

The thesis presented here relates the following topics: The Milky Way
dynamics, Cosmic Ray diffusion in the Galaxy, the CR record in Iron
meteorites, the effect of Cosmic Rays on climatology, and
glaciology. This is achieved by considering the intimation that CRs
can affect the global cloud cover and that the CR flux from the Galaxy
should be variable. Before summarizing  the main conclusions relating
the above topics, ``preliminary'' conclusions on each of the various topics can
be drawn from the background analysis.

\smallskip
\noindent
{\bf Milky Way Dynamics}: By studying the 4-arm structure seen in HI,
extending to roughly $2 R_\odot$, stringent constraints can be placed
on the spiral pattern and dynamics of the Milky Way:
\begin{enumerate}
  \item Since observations of HI {\em outside} the solar circle suffer
  no velocity-distance ambiguity, the observations of 4-spiral arms
  should be considered robust.  {\em If one further assumes} that these arms
  are density waves (which is by far the most consistent explanation
  for spiral arms) then these arms cannot extend further out than the 4
  to 1 Lindblad resonance.  This implies $\Omega_p \lesssim 16 \pm 2.5$
  \omunit.  This number agrees with roughly half the
  determinations of $\Omega_p$, which cluster around this value.

  \item Several additional considerations point to the above value
  being not an upper limit for $\Omega_p$, but the
  actual value of it.

  \item The 4-arm spiral cannot extend much further in from the solar
  galactocentric radius. Several possibilities were raised. Since the
  MW has a bar, at least two different pattern speeds exist in the MW.

  \item On a more speculative note, long term activity in the star formation rate appear to 
  correlates with
  activity in the LMC and possibly with its perigalacticon passages,
  suggesting that long term star formation in the Milky Way and LMC are
  related.

\setcounter{myenum}{\value{enumi}}

\end{enumerate}

\noindent
{\bf CR Diffusion:}

\begin{enumerate}

\setcounter{enumi}{\value{myenum}}

 \item Since CR sources are clearly more common to the Galactic spiral
 arms, models for CR propagation in the MW should take this fact into
 account.  Otherwise, significant discrepancies can arise.  In
 particular, quantities such as the ``age'' of the cosmic rays
 reaching Earth can be distorted.

 \item For typical diffusion model parameters, the CRF is expected to
 vary by ${\cal O}(1)$, which is consistent with radio observations of
 external spiral galaxies.

 \item The distribution of CRs is both lagging behind the spiral
 crossings (when defined for example by the HII regions) and skewed
 towards later times.  The skewness arises from the asymmetry
 introduced by the spiral arm motion.  The lag in the peaks arises
 because the SN explosions are lagging the ionizing photons that
 produce the HII regions, and which were used here (and in
 \cite{Taylor1993}) to define the location of the arms.

\setcounter{myenum}{\value{enumi}}

\end{enumerate}

\noindent
{\bf CR record in Iron meteorites}:

\begin{enumerate}

\setcounter{enumi}{\value{myenum}}

 \item The ``standard'' method for extracting historic variability of
 the CRF, by comparing $^{41}{\rm K}/^{40}$K dating to dating with a short
 lived unstable nucleotide (such as $^{10}\mr{Be}/^{21}\mr{Ne}$) are effective
 only at extracting ``recent'' changes (over several Myrs) or secular
 changes over longer durations, but they are not effective at extracting
 the signal expected from the periodic variability of the spiral arm
 passages.

 \item An effective method for finding the periodic CRF variations is a
 statistical analysis of CR exposure ages.  It assumes that the rate
 at which ``new exposed surfaces'' appear in Iron meteorites does not
 vary rapidly.

 \item This method shows that the exposure age data is consistent with it being periodic, with
 a period of $143 \pm 10$ Myr, and a CRF contrast of $\max (\Phi) /
 \min (\Phi) \gtrsim 3$.  It uses 50 meteorites which were dated only
 with $^{41}{\rm K}/^{40}{\rm K}$. The method reveals that statistically, it is unlikely that the meteor exposure ages were generated by a random process.

\setcounter{myenum}{\value{enumi}}

\end{enumerate}

\noindent
 {\bf The Milky Way -- Climate connection}: By comparing the above
 findings to the appearance of glaciations on Earth, the following
 conclusions can be deduced:

\begin{enumerate}

\setcounter{enumi}{\value{myenum}}

 \item Periodic glaciation epochs on earth in the past 1 Gyr can be
 consistently explained using the proposed scenario:
 Periodic passages through the Galactic spiral arms are responsible
 for an increased CRF, an increased LACC, a reduced global temperature
 and consequent ice-ages.

 \item Long term glaciation activity is apparently related to the
 global SFR activity in the Milky Way.  This may be related to LMC
 perigalacticon passages which would imply that the nearby
 extragalactic environment could be added to the factors affecting the
 global climate.

\setcounter{myenum}{\value{enumi}}

\end{enumerate}

The evidence upon which the above conclusions were based, is the
following:

\begin{enumerate}

\renewcommand{\labelenumi}{\Roman{enumi}}

 \item The period with which spiral arm passages have occurred using
 the HI data alone (and which should also correspond to the CRF
 period), is $163 \pm 50$ Myr.  By combining the result of the 8 total
 measurements that scatter around this value, the predicted spiral arm 
 crossing period is $134 \pm 22$ Myr.  The period with which the CRF appears to vary
 using Fe/Ni meteorites, is $143 \pm 10$ Myr. 
 The period with which glaciations have been observed to reoccur on
 Earth is $145.5 \pm 7$ Myr on average.  Clearly, all three signals
 are consistent with each other.

 \item By comparing the actual prediction for the location of the
 spiral arms to the glaciations, a best fit of $143 \pm 5$ Myr is
 obtained. The phase of all the three signals using this periodicity
 is then found to fit the predictions. In particular:
 \begin{itemize}

     \item The average mid-point of the glaciations is lagging by 
     $33\pm 20$~Myr after the spiral arm crossing, as is predicted 
     ($31\pm 8$~Myr, from a possible range of $\pm 75$~Myr).

     \item The CR exposure ages of Iron meteorites cluster around
     troughs in the glaciations.

 \end{itemize}

% \item The fit is better when the actual asymmetric location (as
% fitted by \cite{Taylor1993}) of the arms is considered.

 \item A random phenomenon for the appearance of glaciations is
 excluded with very high statistical significance.

 \item Long term variability in the appearance glaciations correlates
 with the observed SFR variability of the Milky Way.  In particular,
 the lack of apparent glaciation between 1 and 2 Gyr BP, correlates
 with a particularly low SFR in the milky way (less than half of
 today), while a high SFR rate between 2 and 3 Gyr BP (with a peak
 towards 2 Gyrs), correlate with the glaciations that Earth
 experienced between 2 and 3 Gyrs BP.

\end{enumerate}

\noindent {\bf Additional conclusions}: More conclusions can be
reached by considering the above results.  They do not pertain to
climatic variability, but instead, they can be drawn by assuming that
long term climate variability is indeed a measure of CRF variability, as the above results seem to endorse.

\begin{enumerate}

\setcounter{enumi}{\value{myenum}}

 \item Once more accurate ``Eulerian'' measurements of the Galactic
 pattern speed will be available, interesting constraints could be
 placed on solar migration in the galaxy.  This is achieved by
 comparing the ``Eulerian'' measurements of the pattern speed to the
 ``Lagrangian'' pattern speed (as measured by the moving solar
 system), which is now known to an accuracy of $\pm 3\%$.
 
 \item The variability observed in the CRF record in Iron meteorites
 (see point 10) can be used to place constraints on models for the CR
 diffusion in the Milky Way.  This is in addition to the currently
 used constraint on the survival ratio of spallation products
 (primarily Be).  Thus, from CRF variability, we find $D \lesssim 2
 \times 10^{28} {\rm ~cm}^2/{\rm sec}$ and $l_H \lesssim 2$ kpc.

 \item The 4-arm spiral pattern has been stable for at least a
 billion years. A better understanding of past Glaciations (e.g., 2-3
 Gyrs ago) could place interesting constraints on the lifetime of the
 spiral arms which cannot be done by any other means.

\end{enumerate}

As apparent from the above list, there are several interesting ramifications to the picture presented here. However, one which bears particular interest on global warming and which was not discussed here, is that it now appears even more plausible that cosmic rays indeed affect the global climate, as suggested for example by \citet{Ney1959}, \citet{Tinsley1991}, and \citet{Sven1997}. This implies that it is now more plausible that solar variations are affecting climate through modulation of the cosmic ray flux. This would explain an important part of the global warming observed over the past century. 
It is because of this important aspect, in particular, that we are obligated to iron out the still unclear points  raised in the caveats section, whether it be the paleoclimatological record, the possible mechanism by which cosmic rays affect climate,  or our astronomical understanding of the structure and dynamics of the Milky Way.

\section*{Acknowledgements}

 The author is particularly grateful to Peter Ulmschneider for the
 stimulating discussions which led to the development of this
 idea. The author also wishes to thank Norm Murray, Chris Thompson,
 Joe Weingartner and Chris Matzner for their very helpful comments
 and suggestions, and also Fred Singer, Sidney van den Bergh and in particular the anonymous referee for helping improve the manuscript.

 \section*{Appendix A: The ``Age'' of cosmic rays in the spiral arm diffusion
 model}

The tightest constraint on different types of CR diffusion models in
the galaxy can be placed by comparing their predictions to the expected
ratio between $^{10}$Be, which is a radioactive isotope, to $^{9}$Be,
which is stable. Both nucleotides are formed by CR spallation of mainly
C, N and O.

In a leaky box model, the ratio at steady state between the density of
specie $i$ of a stable isotope to that of an unstable isotope $j$, can
be obtained from eq.~\ref{eq:leakybox}. It is:
\begin{equation}
  {N_j \over N_i} = {{\tilde Q}_j \over {\tilde Q}_i} {\tau_r \over
  \tau_{e,lb} + \tau_r}
\end{equation}
where $\tau_r$ is the decay rate of the unstable isotope and
$\tau_{e,lb}$ is the leaky box decay rate. It is assumed that the
spallation time is much longer than $\tau_{e,lb}$ or $\tau_r$.

Different measurements for the isotope ratio $^{10}$Be/Be are
$6.4\pm1.5$\% \cite{Wieden1980}, $3.9\pm1.4$\% \cite{Simpson1988},
$4.3\pm1.5$\% \cite{Lukasiak1994}.  These ratios respectively yield a
$\tau_{e,lb}$ of $18.1^{+8.1}_{-5.0}$ Myr, $30.3^{+21.5}_{-10.0}$~Myr, and
$27.1^{+18.9}_{-9.0}$ Myr \cite{Lukasiak1994}.  We shall adopt an age
of $24^{+12}_{-6}$ Myr which is the measurements' average and a
somewhat conservative error which is the maximum error divided by
$\sqrt{3}$.

In the spiral arm diffusion model, there is no steady state, so the
calculation of the ratio ${N_j / N_i}$ is more complicated.

To simplify the calculation, we utilize the previously obtained result
that to a reasonable approximation, the effects of the boundaries is
to have the CR distribution decay exponentially, not unlike the leaky
box model.  (Namely, we approximate the $\vartheta_4$-function as an
exponent).  This implies that the number density of the CRs that were
emitted at a given time decays exponentially with a time constant of
$\tau_e=l_H^2/(2.47 D)$.  Thus, the density of specie $k$ which is the
parent nucleotide of the spallation products (namely, C, N and O
producing the Be) and which where generated at time $t'$ will decay as
$\exp(-(t-t')/\tau_e)$.

The two (united) equations that describe the production and decay of the
spallation products formed by parent nucleotides that were generated 
at time $t'$, are therefore:
\begin{equation}
 {d n_{i,j} \over d t} = -{n_{i,j} \over \tau_e} + q_{i,j} \exp
 \left(-{t-t' \over \tau_e}\right) - \left\{ 0~ {\rm or}~ {n_j\over
 \tau_r} \right\}.
\end{equation}
$n_{i,j}(t,t')$ is the density of specie $i$ (stable) or $j$ (unstable) that were
generated from nucleotides of type $k$ at $t=t'$.

The solution to the two equations with the assumption that
$n_{i,j}(t=t')=0$ (namely, that the spallation products are not found
initially in the accelerated CRs) are:
\begin{equation}
 n_i(t,t') = q_i \exp(-\Delta t/\tau_e)\Delta t
\end{equation}
\begin{equation}
 n_j(t,t') = q_j \left[ \exp\left(-{\Delta t\over\tau_e}\right) -
 \exp\left(-{\Delta t \over \tau_\mathrm{eff}}\right)\right] {\tau_e \tau_\mathrm{eff}
 \over \tau_e - \tau_\mathrm{eff}}
\end{equation}
with $\tau_\mathrm{eff}^{-1} \equiv \tau_r^{-1} + \tau_{e}^{-1}$, 
and $\Delta t \equiv t-t'$.
If we write eq.~\ref{eq:difres} as:
\begin{equation}
\Phi(x,z=0,t) = \int_{-\infty}^{t} dt' G(x,t,t')
\vartheta_{4}\left( 0,\exp\left[-\tau/(\Delta t+\delta t)\right]\right)
\end{equation}
with $\delta t \equiv \sigma_z^2/(2 D)$, then:
\begin{equation}
 N_i \approx q_i \int_{-\infty}^{t} dt' G(x,t,t') \exp\left[-(t-t'+\delta
 t)/\tau_e\right]
\end{equation}
and
\begin{eqnarray}
 N_j &\approx& q_j \int_{-\infty}^{t} dt' G(x,t,t') {\tau_e \tau_\mathrm{eff} \over
 \tau_e - \tau_\mathrm{eff}} \nonumber \\ & &\times \left[
 \exp\left(-{\Delta t+\delta t\over\tau_e}\right) - \exp\left(-{\Delta
 t+\delta t \over \tau_\mathrm{eff}}\right)\right]. 
\end{eqnarray}

If we define ${\cal R}\equiv {N_j/N_i}$, and compare the above result
to those of the leaky box model, then the effective leaky box decay
time $\tau_{e,lb}$ is related to $\tau_e$ and ${\cal R}$ through:
\begin{equation}
  \tau_{e,\mathrm{ lb,eff}} = \tau_e \left( {1\over {\cal R}} -
  1\right).
\end{equation}
This allows us to calculate the effective age of the CRs that the Be
isotope ratio would correspond to in a ``leaky box'' model.

 \section*{Appendix B: Systematic errors from diffusion and epicycles
 in the solar orbit}

In this work, we compare two completely different types of
measurements of the spiral pattern speed. On one hand, there are the
present day astronomical measurements of the spiral pattern speed. On
the other hand, there are the temporal records of passages through
spiral arms---either as recorded in the ice age reoccurrence, or as
recorded in the variable CRF in Iron meteorites. The two types of
measurements should yield the same result if the solar orbit is stable
at its current radius. This is not the general case however, we are
therefore required to estimate the error that can arise when
comparing the two.

Although collisions that drastically change the solar orbit  are
statistically unexpected, the Solar system undergoes many small angle
deflection such that its orbit parameters slowly diffuse.  Namely, the
``Lagrangian'' record of spiral arm crossings, as measured by the
moving solar system, need not be exactly the same as a test ``Eulerian''
particle would experience when fixed at the current $R_\odot$.

The solar motion itself can be broken down to a circular motion of the
guiding center at radius $R_m$ and an epicyclic motion at frequency
$\kappa$ and radius $a_R$.

According to \citet{Wielen1977} and \citet{Wielen1996}, both amplitudes
undergo diffusion, according to:
\begin{equation}
\Delta R_\mr{m,rms} \equiv \sqrt{\left< \Delta R_m^2 \right>} = \sqrt{D_m 
t}~~;~~D_m = \mr{(0.72~{kpc})^2 / Gyr,}
\end{equation}
and
\begin{equation}
\Delta a_\mr{R,rms} \equiv \sqrt{\left< \Delta a_R^2 \right>} = \sqrt{D_a 
t}~~;~~D_a = \mr{(0.84~kpc)^2 / Gyr},
\end{equation}
for $t>10^8$ yrs.  Also relevant, is the ``diffusion'' along the 
azimuthal angle
direction. This is given by:
\begin{equation}
\Delta S_\mr{rms} \equiv \sqrt{ \left< \Delta S^2 \right>} = \sqrt{D_s t^3}~~;~~D_s = 
\mr{(12.7~kpc)^2 / Gyr^3}.
\end{equation}

If we look at a given time $t$ before present, the uncertainties in
$R_m$ and $S$ will be given by the above formulae. The uncertainty in
$a_R$ is a trifle more complicated. This is because we have the
prior knowledge that at $t_0=4.5$ Gyr BP, $a_{R,0} = 0$ while today it is
$a_{R,1} = 0.6~\mr{kpc}$. To calculate $\Delta a_\mr{R,rms}$, we calculate the
probability of the solar system having a given $a_R$ at time $t$
before present. It is given by:
\def\Gauss{\rm G}
\begin{eqnarray}
 P(a_R,t) & = & \\ & & \nonumber \hskip -8mm  { \Gauss(a_R,\sqrt{D_a (t_0-t)} )
 \Gauss(a_R-a_{R,1},\sqrt{D_a t} ) \over   \int_{-\infty}^{\infty}
 \Gauss(a_R,\sqrt{D_a (t_0-t)} ) \Gauss(a_R-a_{R,1},\sqrt{D_a t} ) d
 a_R}
\end{eqnarray}
with
\begin{equation}
 \Gauss(x,\sigma) \equiv {1\over \sqrt{2 \pi} \sigma} \exp\left(-x^2
 \over 2 \sigma\right).
\end{equation}
Using this, we find that:
\begin{eqnarray}
 \Delta a_{R,\mr{rms}} &=& \sqrt{ \int_{-\infty}^{\infty} a_R^2 P(a_R, t)
 da_R } \\ \nonumber &=& \sqrt{ (t_0-t) (D_a t_{0} t + (t_0-t)
 a_{R,1}^2) \over t_0^2}.
\end{eqnarray}
This result is the expected amplitude of the epicyclic motion as a
function of time before present.

The diffusion processes will translate into two types of errors.  One
is the measurement error in the location of each spiral arm.  The
second is an error arising when calculating the average pattern speed
over several periods.

The main contribution to the error in each spiral arm location is due
to the epicyclic motion of the solar system around the `guiding
center', at a typical period of 200 Myrs. This implies that each
spiral arm crossing is inaccurately predicted. This translates into a phase
shift within the spiral passages. The r.m.s.~of the variation is:
\begin{equation}
\Delta \phi_\mr{{a,rms}} = { \Delta a_\mr{{R,rms}}/\sqrt{2} \over \pi R_\odot \sin i},
\end{equation}
assuming 4 arms inclined at an angle $i$, and that the r.m.s.~of the
``circular'' epicyclic motions is $1/\sqrt{2}$ of the radius.  For a
140 Myr period (which will be a typical value found), we obtain that
the phase error ranges from $\Delta\phi_\mr{rms}=15\%$ today to
$\Delta\phi_\mr{rms}=22\%$ at 800 Myr BP. For a typical spiral arm
crossing period of 140 Myr, this translates into an error of 21 to 31
Myrs respectively.

 If we measure the pattern speed by dividing the duration it takes for
 $N$ spiral passages, several errors appear. First, the solar orbit
 can diffuse in or out. This introduces two errors by itself. The
 first is a phase error. Averaged over $N$ spiral passages, it
 introduces a relative error of:
\begin{equation}
 \Delta \phi_\mr{m,rms,1} = {2  \Delta R_\mr{{m,rms}} \over \pi N R_\odot \tan
 i},
\end{equation}
where $\Delta R_\mr{m,rms}$ is the value expected after $N$ spiral
passages.
This will give an absolute error of:
\begin{eqnarray}
 \nonumber \Delta \Omega_\mr{m,rms,1} &=& \Delta \phi_\mr{m,rms,1} \left|
 \Omega_\odot - \Omega_p \right| \\ & =& \nonumber { 2 \Delta
 R_\mr{m,rms} \left|\Omega_\odot - \Omega_p\right| \over \pi N R_\odot
 \tan i} \\ \nonumber &\approx& 0.4~\omun {\rm~~for~~}t=800~{\rm
 Myr}
\end{eqnarray}
in the relative pattern speed.  The second error from the diffusion of
the solar orbit arises from the fact that at a different orbit, the
pattern speed it different. Again, averaged over $N$ spiral passages,
it introduces an absolute error in the pattern velocity of:
\begin{eqnarray}
 \nonumber \Delta \Omega_\mr{m,rms,2} &=& { \Delta R_\mr{m,rms} 
 \Omega_\odot
 \over 2 R}\\ &\approx& 1.1~\omun {\rm~~for~~} t=800 {\rm ~Myr}.
\end{eqnarray}
A phase error also arises from the diffusion along the orbit:
\begin{eqnarray}
 \nonumber \Delta \Omega_\mr{S,rms} &=& \Delta \phi_\mr{S,rms,1}
 \left|\Omega_\odot -
 \Omega_p\right| = { 2 \Delta S_\mr{rms} \left|\Omega_\odot -
 \Omega_p\right| \over \pi N R_\odot} \\  & \approx & 0.3~\omun {\rm~~for~~} t=800~{\rm Myr}.
\end{eqnarray}
The first two errors are maximally correlated (they would have been
anti correlated if either $\Omega_p > \Omega_\odot$ or $i<0$, but not
both), and they are almost uncorrelated with the last. Therefore, the
total error is:
\begin{eqnarray}
 \nonumber \Delta \Omega_\mr{rms} & =& \sqrt{\left(\Delta
 \Omega_\mr{m,rms,1}+\Delta \Omega_\mr{m,rms,1} \right)^2 + \Delta
 \Omega_\mr{S,rms}^2 } \\ &\approx& 1.5~\omun {\rm~~for~~}t=800~{\rm
 Myr}.
\end{eqnarray}

Some indication to the radial solar migration can be obtained from its
anomalously higher metallicity than its neighbors. This is
statistically not very significant, but it does bias our expectation
for solar diffusion. By comparing the solar metallicity to the
surroundings and taking into account the apparent radial gradient in
metallicity in the Milky Way, \citet{Wielen1996}  found that the sun
has migrated outwards by $\Delta R_0 = 1.9 \pm 0.9$ kpc. Namely, we
expect that the probability for the formation radius $R_0$ of the sun
be given by $\Gauss(R_\odot-R_0-1.9~{\rm kpc}, \sigma_0 =
0.9~{\mr kpc})$. Therefore, the probability at time $t$ before present, to
find the solar system at a radius $R$ is given by:
\begin{equation}
 P(R,t) = {\int_{-\infty}^{\infty}  K(R_0,R,R_\odot,t) dR_0 \over
\int_{-\infty}^{\infty} K(R_0,R,R_\odot,t) dR_0 dR }
\end{equation}
\begin{eqnarray}
 K(R_0,R,R_\odot,t) &\equiv&  \Gauss(R_\odot-R_0-\Delta R_0,
 \sigma_0)  \nonumber \\ & & \Gauss(R-R_0,\sqrt{D_m (t_0-t)}) \nonumber \\ & & \times  \Gauss(R_\odot-R,\sqrt{D_m
 t})
\end{eqnarray}
We then obtain that the expectation value for $R$ at time $t$. It is:
\begin{eqnarray}
 \left< R \right> &=& \int_{-\infty}^{\infty} R P(R,t) dR \\ \nonumber
 &=& R_\odot - { D_m t \Delta R_0 \over D_m t_0 + \sigma_0^2}
 \approx|_{t=0.8~{\rm Gyr}} ~R_\odot - 0.25 ~\mr{kpc}
\end{eqnarray}
and its r.m.s. error at $t=800$ Myr is:
\begin{eqnarray}
 \Delta R_\mr{rms}^2 &\equiv& \left< R^2\right> -\left< R \right>^2 \\
 \nonumber &=& \int_{-\infty}^{\infty} R^2 P(R,t) dR - \left(
 \int_{-\infty}^{\infty} R P(R,t) dR \right)^2 \nonumber \\ &=& D_m t
 \left( 1 + {D_m t \over D_m t_0 + \sigma_0^2} \right)
 \approx|_{t=0.8~{\rm Gyr}} (0.59~{\rm kpc})^2. \nonumber
\end{eqnarray}
Using the previous results, we obtain that for a measurement baseline
spanning 800 Myrs BP until today, the difference between the ``Lagrangian'' 
and ``Eulerian'' measurements is
\begin{equation}
 (\Omega_\odot - \Omega_p)_{\rm Lag} - (\Omega_\odot -
 \Omega_p)_{\rm Euler} \approx 0.54 \pm 1.5~ {\rm km~s^{-1}~kpc^{-1}}.
\end{equation}

%=-=-=-=-=-=-=-=-=-=-=-=-=-=-=-=-=-=-=-=-=-=-=-=-=-=-=-=-=-=-=-=-=-=-=-=-=-=-=-=-=-=-=-=-=-=-=-=-=-=-
\newcommand{\mnras}{{Mon.~Not.~Roy.~Astr.~Soc.}}
\newcommand{\apjs}{{Astrophys.~J. Supp.}}
\newcommand{\apjl}{{Astrophys.~J.~Letters}}
\newcommand{\aj}{{Astron.~J.}}
\renewcommand{\aa}{{Astron.~Astrophys.}}
\newcommand{\ass}{{Astrophys.~Space~Sci.}}
\newcommand{\apss}{{Astrophys.~Space~Sci.}}
\newcommand{\pasp}{{Pub.~Astro.~Soc.~Pac.}}
\newcommand{\aap}{{Astron.~Astrophy.}}
\newcommand{\azh}{{Astronomicheskii Zhurnal}}
\newcommand{\aapr}{{Astron.~Astrophys.~ Rev.}}
\newcommand{\planss}{{Planet.~Space~Sci.}}

% after bibtex'ing:
% more longice.bbl | grep -v url | grep -v URL | grep -v AST | grep -v PHY > temp;\mv temp longice.bbl

\small

\def\bf{}
\def\em{}

\bibliography{CRWpapers,CRWpaps_long,CRWpapers_PRL}

\begin{thebibliography}{}

\bibitem[\protect\citename{{Alvarez} {\em et~al.}, }1980]{Alvarez1980}
{Alvarez}, L.~W., {Alvarez}, W., {Asaro}, F., \& {Michel}, H.~V. 1980.
\newblock {\em Science}, {\bf 208}, 1095.

\bibitem[\protect\citename{{Amaral} \& {Lepine}, }1997]{Amaral1997}
{Amaral}, L.~H., \& {Lepine}, J.~R.~D. 1997.
\newblock {\em \mnras}, {\bf 286}, 885.

\bibitem[\protect\citename{{Avedisova}, }1989]{Avedisova1989}
{Avedisova}, V.~S. 1989.
\newblock {\em Astrophys.}, {\bf 30}, 83.

\bibitem[\protect\citename{{Barry}, }1988]{Barry1988}
{Barry}, D.~C. 1988.
\newblock {\em \apj}, {\bf 334}, 436.

\bibitem[\protect\citename{{Bazilevskaya}, }2000]{Bazil2000}
{Bazilevskaya}, G.~A. 2000.
\newblock {\em Sp. Sci. Rev.}, {\bf 94}, 25.

\bibitem[\protect\citename{{Beer} {\em et~al.}, }2000]{Beer2000}
{Beer}, J., {Mende}, W., \& {Stellmacher}, R. 2000.
\newblock {\em Quat. Sci. Rev.}, {\bf 19}, 403.

\bibitem[\protect\citename{{Begelman} \& {Rees}, }1976]{Begelman1976}
{Begelman}, M.~C., \& {Rees}, M.~J. 1976.
\newblock {\em \nat}, {\bf 261}, 298.

\bibitem[\protect\citename{{Bell}, }1978]{Bell1978}
{Bell}, A.~R. 1978.
\newblock {\em \mnras}, {\bf 182}, 443.

\bibitem[\protect\citename{{Berezinski\u{\i}} {\em et~al.}, }1990]{Berez1990}
{Berezinski\u{\i}}, V.~S., {Bulanov}, S.~V., {Dogiel}, V.~A., \& {Ptuskin},
  V.~S. 1990.
\newblock {\em {Astrophysics of cosmic rays}}.
\newblock Amsterdam: North-Holland, 1990, edited by Ginzburg, V.L.

\bibitem[\protect\citename{{Binney} \& {Tremaine}, }1988]{Binney1988}
{Binney}, J., \& {Tremaine}, S. 1988.
\newblock {\em {Galactic Dynamics}}.
\newblock Princeton University Press, Princeton.

\bibitem[\protect\citename{{Blitz} {\em et~al.}, }1983]{NewMW}
{Blitz}, L., {Fich}, M., \& {Kulkarni}, S. 1983.
\newblock {\em Science}, {\bf 220}, 1233.

\bibitem[\protect\citename{{Christensen-Dalsgaard} {\em et~al.},
  }1974]{Dilke1974}
{Christensen-Dalsgaard}, J., {Dilke}, F.~W.~W., \& {Gough}, D.~O. 1974.
{\em \mnras}, {\bf 169}, 429.

\bibitem[\protect\citename{{Comeron} \& {Torra}, }1991]{Comeron1991}
{Comeron}, F., \& {Torra}, J. 1991.
\newblock {\em \aap}, {\bf 241}, 57.

\bibitem[\protect\citename{{Compton}, }1933]{Compton33}
{Compton}, A.~H. 1933.
\newblock {\em Phys. Rev.}, {\bf 43}, 387.

\bibitem[\protect\citename{{Creze} \& {Mennessier}, }1973]{Creze1973}
{Creze}, M., \& {Mennessier}, M.~O. 1973.
\newblock {\em \aap}, {\bf 27}, 281.

\bibitem[\protect\citename{{Crowell}, }1999]{Crowell1999}
{Crowell}, J.~C. 1999.
\newblock {\em Pre-Mesozoic Ice Ages: Their Bearing on Understanding the
  Climate System}.
\newblock  Vol. 192.
\newblock Memoir Geological Society of America.

\bibitem[\protect\citename{{Dame} {\em et~al.}, }2001]{Dame2001}
{Dame}, T.~M., {Hartmann}, D., \& {Thaddeus}, P. 2001.
\newblock {\em \apj}, {\bf 547}, 792.

\bibitem[\protect\citename{{Dickinson}, }1975]{Dickinson1975}
{Dickinson}, R.~E. 1975.
\newblock {\em Bull. Amer. Met. Soc.}, {\bf 56}, 1240.

\bibitem[\protect\citename{{Dilke} \& {Gough}, }1972]{Dilke1972}
{Dilke}, F.~W.~W., \& {Gough}, D.~O. 1972.
\newblock {\em \nat}, {\bf 240}, 262.

\bibitem[\protect\citename{{Dopita} {\em et~al.}, }1997]{Dopita1997}
{Dopita}, M.~A. {et al.}
  1997.
\newblock {\em \apj}, {\bf 474}, 188.

\bibitem[\protect\citename{{Dragicevich} {\em et~al.}, }1999]{Drag1999}
{Dragicevich}, P.~M., {Blair}, D.~G., \& {Burman}, R.~R. 1999.
\newblock {\em \mnras}, {\bf 302}, 693.

\bibitem[\protect\citename{{Duric}, }2000]{SNRs}
{Duric}, N. 2000.
\newblock Supernova remnants and cosmic rays in M31 and M33.
\newblock {\em Pages  179--186 of:} {\em Proceedings 232. WE-Heraeus Seminar,
  22-25 May 2000, Bad Honnef, Germany. Edited by Elly M. Berkhuijsen, Rainer
  Beck, and Rene A. M. Walterbos. Shaker, Aachen, 2000, p. 179}.

\bibitem[\protect\citename{{Efremov}, }1983]{Efremov1983}
{Efremov}, Y.~N. 1983.
\newblock {\em Sov. Astron. Lett.}, {\bf 9}, 51.

\bibitem[\protect\citename{{Egorova} {\em et~al.}, }2000]{Egorova}
{Egorova}, L.~Y., {Vovk}, V.~Ya., \& {Troshichev}, O.~A. 2000.
\newblock {\em J.~Atmos. Solar-Terr.~Phys.}, {\bf 62}, 955.

\bibitem[\protect\citename{{Elmegreen}, }1998]{Elmegreenbook}
{Elmegreen}, D.~M. 1998.
\newblock {\em {Galaxies and Galactic Structure}}.
\newblock Prentice Hall.

\bibitem[\protect\citename{{Evans}, }1991]{Evans1991}
{Evans}, N.~J. 1991.
\newblock in {\em ASP Conf. Ser. 20: Frontiers of Stellar
  Evolution}, p. 45.

\bibitem[\protect\citename{{Fastrup} {\em et~al.}, }2001]{CloudExperiment}
{Fastrup}, B. {et al.}
  2001.
\newblock A study of the link between cosmic rays and clouds with a cloud
  chamber at the CERN PS.
\newblock {\em CERN-SPSC-2000-021}.
\newblock Also as: LANL physics/0104048.

\bibitem[\protect\citename{{Feitzinger} \& {Schwerdtfeger}, }1982]{Feitz82}
{Feitzinger}, J.~V., \& {Schwerdtfeger}, H. 1982.
\newblock {\em \aap}, {\bf 116}, 117.

\bibitem[\protect\citename{{Fern{\' a}ndez} {\em et~al.}, }2001]{Fernandez2001}
{Fern{\' a}ndez}, D., {Figueras}, F., \& {Torra}, J. 2001.
\newblock {\em \aap}, {\bf 372}, 833.

\bibitem[\protect\citename{{Frakes} {\em et~al.}, }1992]{Frakes1992}
{Frakes}, L.~A., {Francis}, E., \& {Syktus}, J.~I. 1992.
\newblock {\em {Climate modes of the Phanerozoic; the history of the Earth's
  climate over the past 600 million years}}.
\newblock Cambridge: Cambridge University Press, 1992.

\bibitem[\protect\citename{{Friis-Christensen} \& {Lassen}, }1991]{Christ1991}
{Friis-Christensen}, E., \& {Lassen}, K. 1991.
\newblock {\em Science}, {\bf 254}, 698.

\bibitem[\protect\citename{{Gallagher} {\em et~al.}, }1996]{Gallagher1996}
{Gallagher}, J.~S. {et al.}
  1996.
\newblock {\em \apj}, {\bf 466}, 732.

\bibitem[\protect\citename{{Gardiner} {\em et~al.}, }1994]{Gardiner1994}
{Gardiner}, L.~T., {Sawa}, T., \& {Fujimoto}, M. 1994.
\newblock {\em \mnras}, {\bf 266}, 567.

\bibitem[\protect\citename{{Georgelin} \& {Georgelin}, }1976]{Georgelin1976}
{Georgelin}, Y.~M., \& {Georgelin}, Y.~P. 1976.
\newblock {\em \aap}, {\bf 49}, 57.

\bibitem[\protect\citename{{Gordon}, }1978]{Gordon1978}
{Gordon}, M.~A. 1978.
\newblock {\em \apj}, {\bf 222}, 100.

\bibitem[\protect\citename{{Grivnev}, }1981]{Grivnev1981}
{Grivnev}, E.~M. 1981.
\newblock {\em Sov. Astron. Lett.}, {\bf 7}, 303.

\bibitem[\protect\citename{{Grivnev}, }1983]{Grivnev1983}
{Grivnev}, E.~M. 1983.
\newblock {\em Sov. Astron. Lett.}, {\bf 9}, 287.

\bibitem[\protect\citename{{Guyodo} \& {Valet}, }1996]{Guyodo1996}
{Guyodo}, Y., \& {Valet}, J. 1996.
\newblock {\em Earth Planet. Sci. Lett.}, {\bf 143}, 23.

\bibitem[\protect\citename{{Haigh}, }1996]{Haigh1996}
{Haigh}, J.~D. 1996.
\newblock {\em Science}, {\bf 272}, 981.

\bibitem[\protect\citename{{Hambrey} \& {Harland}, }1985]{Hambrey1985}
{Hambrey}, M.~J., \& {Harland}, W.~B. 1985.
\newblock {\em Palaeogeography, Palaeoclimatorlogy, Palaeoecology}, {\bf 51},
  255.

\bibitem[\protect\citename{{Hampel} \& {Schaeffer}, }1979]{Hampel1979}
{Hampel}, W., \& {Schaeffer}, O.~A. 1979.
\newblock {\em Earth. Planet. Sci. Lett.}, {\bf 42}, 348.

\bibitem[\protect\citename{{Harrison}, }2000]{Harrison2000}
{Harrison}, R.~G. 2000.
\newblock {\em Sp. Sci. Rev.}, {\bf 94}, 381.

\bibitem[\protect\citename{{Hernandez} {\em et~al.}, }2000]{HippSFR}
{Hernandez}, X., {Valls-Gabaud}, D., \& {Gilmore}, G. 2000.
\newblock {\em \mnras}, {\bf 316}, 605.

\bibitem[\protect\citename{{Herschel}, }1796]{Herschel}
{Herschel}, W. 1796.
\newblock {\em Phil. Trans. Roy. Soc., London},
  166.

\bibitem[\protect\citename{{Herschel}, }1801]{Herschel2}
{Herschel}, W. 1801.
\newblock {\em Phil. Trans. Roy. Soc., London, Part
  1}, 265.

\bibitem[\protect\citename{{Heyer} {\em et~al.}, }1998]{Heyer1998}
{Heyer}, M.~H., {Brunt}, C., {Snell}, R.~L., {Howe}, J.~E., {Schloerb}, F.~P.,
  \& {Carpenter}, J.~M. 1998.
\newblock {\em \apjs}, {\bf 115}, 241.

\bibitem[\protect\citename{{Hodell} {\em et~al.}, }2001]{Hodell2001}
{Hodell}, D.~A., {Brenner}, M., {Curtis}, J.~H., \& {Guilderson}, T. 2001.
\newblock {\em Science}, {\bf 292}, 1367.

\bibitem[\protect\citename{{Hoffman} {\em et~al.}, }1995]{Hoffman1995}
{Hoffman}, P.~F., {Kaufman}, A.~J., {Halverson}, G.~P., \& {Schrag}, D.~P.
  1995.
\newblock {\em Science}, {\bf 281}, 1342.

\bibitem[\protect\citename{{Hoyle} \& {Lyttleton}, }1939]{Hoyle1939}
{Hoyle}, F., \& {Lyttleton}, R.~A. 1939.
\newblock {\em Proc. Cambridge Phil. Soc.}, {\bf 35}, 405.

\bibitem[\protect\citename{{Hoyle} \& {Wickramasinghe}, }1978]{Hoyle1978}
{Hoyle}, F., \& {Wickramasinghe}, C. 1978.
\newblock {\em \apss}, {\bf 53}, 523.

\bibitem[\protect\citename{{Huber}, }1998]{Huber1998}
{Huber}, B.~T. 1998.
\newblock {\em Science}, {\bf 282}, 2199.

\bibitem[\protect\citename{{IPCC}, }1995]{IPCC1995}
{IPCC}. 1995.
\newblock {\em Climate Change 1995, Intergovernmental Panel on Climate Change}.
\newblock Cambridge University Press.

\bibitem[\protect\citename{{Ivanov}, }1983]{Ivanov1983}
{Ivanov}, G.~R. 1983.
\newblock {\em Pis ma Astronomicheskii Zhurnal}, {\bf 9}(Apr.), 200.

\bibitem[\protect\citename{{Kirkby} \& {Laaksonen}, }2000]{Kirkby2000}
{Kirkby}, J., \& {Laaksonen}, A. 2000.
\newblock {\em Sp. Sci. Rev.}, {\bf 94}, 397.

\bibitem[\protect\citename{{Labitzke} \& {van Loon}, }1992]{Labitzke1992}
{Labitzke}, K., \& {van Loon}, H. 1992.
\newblock {\em J. Clim.}, {\bf 5}, 240.

\bibitem[\protect\citename{{Lavielle} {\em et~al.}, }1999]{Lavielle1999}
{Lavielle}, B., {Marti}, K., {Jeannot}, J., {Nishiizumi}, K., \& {Caffee}, M.
  1999.
\newblock {\em Earth Planet. Sci. Lett.}, {\bf 170}, 93.

\bibitem[\protect\citename{{Lees} \& {Lo}, }1990]{Lees1990}
{Lees}, J.~F., \& {Lo}, K.~Y. 1990 (July).
\newblock Dissociation and ionization of molecular gas in the spiral arms of
  M51.
\newblock in {\em NASA, Ames Research Center, The
  Interstellar Medium in External Galaxies: Summaries of Contributed Papers, p
  296 (SEE N91-14100 05-90)}.

\bibitem[\protect\citename{{Leitch} \& {Vasisht}, }1998]{Leitch1998}
{Leitch}, E.~M., \& {Vasisht}, G. 1998.
\newblock {\em New Astron.}, {\bf 3}, 51.

\bibitem[\protect\citename{{Leitherer} {\em et~al.}, }1999]{starburst}
{Leitherer}, {et al.}
   1999.
\newblock {\em \apjs}, {\bf 123}, 3.

\bibitem[\protect\citename{{Lin} {\em et~al.}, }1995]{Lin1995}
{Lin}, D.~N.~C., {Jones}, B.~F., \& {Klemola}, A.~R. 1995.
\newblock {\em \apj}, {\bf 439}, 652.

\bibitem[\protect\citename{{Lisenfeld} {\em et~al.}, }1996]{Lisenfeld1996}
{Lisenfeld}, U., {Alexander}, P., {Pooley}, G.~G., \& {Wilding}, T. 1996.
\newblock {\em \mnras}, {\bf 281}, 301.

\bibitem[\protect\citename{{Lo} {\em et~al.}, }1987]{Lo1987}
{Lo}, K.~Y., {Ball}, R., {Masson}, C.~R., {Phillips}, T.~G., {Scott}, S., \&
  {Woody}, D.~P. 1987.
\newblock {\em \apjl}, {\bf 317}, L63.

\bibitem[\protect\citename{{Loktin} {\em et~al.}, }1994]{OpenClusters}
{Loktin}, A.~V., {Matkin}, N.~V., \& {Gerasimenko}, T.~P. 1994.
\newblock {\em Astron. Astrophys. Trans.}, {\bf 4}, 153.

\bibitem[\protect\citename{{Longair}, }1994]{Longair1994}
{Longair}, M.~S. 1994.
\newblock {\em {High energy astrophysics. Vol.2: Stars, the galaxy and the
  interstellar medium}}.
\newblock Cambridge: Cambridge University Press, 1994, 2nd ed.

\bibitem[\protect\citename{{Lukasiak} {\em et~al.}, }1994]{Lukasiak1994}
{Lukasiak}, A., {Ferrando}, P., {McDonald}, F.~B., \& {Webber}, W.~R. 1994.
\newblock {\em \apj}, {\bf 423}, 426.

\bibitem[\protect\citename{{Lynds}, }1980]{Lynds1980}
{Lynds}, B.~T. 1980.
\newblock {\em \aj}, {\bf 85}, 1046.

\bibitem[\protect\citename{{Marsh} \& {Svensmark}, }2000]{Marsh2000}
{Marsh}, N., \& {Svensmark}, H. 2000.
\newblock {\em Sp. Sci. Rev.}, {\bf 94}, 215.

\bibitem[\protect\citename{{Michel} {\em et~al.}, }1995]{Michel1995}
{Michel}, R. {et al.}
1995.
\newblock {\em Nuc. Inst. Meth. Phys. Res. B}, {\bf
  103}, 183.

\bibitem[\protect\citename{{Mishurov} {\em et~al.}, }1979]{Mishurov1979}
{Mishurov}, I.~N., {Pavlovskaia}, E.~D., \& {Suchkov}, A.~A. 1979.
\newblock {\em \azh}, {\bf 56}, 268.

\bibitem[\protect\citename{{Mishurov} \& {Zenina}, }1999]{Mishurov1999}
{Mishurov}, Y.~N., \& {Zenina}, I.~A. 1999.
\newblock {\em \aap}, {\bf 341}, 81.

\bibitem[\protect\citename{{Napier} \& {Clube}, }1979]{Napier1979}
{Napier}, W.~M., \& {Clube}, S. V.~M. 1979.
\newblock {\em \nat}, {\bf 282}, 455.

\bibitem[\protect\citename{{Neff} {\em et~al.}, }2001]{Neff2001}
{Neff}, U., {Burns}, S.~J., {Mangnini}, A., {Mudelsee}, M., {Fleitmann}, D., \&
  {Matter}, A. 2001.
\newblock {\em Nature}, {\bf 411}, 290.

\bibitem[\protect\citename{{Nelson} \& {Matsuda}, }1977]{Nelson1977}
{Nelson}, A.~H., \& {Matsuda}, T. 1977.
\newblock {\em \mnras}, {\bf 179}, 663.

\bibitem[\protect\citename{{Ney}, }1959]{Ney1959}
{Ney}, E.~P. 1959.
\newblock {\em Nature}, {\bf 183}, 451.

\bibitem[\protect\citename{{Olling} \& {Merrifield}, }1998]{Olling1998}
{Olling}, R.~P., \& {Merrifield}, M.~R. 1998.
\newblock {\em \mnras}, {\bf 297}, 943.

\bibitem[\protect\citename{{Palle Bago} \& {Butler}, }2000]{SecondAnalysis}
{Palle Bago}, E., \& {Butler}, J. 2000.
\newblock {\em Astron. \& Geophys.}, {\bf 41}, 18.

\bibitem[\protect\citename{{Palous} {\em et~al.}, }1977]{Palous1977}
{Palous}, J., {Ruprecht}, J., {Dluzhnevskaia}, O.~B., \& {Piskunov}, T. 1977.
\newblock {\em \aap}, {\bf 61}, 27.

\bibitem[\protect\citename{{Perko}, }1987]{Perko1987}
{Perko}, J.~S. 1987.
\newblock {\em \aap}, {\bf 184}, 119.

\bibitem[\protect\citename{{Pudovkin} \& {Veretenenko}, }1995]{Pudovkin}
{Pudovkin}, M.~I., \& {Veretenenko}, S.~V. 1995.
\newblock {\em J Atmos. Terr. Phys.}, {\bf
  57}, 1349.

\bibitem[\protect\citename{{Rind} \& {Overpeck}, }1993]{Rind1993}
{Rind}, D., \& {Overpeck}, J. 1993.
\newblock {\em Quat.~Sci.~Rev.}, {\bf 12}, 357.

\bibitem[\protect\citename{{Robin} {\em et~al.}, }1992]{Robin1992}
{Robin}, A.~C., {Creze}, M., \& {Mohan}, V. 1992.
\newblock {\em \apjl}, {\bf 400}, L25.

\bibitem[\protect\citename{{Rocha-Pinto} {\em et~al.}, }2000a]{RochaPinto2000}
{Rocha-Pinto}, H.~J., {Scalo}, J., {Maciel}, W.~J., \& {Flynn}, C. 2000a.
\newblock {\em \aap}, {\bf 358}, 869.

\bibitem[\protect\citename{{Rocha-Pinto} {\em et~al.}, }2000b]{RochaPinto2000a}
{Rocha-Pinto}, H.~J., {Scalo}, J., {Maciel}, W.~J., \& {Flynn}, C. 2000b.
\newblock {\em \apjl}, {\bf 531}, L115.

\bibitem[\protect\citename{{Ruphy} {\em et~al.}, }1996]{Ruphy1996}
{Ruphy}, S., {Robin}, A.~C., {Epchtein}, N., {Copet}, E., {Bertin}, E.,
  {Fouque}, P., \& {Guglielmo}, F. 1996.
\newblock {\em \aap}, {\bf 313}, L21.

\bibitem[\protect\citename{{Scalo}, }1987]{Scalo1987}
{Scalo}, J.~M. 1987.
\newblock in {\em Starbursts and Galaxy Evolution}, p. 445.

\bibitem[\protect\citename{{Schaeffer} {\em et~al.}, }1981]{Sch1981}
{Schaeffer}, O.~A., {Nagel}, K., {Fechtig}, H., \& {Neukum}, G. 1981.
\newblock {\em \planss}, {\bf 29}, 1109.

\bibitem[\protect\citename{{Shaviv}, }2002]{Shaviv2001}
{Shaviv}, N.~J. 2002.
\newblock {\em \prl}, {\bf 89}, 051102.

\bibitem[\protect\citename{{Simpson} \& {Garcia-Munoz}, }1988]{Simpson1988}
{Simpson}, J.~A., \& {Garcia-Munoz}, M. 1988.
\newblock {\em Sp. Sci. Rev.}, {\bf 46}, 205.

\bibitem[\protect\citename{{Singer}, }1954]{Singer1954}
{Singer}, S.~F. 1954.
\newblock {\em \apj}, {\bf 119}, 291.

\bibitem[\protect\citename{{Soon} {\em et~al.}, }1996]{Soon1996}
{Soon}, W.~H., {Posmentier}, E.~S., \& {Baliunas}, S.~L. 1996.
\newblock {\em \apj}, {\bf 472}, 891.

\bibitem[\protect\citename{{Soon} {\em et~al.}, }2000]{Soon2000}
{Soon}, W.~H., {Posmentier}, E.~S., \& {Baliunas}, S.~L. 2000.
\newblock {\em Annales Geophysicae}, {\bf 18}, 583.

\bibitem[\protect\citename{{Steiner} \& {Grillmair}, }1973]{SG1973}
{Steiner}, J., \& {Grillmair}, E. 1973.
\newblock {\em Geol.~Soc.~Am.~Bull.}, {\bf 84}, 1003.

\bibitem[\protect\citename{{Stozhkov} \& {et al.}, }1995]{Lebvedev}
{Stozhkov}, Yu.~I., \& {et al.} 1995.
\newblock {\em Il Nuovo Cimento C}, {\bf 18}, 335.

\bibitem[\protect\citename{{Svensmark}, }1998]{Sven1998}
{Svensmark}, H. 1998.
\newblock {\em \prl}, {\bf 81}, 5027.

\bibitem[\protect\citename{{Svensmark} \& {Friis-Christensen}, }1997]{Sven1997}
{Svensmark}, H., \& {Friis-Christensen}, E. 1997.
\newblock {\em J. Atmos. Terr. Phys.}, {\bf 59},
  1225.

\bibitem[\protect\citename{{Tammann} {\em et~al.}, }1994]{Tammann1994}
{Tammann}, G.~A., {L\"offler}, W., \& {Schr\"oder}, A. 1994.
\newblock {\em \apjs}, {\bf 92}, 487.

\bibitem[\protect\citename{{Taylor} \& {Cordes}, }1993]{Taylor1993}
{Taylor}, J.~H., \& {Cordes}, J.~M. 1993.
\newblock {\em \apj}, {\bf 411}, 674.

\bibitem[\protect\citename{{Tinsley} \& {Deen}, }1991]{Tinsley1991}
{Tinsley}, B.~A., \& {Deen}, G.~W. 1991.
\newblock {\em J.~Geophys. Res.}, {\bf 12}, 22,283.

\bibitem[\protect\citename{{Vall{\' e}e}, }1995]{Vallee}
{Vall{\' e}e}, J.~P. 1995.
\newblock {\em \apj}, {\bf 454}, 119.

\bibitem[\protect\citename{{Vall{\' e}e}, }2002]{Vallee2002}
{Vall{\' e}e}, J.~P. 2002.
\newblock {\em \apj}, {\bf 566}, 261.

\bibitem[\protect\citename{{Vallenari} {\em et~al.}, }1996]{Vallenari1996}
{Vallenari}, A., {Chiosi}, C., {Bertelli}, G., \& {Ortolani}, S. 1996.
\newblock {\em \aap}, {\bf 309}, 358.

\bibitem[\protect\citename{{van den Bergh} \& {McClure}, }1994]{vandBergh1994}
{van den Bergh}, S., \& {McClure}, R.~D. 1994.
\newblock {\em \apj}, {\bf 425}, 205.

\bibitem[\protect\citename{{Veizer} {\em et~al.}, }2000]{Veizer}
{Veizer}, J., {Godderis}, Y., \& {Francois}, L.~M. 2000.
\newblock {\em Nature}, {\bf 408}, 698.

\bibitem[\protect\citename{{Voshage}, }1967]{Voshage0}
{Voshage}, H. 1967.
\newblock {\em Zeitschrift Naturforschung Teil A}, {\bf 22}, 477.

\bibitem[\protect\citename{{Voshage} \& {Feldmann}, }1979]{Voshage1}
{Voshage}, H., \& {Feldmann}, H. 1979.
\newblock {\em Earth Planet. Sci. Lett.}, {\bf 45}, 293.

\bibitem[\protect\citename{{Voshage} {\em et~al.}, }1983]{Voshage2}
{Voshage}, H., {Feldmann}, H., \& {Braun}, O. 1983.
\newblock {\em Zeitschrift Naturforschung Teil A}, {\bf 38}, 273.

\bibitem[\protect\citename{{Wagner} {\em et~al.}, }1999]{Wagner1999}
{Wagner}, G., {Beer}, J., {Kubik}, P.~W., \& {Synal}, H.-A. 1999.
\newblock in Annul Report of the Paul Scherrer Institute, Swiss Federal
  Institute of Technology (ETH), Zurich.

\bibitem[\protect\citename{{Webber} \& {Soutoul}, }1998]{Webber1998}
{Webber}, W.~R., \& {Soutoul}, A. 1998.
\newblock {\em \apj}, {\bf 506}, 335.

\bibitem[\protect\citename{{Westerlund}, }1990]{Westerlund1990}
{Westerlund}, B.~E. 1990.
\newblock {\em \aapr}, {\bf 2}, 29.

\bibitem[\protect\citename{{Wiedenbeck} \& {Greiner}, }1980]{Wieden1980}
{Wiedenbeck}, M.~E., \& {Greiner}, D.~E. 1980.
\newblock {\em \apjl}, {\bf 239}, L139.

\bibitem[\protect\citename{{Wielen}, }1977]{Wielen1977}
{Wielen}, R. 1977.
\newblock {\em \aap}, {\bf 60}, 263.

\bibitem[\protect\citename{{Wielen} {\em et~al.}, }1996]{Wielen1996}
{Wielen}, R., {Fuchs}, B., \& {Dettbarn}, C. 1996.
\newblock {\em \aap}, {\bf 314}, 438.

\bibitem[\protect\citename{{Williams}, }1975]{Williams1975}
{Williams}, G.~E. 1975.
\newblock {\em Earth Planet. Sci. Lett.}, {\bf 26}, 361.

\bibitem[\protect\citename{{Yabushita} \& {Allen}, }1985]{Yabushita1985}
{Yabushita}, S., \& {Allen}, A.~J. 1985.
\newblock {\em The Observatory}, {\bf 105}, 198.

\bibitem[\protect\citename{{Yu}, }2002]{Yu}
{Yu}, F. 2002.
\newblock {\em J.~Geophys.~Res.}, {\bf in press}, .

\bibitem[\protect\citename{{Yuan}, }1969a]{Yuan1969a}
{Yuan}, C. 1969a.
\newblock {\em \apj}, {\bf 158}, 871.

\bibitem[\protect\citename{{Yuan}, }1969b]{Yuan1969b}
{Yuan}, C. 1969b.
\newblock {\em \apj}, {\bf 158}, 889.

\bibitem[\protect\citename{{Zachos} {\em et~al.}, }2001]{Zachos2001}
{Zachos}, J., {Pagani}, M., {Sloan}, L., {Thomas}, E., \& {Billups}, K. 2001.
\newblock {\em Science}, {\bf 292}, 686.

\end{thebibliography}

\end{document}